\documentclass[11pt]{article}
\usepackage[dvipsnames]{xcolor}
\usepackage[colorlinks=true,citecolor=ForestGreen,urlcolor=blue,linkcolor=blue]{hyperref}

\usepackage{cite}
\usepackage{comment}
\usepackage{amsfonts}
\usepackage{amsmath,amssymb}
\usepackage{dsfont}
\usepackage{mathrsfs} 
\usepackage[active]{srcltx}
\usepackage[all]{xy}
\input diagxy

\usepackage[a4paper, left=23mm, right=23mm, top=25mm, bottom=20mm]{geometry}

\usepackage[nottoc]{tocbibind}
\usepackage[colorinlistoftodos]{todonotes} 

\usepackage{enumerate}

\numberwithin{equation}{section}

\usepackage{mathtools}

\font\fr=eufm10 scaled \magstep 1 
\def\beq{\begin{equation}}
\def\eeq{\end{equation}}
\def\bea{\begin{eqnarray}}
\def\eea{\end{eqnarray}}
\def\beann{\begin{eqnarray*}}
\def\eeann{\end{eqnarray*}}
\def\beasn{\begin{sneqnarray}}
\def\eeasn{\end{sneqnarray}}
\def\ben{\begin{enumerate}}
\def\een{\end{enumerate}}
\def\bit{\begin{itemize}}
\def\eit{\end{itemize}}
\newcommand{\ds}{\displaystyle}
\def\derpar#1#2{\frac{\partial{#1}}{\partial{#2}}}

\def\bm#1{\hbox{\boldmath$#1$}}

\def\qed{\ifvmode\Realemovelastskip\fi
{\unskip\nobreak\hfil\penalty50\hbox{}\nobreak\hfil \hbox{\vrule
height1.2ex width1.2ex}\parfillskip=0pt \finalhyphendemerits=0
\par\smallskip}}

\def\vf{\mbox{\fr X}}
\def\df{{\mit\Omega}}
\def\Lag{{\cal L}}
\def\L{{\cal L}}

\def\d{{\rm d}}
\def\H{{\cal H}}

\def\Nat{\mathbb{N}}

\def\Real{\mathbb{R}}

\newcommand{\Cinfty}{\mathscr{C}^\infty}
\def\Tan{{\rm T}}

\newcommand{\innp}[1]{\iota\left({#1}\right)}
\def\tabaddress#1{{\small\it\begin{tabular}[t]{c}#1
\\begin{equation}1.2ex]\end{tabular}}}

\newcommand*{\dd}{\mathrm{d}}

\newcommand*{\F}{\mathcal{F}}
\renewcommand*{\P}{\mathcal{P}}

\makeatletter
\DeclareOldFontCommand{\rm}{\normalfont\rmfamily}{\mathrm}
\DeclareOldFontCommand{\sf}{\normalfont\sffamily}{\mathsf}
\DeclareOldFontCommand{\tt}{\normalfont\ttfamily}{\mathtt}
\DeclareOldFontCommand{\bf}{\normalfont\bfseries}{\mathbf}
\DeclareOldFontCommand{\it}{\normalfont\itshape}{\mathit}
\DeclareOldFontCommand{\sl}{\normalfont\slshape}{\@nomath\sl}
\DeclareOldFontCommand{\sc}{\normalfont\scshape}{\@nomath\sc}
\makeatother

\usepackage[english]{babel}
\usepackage{microtype}
\usepackage{csquotes}
\usepackage{amsthm}
\usepackage{mathtools}
\usepackage{amssymb,amsfonts,stmaryrd}
\usepackage{tikz}
\usetikzlibrary{cd}
\usepackage{xparse}

\RequirePackage{hyperref}

\theoremstyle{plain}
\newtheorem{theorem}{Theorem}[section]

\newtheorem{proposition}[theorem]{Proposition}

\theoremstyle{definition}
\newtheorem{definition}[theorem]{Definition}
\newtheorem{remark}[theorem]{Remark}

\newcommand*{\R}{\mathbb{R}}

\renewcommand{\L}{\mathcal{L}}
\newcommand{\bfX}{\mathbf{X}}

\newcommand{\X}{\mathfrak{X}}
\renewcommand{\d}{\mathrm{d}}
\newcommand*{\bd}{\overline{\mathrm{d}}}

\newcommand{\parder}[2]{\frac{\partial #1}{\partial #2}}
\newcommand{\dparder}[2]{\dfrac{\partial #1}{\partial #2}}

\newcommand{\parderr}[3]{\frac{\partial^2 #1}{\partial #2\partial #3}}

\title{\sc Practical Introduction to 
\\
Action-Dependent Field Theories}

\author{\sffamily 
\sc $^a$Manuel de Le\'on
\thanks{mdeleon@icmat.es\quad ORCID: 0000-0002-8028-2348}\, ,
$^b$Jordi Gaset Rif\`a
\thanks{jordi.gaset@cunef.edu\quad ORCID: 0000-0001-8796-3149}\, ,
\sc $^c$Miguel C. Mu\~noz-Lecanda
\thanks{miguel.carlos.munoz@upc.edu\quad ORCID: 0000-0002-7037-0248} \, , \\
\sc $^d$Xavier Rivas
\thanks{xavier.rivas@urv.cat\quad ORCID: 0000-0002-4175-5157}\, ,
\sc $^c$Narciso Rom\'an-Roy
\thanks{narciso.roman@upc.edu\quad ORCID: 0000-0003-3663-9861}\, .
\\[1ex]
\normalsize\itshape\sffamily 
$^a$Instituto de Ciencias Matem\'aticas,
Consejo Superior de Investigaciones Cient\'ificas\\
\normalsize\itshape\sffamily 
and Real Academia de Ciencias, Madrid, Spain.
\\[1ex]
\normalsize\itshape\sffamily 
$^b$Department of Mathematics,
CUNEF Univ.,
Madrid, Spain.
\\[1ex]
\normalsize\itshape\sffamily 
$^c$Department of Mathematics,
Univ. Polit\`ecnica de Catalunya,
Barcelona, Spain.
\\[1ex]
\normalsize\itshape\sffamily 
$^d$Dept. Computer Engineering and Mathematics, Univ. Rovira i Virgili, Tarragona, Spain.
}

\begin{document}

\maketitle

\begin{abstract}
Action-dependent field theories are systems where the Lagrangian or Hamiltonian depends on new variables that encode the action. They model a larger class of field theories, including non-conservative behavior, while maintaining a well-defined notion of symmetries and a Noether theorem. This makes them especially suited for open systems.
After a conceptual introduction, we make a quick presentation of a new mathematical framework for action-dependent field theory: {\sl multicontact geometry}.
The formalism is illustrated with a variety of action-dependent Lagrangians, some of which are regular and others singular, derived from well-known theories whose Lagrangians have been modified to incorporate action-dependent terms. Detailed computations are provided, including the constraint algorithm for the singular cases, in both the Lagrangian and Hamiltonian formalisms.
These are the one-dimensional wave equation, the Klein--Gordon equation and the telegrapher equation, Maxwell’s electromagnetism, 
Metric-affine gravity, the heat equation and Burgers' equation, the Bosonic string theory, and $(2+1)$-dimensional gravity and Chern--Simons equation.
\end{abstract}


\noindent\textbf{MSC\,(2020):}
{\sl Primary:}
70S05, 
70S10,
53D10
35R01. \\
\indent\indent\indent\indent\
{\sl Secondary:}
35Q99, 53C15, 53Z05, 58A10.

\noindent \textbf{PACS\,(2022):}
03.50.-z, 11.10.Ef, 02.40.-k, 04.50.Kd, 04.20.Fy, 11.25.-w, 11.30.-j, 02.30.Jr.

\pagestyle{myheadings}
\markright{\small\itshape\sffamily 
{\rm M. de Le\'on} {\it et al} ---
Practical introduction to action-dependent field theories}

\newpage

{\setcounter{tocdepth}{2}
\def\baselinestretch{1}
\small
\def\addvspace#1{\vskip 1pt}
\parskip 0pt plus 0.1mm
\tableofcontents
}

\section{Introduction}

An action-dependent field theory is a system where the Lagrangian or the Hamiltonian depends on new variables that represent the flow of action. They are a natural generalization of classical field theories with an array of appealing properties that make them exceptionally interesting.

To begin with, action-dependent field theories provide terms in the field equations that cannot be generated with standard classical field theories, even considering the addition of new particles. These terms have been used to model dissipative behavior, but their vast potential goes far beyond. Moreover, the new variables related to the action create a bridge between the boundary and the bulk.

In addition, action-dependent field theories are non-conservative, which makes them especially suited for open systems. Remarkably, the conservation failure is tightly controlled. Noether's theorem relates symmetries to dissipated quantities, which are not conserved but have a very specific behavior. This is encoded by a differential $1$-form (called dissipation form) that is easily computed from the Lagrangian or Hamiltonian, giving a clear control over the non-conservation of the system.

Importantly, action-dependent field theories are grounded in a rigorous geometric formulation: {\sl multicontact geometry}. A geometric description also provides the generality to incorporate singular and higher-order theories, as well as the underlying bracket structure, a prerequisite for quantization. The benefits of such a mathematical approach are profuse, as evidenced by the successful history of symplectic geometry.

\subsection*{Symplectic mechanics}

Mechanics experienced a drastic change as soon as it was able to use symplectic geometry in its description. This occurred in the 50's and 60's of the last century, and made it possible to obtain Hamilton's equations as the integral curves of a vector field on a symplectic manifold, in fact, in the cotangent bundle of the configuration space of the system. This implied a liberation from coordinates and the possibility of obtaining the usual properties of mechanical systems (conservation of energy, other conserved quantities, integration, \ldots) in a simple and elegant way. As it was not only a different way of approaching the study of mechanical systems, but this formulation served to go further in this task. For example, the definition of the momentum map made possible to create a bridge between geometry and dynamics with algebra and combinatorics through the use of polytopes \cite{Atiyah}, obtained as images of the phase space in the dual of the Lie algebra of the Lie group of symmetries of the system.

Another subject in which symplectic geometry has been used to the benefit of mechanics is integration problems, by exploiting the symmetries of the system in the so-called {\sl Meyer--Marsden--Weinstein Reduction Theorem} \cite{MW}. Or, for example, the development of the {\sl Hamilton--Jacobi equation} \cite{AM-78}, which has led to a better understanding of the problem and extensions of the procedure to more general situations. It has also been possible to study integrability problems and to develop the {\sl Liouville--Arnold Theorem} \cite{Ar}, with the corresponding action-angle coordinates. Other important developments achieved through this geometric vision are the applications to numerical integration (geometric integrators), the study of {\sl Noether's Theorem} and its generalizations developing the different types of infinitesimal symmetries and the corresponding conserved quantities \cite{N2020}, or the applications to geometric quantization \cite{EMRV}. Another subject that has received a boost from the geometric description is systems subject to non-holonomic constraints \cite{LD-1996}, which have been extensively studied since the 1980s in this new light. This has made it possible, for example, to obtain an appropriate Hamilton--Jacobi equation.

The above items are just some of the milestones that symplectic geometry has achieved in the study of what is now called geometric mechanics. There are countless references to these results, that is why we cite a series of books (some of them already classics) where these developments and the original references to them can be found \cite{AM-78,Ar,LR-85,LR-89,Go-69,holm1,LM-87,MR-2024}.

\subsection*{Contact mechanics and dissipation}

Recently, interest has grown in the use of {\sl contact geometry} in the description of mechanical systems that experience dissipation phenomena. Although this was known and used in the case of, for example, thermodynamics, it has now been used in the general context, and this geometry has been identified as a particular case of Jacobi structures. These types of systems are what in physics and in their Lagrangian version are called action-dependent Lagrangian systems. The interest in this description has led to the publication of numerous papers that extend to this scenario what had already been done in the symplectic case. For a motivated introduction see \cite{Bravetti2017,CG-2019}. Contact Hamiltonian systems have been extensible studied \cite{BCT-2017,BLMP-2020,GG-2022,Lainz2018} and the description of contact Lagrangian systems can be found in \cite{GGMRR-2019b}. For contact systems with an explicit time dependence, see \cite{LGGMR-2022,RT-2023,LLLR-2022,GRL-2023}.
For the quantization of contact systems, we recommend  \cite{CCM-2018}. Of course, there is a variational description that comes back to the original papers by G. Herglotz and generalizes the well-known Hamilton's principle \cite{He-1930,Her-1985}. Generalities about contact geometry can be found in \cite{Go-69,LM-87,Banyaga2016,Geiges2008,Kholodenko2013}).

One of the key contributions of both symplectic and contact geometry is that in both cases a bracket of (observable) functions is obtained, which in the first case is Poisson, and in the second Jacobi, the difference being precisely that the Jacobi bracket does not satisfy Leibniz's identity (in other words, it is not a derivation in its arguments but a first-order differential operator). Let us not forget that the existence of a function bracket is essential for the quantization process.

\subsection*{Multisymplectic field theories}

At the end of the 60's of the last century, something similar happened in the case of classical field theories. Three groups of physicist-mathematicians independently developed a formalism called {\sl multisymplectic}, which sought to extend the symplectic of mechanics to this case \cite{Gc-73,GS-73,Kij1979}. The difficulty of this new geometry is that, while symplectic geometry is very rigid (it is always locally equivalent to the canonical of a cotangent bundle via the {\sl Darboux Theorem\/}), the situation is very different in the multisymplectic case. In the last 50 years, much effort has been made to achieve progress in this direction, and despite many achievements, a theory as satisfactory as for mechanics has not yet been achieved, being still a field of research in full development. 
For Lagrangian formalism of classical field theories we refer to 
\cite{Gc-73,GS-73,Ald1980,LMS-2004,EMR-96,art:Echeverria_Munoz_Roman98,art:Roman09,book:Saunders89,LeSaVi2016},
and for the Hamiltonian formalism, see
\cite{art:Roman09,LeSaVi2016,CCI-91,LMM-96,EMR-99b,
EMR-00b,GMM-2022,GIMMSY,HK-04,Kru2002,MS-98,Pau2002}. These are only some references, since the literature about this subject is really large.
In any case, a brief historical account on the subject can be found in
\cite{book:binz}.

\subsection*{Multicontact field theories}

In the same way that there are mechanical systems with dissipation, the same occurs in the case of classical field theories. 
Echoing contact geometry, this leads to action-dependent field theory, that is, field theories where the Lagrangian or the Hamiltonian depends on new variables that represent the flow of the action. 
This leads to a modification of the field equations.
And, again, there is a geometric structure, called {\sl multicontact}, that allows us to obtain these field equations. This has been developed in a recent paper \cite{LGMRR-2022}. Let us notice that in \cite{Vi-2015}, L. Vitagliano introduces higher codimensional versions of contact manifolds that he calls `multicontact manifolds'. This notion is different from ours. 
As in the contact case, one can derive the field equations from a variational principle, and we refer to several papers on this subject \cite{Geor-2003,LGMRR-2022,GLMR-2022,GasMas2022,LPAF2018} (for a precise and general derivation, including higher-order, we recommend \cite{GLMR-2022}). We would like also to refer to an alternative (but related) approach for action-dependent field theories using the so-called
{\sl $k$-contact structures} \cite{GGMRR-2019,GGMRR-2020,Ri-2022,LRS-2024}.

The multicontact geometry was derived as a mathematical generalization of other approaches, motivated by the capacity to formalize phenomenology outside the scope of the usual Hamilton's principle. So far, there are only a handful of examples developed with any detail. This paper aims to provide an easy introduction to action-dependent field theories and the main associated geometric tools. We lay out the field equations for regular systems, and provide an explicit guide for the multicontact formalism. The method is showcased with a diverse list of fully worked-out examples. We hope this paper will facilitate the work of those interested in action-dependent field theory and the new geometric objects underlying it.

\subsection*{Structure of the paper}

The paper is structured as follows. In section \ref{0} we introduce the so-called action-dependent field theories, and in Section \ref{1}, we develop the Lagrangian and Hamiltonian formalism for these theories, making use of the multicontact geometry. The rest of the paper is devoted to applying the above framework to particular field theories. In Section \ref{ex1} we consider general Lagrangians that are quadratic (regular and singular) and affine, and that include terms depending on ``action variables''. In Section \ref{ex2}, a number of particular cases are discussed: 
one-dimensional wave equation with first-order terms and external force, Klein--Gordon equation and the telegrapher equation, Metric-affine gravity, Maxwell’s electromagnetism, and Burgers’ equation from the heat equation;
all of them including terms depending on ``action variables''.
Finally, we enumerate some relevant topics for further investigation.
Two additional simple examples, 
Bosonic string theory and $(2+1)$-dimensional gravity and Chern--Simons equation,
are added in a first appendix \ref{apendA}.
To keep the paper self-contained, we include a second appendix \ref{append} on premulticontact and multicontact structures that is not necessary to follow the rest of the paper but may be useful to the reader.

\subsection*{Some general conventions and notation}

All the manifolds are real, second-countable and of class $\Cinfty$, and the mappings are assumed to be smooth.
Sum over crossed repeated indices is understood.
The signature of the space-time metric that has been taken is $(-+++)$.

A brief summary of conventional notation is the following: 
\vspace{-10pt}

\begin{itemize}\itemsep1pt
    \item $\Cinfty({\cal M})$: Smooth functions in a manifold ${\cal M}$.

    \item $\df^k({\cal M})$: Module of differential forms of degree $k$ in a manifold ${\cal M}$.

    \item $\vf({\cal M})$: Module of vector fields in a manifold ${\cal M}$.

    \item $\vf^m({\cal M})$: Module of $m$-multivector fields in a manifold ${\cal M}$.
    
    \item $\innp{X}\Omega$: Natural contraction of a vector field $X\in\vf({\cal M})$ and a $k$-form $\Omega\in\df^k({\cal M})$.


    \item $J^1\pi$: $1$-jet bundle over a bundle $\pi\colon E\to M$.
Coordinates: $(x^\mu,y^a,y^a_\mu)$, $0\leq\mu\leq m-1$, $1\leq a\leq n$.

    \item $J^{1*}\pi$: ``Dual'' jet bundle.
Coordinates: $(x^\mu,y^a,p_a^\mu)$.

    \item $\Lambda^{m-1}(\Tan^*{\cal M})$: Bundle of $(m-1)$-differential forms on a manifold ${\cal M}$ (multicotangent bundle).

    \item $\d$: Exterior differential of differential forms.

    \item $\wedge$: Exterior product of differential forms or vector fields.

    \item $\d^mx\equiv\d x^0\wedge\cdots\wedge\d x^{m-1}$: Coordinate expression of a volume form.

    \item $\ds\d^{m-1}x_\mu=\innp{\derpar{}{x^\mu}}\d^mx
    \equiv(-1)^{\mu}\d x^0\wedge\dotsb\wedge\d x^{\mu-1}\wedge\d x^{\mu+1}\wedge\dotsb\wedge\d x^{m-1}$.

    \item $\ds \partial_\mu=\derpar{}{x^\mu}$.
\end{itemize}

\section{Action-dependent field theories}
\label{0}

In this preliminary section, we discuss the main features of the so-called action-dependent classical field theories.

\subsection{General aspects}

An {\sl action-dependent} first-order field theory is a system where the Lagrangian or the Hamiltonian functions depend on the action in an implicit way.
Since the action is not a local operator, we consider a localized version of the action, given by an $(m-1)$-form $S$ over space-time. 
It is related to the action in the following way: in classical field theories, the action of a field $y(x)=(y^a(x))$ is an integral over a domain in space-time, $D\subset M$,
\begin{equation}
\mathcal{A}(y(x))=\int_D L(y(x))\d^mx\,.
\end{equation}
The action $\mathcal{A}$ is not local in the sense that it depends on the domain $D$. The density of action is the Lagrangian density $L\,\d^mx$. 
Evaluated on a field, $L(y(x))\d^mx$ is an $m$-form over space-time and, therefore, it is a closed form. Locally, we can consider a form whose differential is the Lagrangian density, which is the $(m-1)$-form represented by $S=s^\mu(x^\nu)\d^{m-1}x_\mu$,
\begin{equation}\label{eq:dSactionDep}
 \d S=\d \left(s^\mu\d^{m-1}x_\mu\right)= \frac{\partial s^\mu}{\partial x^\mu}\d^mx=L(y(x))\d^mx\,.
 \end{equation}
Therefore, $S$ plays the role of the potential of the density of action. The value of $S$ depends on the field $y(x)$; hence,  the components $s^\mu$ are new fields; that is, fiber coordinates to be determined as functions of $(x^\mu)$.  

To be more precise, an action-dependent field theory is a system where the Lagrangian or Hamiltonian depends on the components $s^\mu$ of a $(m-1)$ form over space-time, with the relation \eqref{eq:dSactionDep}. 
This leads to an implicit variational principle called 
{\sl Herglotz's variational principle} \cite{GLMR-2022,LPAF2018}. 
A pair $(y(x),s(x))\equiv(y^a(x),s^\mu(x))$ is a solution to the Herglotz variational principle if they are critical for the action
\begin{equation}
\mathcal{A}(y(x),s(x))=\int_D L(x^\mu,y^a,\partial_\mu{y^a},s^\mu)\d^mx\,,
\end{equation}
under the constraint
\begin{equation}\label{eq:dSactionDep2}
\frac{\partial s^\mu}{\partial x^\mu}=L(x^\mu,y^a,\partial_\mu y^a,s^\mu)\,.
\end{equation}
To solve this principle one should first solve \eqref{eq:dSactionDep2} for the unknowns $(y^a(x),s^\mu(x))$, if possible, or consider \eqref{eq:dSactionDep2} as a non-holonomic constraint, which establishes non-trivial restrictions on the possible variations.  For more details on Herglotz's variational principle for fields, see \cite{GLMR-2022}. 

The constraint \eqref{eq:dSactionDep2} is crucial for the theory because it is the difference between Hamilton's and Herglotz's variational principles. Due to it, the $s^\mu$ variables generate the action; a property that differentiates them from other fields. Explicitly, as a consequence of the constraint \eqref{eq:dSactionDep2} we can write the action as,
\begin{align}\label{eq:saction}
\mathcal{A}(y(x),s(x))=\int_D L(x^\mu,y^a,\partial_\mu y^a,s^\mu)\d^mx=\int_{\partial D} s^\mu\d^{m-1}x_\mu\,.
\end{align}
Namely, given a solution $(y^a(x),s^\mu(x))$, the integral of the $(m-1)$-form $s^\mu\d^{m-1}x_\mu$ over the boundary of the domain is actually the value of the action on that solution. The unique character of the $s^\mu$ variables is reflected in its particular geometric description and are sometimes called ``action variables''.
This relation suggests that the frontier could play a more relevant role in action-dependent field theories, although this is a feature that has not been explored so far.

\subsection{Lagrangian formalism}

Given a Lagrangian $L(x^\mu,y^a,y^a_\mu,s^\mu)$, the functions $(y^a(x),s^\mu(x))$ are a solution
 to the Herglotz variational principle for fields if, and only if,
\begin{equation}
\label{eq:HELActionDep}
\frac{\partial L}{\partial y^a}-\frac{\d}{\d x^\mu}
\frac{\displaystyle\partial L}{\partial
y^a_\mu}+\frac{\partial L}{\partial s^\mu}\frac{\partial L}{\partial y^a_\mu}=0
\,, \qquad \derpar{s^\mu}{x^\mu}=L\,.
\end{equation}
These are the {\sl Herglotz--Euler--Lagrange equations} \cite{LPAF2018,GGMRR-2020} (see also Section \ref{mlf}). 
The first group of equations \eqref{eq:HELActionDep} are the usual Euler--Lagrange equations with two extra terms: 
$\displaystyle\frac{\partial L}{\partial s^\mu}\frac{\partial L}{\partial y^a_\mu}$, which is a term proportional to the momenta, 
and $\displaystyle\frac{\partial^2 L}{\partial s^\nu y^a_\mu}\frac{\partial s^\nu}{\partial x^\mu}$, from the total derivative with respect to $x^\mu$. 
Notice that the above equations are not linear on the Lagrangian,
as in the standard Euler--Lagrange equations.
Thus, any new term added to the Lagrangian could interact with all the other ones. The last equation of \eqref{eq:HELActionDep} is the constraint \eqref{eq:dSactionDep2}.

In most applications so far considered, the action-dependent Lagrangian density is the standard one of the system of interest,
$L_0\,\d^mx$, plus a term linear on the $s^\mu$ of the form $\lambda_\mu(x^\nu) s^\mu \d^mx$. Therefore, $L\,\d^mx=(L_0+\lambda_\mu s^\mu)\d^mx$. In these cases, we have that,
\begin{equation}
\frac{\partial L}{\partial s^\mu}\frac{\partial L}{\partial y^a_\mu}=\lambda_\mu\frac{\partial L}{\partial y^a_\mu}
\,,\qquad \frac{\partial^2 L}{\partial s^\nu y^a_\mu}\frac{\partial s^\nu}{\partial x^\mu}=0\,,
\end{equation}
that is, a linear term on the momenta, $\ds\frac{\partial L}{\partial y^a_\mu}$, has been added to the equations. For Lagrangians with kinetic energy, this is generally a term linear in velocities, which is related to dissipative phenomena.

A more general dependence of the Lagrangian on the action usually leads to non-linear terms of this kind in the field equations (for instance, see Section \ref{ex:burguer} for a non-linear example).

\subsection{Hamiltonian formalism}

There are two main approaches to Hamiltonian field theory: the space-time split formalism on an infinite-dimensional phase space \cite{Gotay}, and the covariant formalism, also known as the {\sl De Donder--Weyl theory} \cite{dD-1930,Weyl-1935}. 
The last one has been developed for action-dependent field theories using {\sl $k$-contact} and {\sl multicontact geometries}
\cite{LGMRR-2022,GGMRR-2019,GGMRR-2020,Ri-2022}. 
We now motivate the Herglotz--Hamilton--De Donder--Weyl equations from the Lagrangian equations \eqref{eq:HELActionDep}
with a simple manipulation in coordinates. The reader can find a more detailed derivation in Section \ref{1} and in \cite{LGMRR-2022,GGMRR-2019}.

In the covariant approach, all the space-time coordinates are considered on the same footing. Therefore, we have momenta in each of the space-time coordinates, which are defined as:
\begin{equation}
p^\mu_a=\frac{\partial L}{\partial y^a_\mu}\,.
\end{equation}
The quantities $p^\mu_a$ are usually called the {\sl multimomenta}. 
We can express the velocities in terms of the multimomenta as long as the matrix
$\left(\displaystyle\frac{\partial^2 L}{\partial y^a_\mu\partial y^b_\nu}\right)$ is regular. In this case, we say that the Lagrangian is {\sl regular}, and we can define the Hamiltonian as
\begin{equation}
H(x^\mu,y^a,p^\mu_a,s^\mu)=y^a_\mu(x^\mu,y^a,p^\mu_a,s^\mu)p^\mu_a-L(x^\mu,y^a,y^a_\mu(x^\mu,y^a,p^\mu_a,s^\mu),s^\mu)\,,
\end{equation}
and, by direct calculation, we derive that,
\begin{equation}\label{eq:Ham1Sect2}
    y^a_\mu=\frac{\partial H}{\partial p^\mu_a}\,.
\end{equation}
From the above definition of $H$ we obtain, 
\begin{equation}
L(x^\mu,y^a,y^a_\mu,s^\mu)=y^a_\mu p^\mu_a(x^\mu,y^a,y^a_\mu,s^\mu)-H(x^\mu,y^a,p^\mu_a(x^\mu,y^a,y^a_\mu,s^\mu),s^\mu) \,.
\end{equation}
Using these transformations we can see that the equations \ref{eq:HELActionDep} are equivalent to
\begin{align}\label{eq:hamiltonintro}
\frac{\partial p^\mu_a}{\partial x^\mu}&=-\frac{\partial H}{\partial y^a}-p^\mu_a\frac{\partial H}{\partial s^\mu}\,,
\\
\derpar{s^\mu}{x^\mu}&=\frac{\partial H}{\partial p^\mu_a} p^\mu_a-H \,.
\end{align}
These equations, together with \eqref{eq:Ham1Sect2}, are the 
{\sl Herglotz--Hamilton--de Donder-Weyl equations} for action-dependent field theories. One can consider these equations for an arbitrary Hamiltonian, even if it does not arise from a Lagrangian.

Most of the relevant field theories, like Einsteins' equations and Yang--Mills, are described by singular Lagrangian functions. In that case, one cannot find a one-to-one correspondence between velocities and multimomenta. Moreover, the equations are usually not compatible in the full space and constraints appear. 
To solve these problems, the geometric approach to field theories, in this case premulticontact geometry, gives us a method to obtain the set where solutions exist and which equations they actually satisfy. This procedure is explained in Section \ref{mlf} for Lagrangian systems and in \ref{Hamsing} for the Hamiltonian counterpart. We will see some examples in the second part of the paper.

The space-time split formalism considers the evolution of a field with respect to only one coordinate, usually time. This approach is preferred when quantizing field theories. Moreover, it is more natural to incorporate boundary conditions for the field equations, using the Cartan formalism in the space of Cauchy data \cite{LMS-2004,book:binz,Gimmsy2}. Furthermore, this formulation of the problem has been useful to numerically simulate solutions of the equations \cite{AB-1992,cook-2000}. This approach benefits from a different constraint algorithm (see \cite{ErrastiDiez:2023gme} for a complete algorithm including functional independence). Nevertheless, to our knowledge, the space-time split formalism for action-dependent field theories has not been developed.

\subsection{Properties of action-dependent field theories}

\paragraph{Non-conservation and symmetries.}
A relevant feature of action-dependent field theories is that they are non-conservative in the following sense:
a conserved quantity in field theories is a $(m-1)$-form \,$\xi^\mu(x^\nu)\,\d^{m-1}x_\mu$ such that, on a solution, $\displaystyle\frac{\partial\xi^\mu}{\partial x^\mu}=0$
(see, for instance, \cite{LMS-2004,EMR-99b,AA-78,FS-2012,GPR-2016,GR-2023,GR-2024,RWZ-2016}). 
In integral form, this translates to the conservation of flow through the boundary of a closed region. In action-dependent field theories, we have dissipated quantities, which are governed by the following relation, which holds on any solution:
\begin{equation}\label{eq:dissLaw}
  \frac{\partial\xi^\mu}{\partial x^\mu}=-\frac{\partial H}{\partial s^\mu}\xi^\mu  \,,
\end{equation}
in the Hamiltonian formalism. In the Lagrangian formalism it is:
\begin{equation}\label{eq:dissLaw1}
  \frac{\partial\xi^\mu}{\partial x^\mu}=\frac{\partial L}{\partial s^\mu}\xi^\mu  \,.
\end{equation}
This relation is called the {\sl dissipation law}~\cite{GGMRR-2020}. Despite its name, it does not always describe a decrease of an observable, but a non-conservative behavior in a particular way. The name originates from the use of contact geometry to study friction in mechanical systems. All the dissipated quantities of a system have the same dissipation law, which is given by the Hamiltonian (or the Lagrangian). 

The dissipation law \eqref{eq:dissLaw} (or \eqref{eq:dissLaw1}) has a rich behavior in field theories, far beyond dissipation. For instance, in Electromagnetism only a particular selection of covectors $(\lambda_\mu)$ lead to a clear dissipation of energy \cite{GasMar21}. Meanwhile, one can construct cosmological models with an action-dependent Lagrangian where the energy-momentum tensor is conserved \cite{Lazo2022}.

The notions of symmetries and a Noether theorem exist for action-dependent field theories, that is, a way to construct a dissipated quantity from a symmetry \cite{LPF2019,GGMRR-2019,GGMRR-2020}. Nevertheless, the topic is far less developed than in contact mechanical systems \cite{GGMRR-2019b,GRL-2023}.

\paragraph{New terms in the equations.} 

Another relevant reason to study action-dependent field theories is their capacity to model a larger pool of PDE's. It is clear that the Herglotz–Euler–Lagrange equations \ref{eq:HELActionDep} contain more terms than the Euler–Lagrange equations, but the question remains if those terms could be recovered with a different Lagrangian without invoking action terms. This is a challenging question because it is closely related to the \emph{inverse problem}, a classical problem only solved in particular cases \cite{C1981} (see \cite{LGL-2021} for recent advances in contact mechanics). Nevertheless, several results indicate that action-dependent Lagrangians, and the corresponding contact and multicontact geometry, are the adequate description for these systems.

As a first approach, consider the quadratic Lagrangian of a scalar field $y$ with a linear dependence on the action (see \ref{sec:quadratic} for a more general case)
\begin{equation}
    L=\frac12f^{\mu\nu}y_\mu y_\nu-V(y)-\lambda_\mu(x^\rho)s^\mu\,.
\end{equation}
The corresponding Herglotz--Euler--Lagrange equations are
\begin{equation}
    \frac{\partial V}{\partial y}+f^{\mu\nu}y_{\mu\nu}=-\lambda_\nu f^{\mu\nu}y_\mu\,.
\end{equation}
Assuming $f^{\mu\nu}$ is a regular metric, the right-hand side is a generic linear term in velocities. This term can be partially recovered without the action-dependent term by considering a space-time-depending global factor:
\begin{equation}
    L_b=b(x^\rho)\left[\frac12f^{\mu\nu}y_\mu y_\nu-V(y)\right]\,.
\end{equation}
The resulting equations are
\begin{equation}
    b\left[\frac{\partial V}{\partial y}+f^{\mu\nu}y_{\mu\nu}\right]=-\frac{\partial b}{\partial x^\nu}f^{\mu\nu}y_\mu\,.
\end{equation}
We recover the same equations if, and only if, $\d \ln b=\lambda_\mu\d x^\mu$, which is only possible if $\lambda_\mu\d x^\mu$ is a closed form. In other words, not all terms linear in velocities can be recovered by a global factor.

The problem of the particularity of action-dependent Lagrangians has been studied in more generality in contact mechanical systems. The dynamics of a contact mechanical system is of Reeb type (instead of symplectic) outside the set where the energy vanishes \cite{BLMP-2020}. In some cases, is it possible to recast the contact dynamics as a subset of a symplectic system, although there are obstructions for a general result \cite{GGKM-2024,Sloan:2024kzb}. In summary, the dynamics of an action-dependent mechanical system are associated with contact geometry, which is, in general, qualitatively different from symplectic geometry.


There are a lot of topics to explore, but one should advance with care because our intuition about the standard classical field theories does not always translate to action-dependent field theories. To begin with, the equations are not linear in the Lagrangian. 
Another important case is that of equivalent Lagrangians: if we add a total derivative to an action-dependent Lagrangian, we do not obtain the same equations. To construct equivalent Lagrangians one usually needs to also perform a change of variables in the $s^\mu$ coordinates. This topic has been studied for mechanical systems in \cite{LGL-2021}. A notorious example in field theories is General Relativity, where different equivalent Lagrangians lead to different dynamics when an action-dependent term is added~\cite{GasMas2022,Lazo2022} (see also example \ref{ex:EP}).




\section{Lagrangian and Hamiltonian formalisms for action-dependent field theories}
\label{1}

This section is devoted to reviewing the main features of the multicontact formulation for action-dependent field theories introduced in \cite{LGMRR-2022}.
For this, it is necessary to make a minimal presentation of the inherent geometric structures; 
although our attention will be focused on obtaining and discussing the field equations.
As this framework is used in the following sections to describe both regular and singular field theories, both cases are explained.

\subsection{Multivector fields}
\label{ap:multivector}

(For details on multivector fields see, for instance, \cite{art:Echeverria_Munoz_Roman98,Ca96a,IEMR-2012}). 

Let ${\cal M}$ be a manifold with $\dim{\cal M}=n$.
The {\sl\textbf{$m$-multivector fields}} or {\sl\textbf{multivector fields of order $m$}} on ${\cal M}$ are the contravariant skew-symmetric tensor fields of order $m$ in ${\cal M}$. 
The set of $m$-multivector fields in ${\cal M}$ is denoted $\vf^m({\cal M})$.
A multivector field $\textbf{X}\in \vf^m({\cal M})$ is {\sl\textbf{locally decomposable}} if, for every point ${\rm p}\in{\cal M}$, there exists an open neighbourhood $U_{\rm p} \subset{\cal M}$ such that 
\begin{equation}
\left.\bfX\right\vert_{U_{\rm p}} = X_0\wedge\dotsb\wedge X_{m-1} \,,\qquad \text{for some }X_0,\dotsc,X_{m-1} \in\vf(U_{\rm p}) \,.
\end{equation}
The {\sl contraction} of a locally decomposable multivector field $\textbf{X}\in \vf^m({\cal M})$ and a differentiable form $\Omega\in\df^k({\cal M})$
is the natural contraction between contravariant and covariant tensor fields; i.e., 
$\innp{\textbf{X}}\left.\Omega \right \vert_{U_p}=0$, if $k < m$, and
\begin{equation}
\innp{\textbf{X}}\left.\Omega \right \vert_{U_p}=\innp{X_0\wedge\dotsb \wedge X_{m-1}}\Omega =
\innp{X_{m-1}}\dotsb\innp{X_0}\Omega \,, \qquad
\text{if }k\geq m \, .
\end{equation}

Let $\kappa\colon {\cal M}\rightarrow M$ be a fiber bundle
with $\dim{M}=m$ and $\dim{{\cal M}}=N+m$,
and denote $(x^\mu,z^a)$ the local coordinates on ${\cal M}$;
where $x^\mu$ are coordinates on the base $M$ and $z^a$ are coordinates on the fibers (with $0\leq\mu\leq {m-1}$ and $1\leq I\leq N$).
A multivector field $\textbf{X} \in \vf^m({\cal M})$  is $\kappa$-\textsl{transverse} if
$\innp{\X}(\kappa^*\beta)\vert_{\rm p} \neq 0$,
for every ${\rm p} \in{\cal M}$ and $\beta \in \df^m({\cal M})$.
The local expression for a locally decomposable $\kappa$-transverse multivector field $\textbf{X}\in\vf^m({\cal M})$ is 
\beq
\textbf{X}= \bigwedge^{m-1}_{\mu=0} \textbf{X}_\mu= 
f\bigwedge^{m-1}_{\mu=0} \left( \frac{\partial}{\partial x^\mu} + F_\mu^a\frac{\partial}{\partial z^a}\right)\,, \qquad \text{where }f\in \Cinfty({\cal M}) \,.
\label{mvf}
\eeq
 If $M$ is an orientable manifold with volume form $\omega\in\df^m({\cal M})$, then
the condition for $\textbf{X} \in \mathfrak{X}^m({\cal M})$ to be $\kappa$-transverse can be expressed as
$\innp{\bfX}(\kappa^*\omega)\neq 0$.
This condition can be fixed by taking
$\innp{\bfX}(\kappa^*\omega)=1$,
which implies $f=1$ in \eqref{mvf}.

 If $\textbf{X} \in \vf^m({\cal M})$ is a locally decomposable and $\kappa$-transverse multivector field
and $\psi\colon U\subset M \rightarrow{\cal M}$
is a local section of $\kappa$,
with local expression
$\psi(x^\mu)=(x^\mu,z^a(x^\nu))$;
then $\psi$ is an {\sl\textbf{integral section}} of $\textbf{X}$ if $\displaystyle\derpar{z^a}{x^\mu}=F_\mu^a$.
Then, $\textbf{X}$ is {\sl \textbf{integrable}} if,
for ${\rm p}\in{\cal M}$, there exist $x\in M$ and an integral section, $\psi\colon U_x\subset M\to {\cal M}$, of $\textbf{X}$ such that ${\rm p}=\psi(x)$.

\subsection{Multicontact Lagrangian formalism}
\label{mlf}


In the multisymplectic formulation of conservative classical field theories 
\cite{Gc-73,GS-73,Ald1980,LMS-2004,EMR-96,art:Echeverria_Munoz_Roman98,
art:Roman09,book:Saunders89,Aldaya_Azcarraga78_2},
a first-order Lagrangian field theory is described by the following elements.
First we have the {\sl configuration bundle} $\pi\colon E\to M$,
where $\dim M=m$ and $\dim E=n+m$,
and $M$ is an orientable manifold with volume form
$\omega\in\df^m(M)$,
which usually represents space-time.
Then, the theory is developed on the first-order jet bundle
$J^1\pi\to E\to M$, with $\dim J^1\pi=nm+n+m$.
Natural coordinates in $J^1\pi$ adapted to the bundle structure
are denoted by $(x^\mu,y^a,y^a_\mu)$
($\mu = 0,\ldots,m-1$ and $a=1,\ldots,n$),
and are such that
$\omega=\d x^0\wedge\cdots\wedge\d x^{m-1}\equiv\d^mx$.

For the Lagrangian formulation of non-conservative (or action-dependent) first-order field theories (see \cite{LGMRR-2022}), consider the bundle, 
\begin{equation}
{\cal P}=J^1\pi\times_M\Lambda^{m-1}(\Tan^*M)\,,
\end{equation}
where $\Lambda^{m-1}(\Tan^*M)$ denotes the bundle of $(m-1)$-forms on $M$.
This bundle can be identified with $J^1\pi\times\Real^m$,
and so we shall do henceforth. 
We denote by $\tau\colon{\cal P}\to M$ the bundle projection.
Natural coordinates in ${\cal P}$ are $(x^\mu,y^a,y^a_\mu,s^\mu)$,
and so $\dim\,{\cal P}=2m+n+nm$.
\begin{equation}
\xymatrix{
&\ &  \  &{\cal P}=J^1\pi\times_M\Lambda^{m-1}(\Tan^*M)  \ar[rrd]_{\tau_1}\ar[lld]^{\rho}\ar[ddd]_{\tau}\ &  \   &
\\
&J^1\pi \ar[ddrr]^{\bar{\pi}^1}\ar[d]^{\pi^1}\ & \ & \ & \ &\ar[ddll]_{\tau_o}\Lambda^{m-1}(\Tan^*M) 
\\
&E\ar[drr]^{\pi}\ & \ & \ & \ &  
 \\
&\ & \ &M \ & \ & 
}\nonumber
\end{equation}

Now, a {\sl\textbf{Lagrangian density}} is introduced on ${\cal P}$
as a $m$-form $\mathcal{L}\in\df^m({\cal P})$,
whose expression is
\begin{equation} \Lag(x^\mu,y^a,y^a_\mu,s^\mu)=L(x^\mu, y^a,y^a_\mu,s^\mu)\,\d^mx\,,\end{equation}
where $L\in\Cinfty({\cal P})$ is the
\textsl{Lagrangian function} associated with $\Lag$.
Then, the {\sl \textbf{Lagrangian $m$-form}} associated with $\L$ on ${\cal P}$ is
\begin{equation}
\label{thetacoor1}
    \Theta_{\mathcal{L}}=
    -\frac{\partial L}{\partial y^a_\mu}\d y^a\wedge\d^{m-1}x_\mu +\left(\frac{\partial L}{\partial y^a_\mu}y^a_\mu-L\right)\d^m x+\d s^\mu\wedge \d^{m-1}x_\mu \,.
\end{equation}
The local function
$\displaystyle E_\Lag=\frac{\partial L}{\partial y^a_\mu}y^a_\mu-L$
is called the \textsl{\textbf{energy Lagrangian function}} associated with $L$.

A Lagrangian function $L\in\Cinfty({\cal P})$ is {\sl \textbf{regular}} if the Hessian matrix
\begin{equation} (W_{ab}^{\mu\nu})= 
\left(\frac{\partial^2L}{\partial y^a_\mu\partial y^b_\nu}\right) \end{equation}
is regular everywhere; then $\Theta_\L$ is a {\sl variational multicontact form} on ${\cal P}$
and the triple $({\cal P},\Theta_\L,\omega)$ is called a {\sl \textbf{multicontact Lagrangian system}} (see Appendix \ref{append} for a quick review of multicontact and premulticontact structures and their terminology).
Otherwise, $L$ is a {\sl \textbf{singular}} Lagrangian.

\begin{remark}
\label{rem1}
It is important to note that, given a singular Lagrangian, the associated Lagrangian form 
$\Theta_\L$ is not always a {\sl premulticontact form} on ${\cal P}$.
Consider, for instance, the example given by the singular Lagrangian $\displaystyle L=\sum_{a=1}^n y^a_\mu s^\mu$
which yields a structure $(\Theta_\L,\omega)$
which has no Reeb distribution associated with it, and then 
it is neither a multicontact, nor a premulticontact structure
(see Appendix \ref{append} and \cite{LGMRR-2022} for details on these definitions and results). 
Nevertheless, in this paper we  only consider singular Lagrangians for which $\Theta_{\mathfrak L}$ is a premulticontact form,
and then the triple $({\cal P},\Theta_\mathfrak L,\omega)$ will be called a {\sl \textbf{premulticontact Lagrangian system}}.
\end{remark}

To set the field equations for action-dependent field theories in an intrinsic way, 
we need first to introduce the so-called {\sl \textbf{dissipation form}} whose expression in coordinates is
\begin{align}\label{sigmaL}
    \sigma_{\Theta_\L}=-\derpar{L}{s^\mu}\,\d x^\mu \,.
\end{align}
Then we define the {\sl \textbf{Lagrangian $(m+1)$-form}}
\begin{align}
\overline{\Omega}_\Lag &=\d\Theta_\Lag+\sigma_{\Theta_\L}\wedge\Theta_\Lag
 = \d\Theta_\Lag-\derpar{L}{s^\nu}\d x^\nu\wedge\Theta_\Lag
\nonumber \\ &=
\d\left(-\frac{\partial L}{\partial y^a_\mu}\d y^a\wedge\d^{m-1}x_\mu +\Big(\frac{\partial L}{\partial y^a_\mu}y^a_\mu-L\Big)\d^m x\right)
-\left(\derpar{L}{s^\mu}\frac{\partial L}{\partial y^a_\mu}\d y^a
-\derpar{L}{s^\mu}\d s^\mu\right)\wedge\d^mx\,.
\label{Omegabar}
\end{align}

Now, a section $\bm{\psi}\colon M\rightarrow {\cal P}$ of the projection $\tau$
is said to be a {\sl\textbf{holonomic section}} on ${\cal P}$ if it is locally expressed as
\begin{equation}
\bm{\psi}(x^\mu)= \left(x^\mu,y^a(x^\nu),\derpar{y^a}{x^\mu}(x^\nu),s^\mu(x^\nu) \right) \,.
\end{equation}
Then $\bm{X}\in\vf^m({\cal P})$ is a {\sl \textbf{holonomic $m$-multivector field}} on ${\cal P}$ if it is $\tau$-transverse, integrable,
and its integral sections are holonomic on ${\cal P}$.
The local expression of a holonomic $m$-multivector field on ${\cal P}$ verifying the condition $\innp{\bfX}\omega=1$ is
\begin{equation}
\label{semiholmvf}
\bm{X}=\bigwedge^{m-1}_{\mu=0}\Big(\derpar{}{x^\mu}+y^a_\mu\frac{\displaystyle\partial} {\displaystyle\partial y^a}+F_{\mu\nu}^a\frac{\displaystyle\partial}{\displaystyle \partial y^a_\nu}+g^\nu_\mu\,\frac{\partial}{\partial s^\nu}\Big)\,, \qquad\text{where } F_{\mu\nu}^a,g^\nu_\mu\in\Cinfty({\cal P}) \,.
\end{equation}
Holonomic multivector fields are also called \textsc{sopde}s because the equations of their integral sections are the solutions to a system of second-order partial differential equations of the form
\begin{equation} y^a_\mu=\derpar{y^a}{x^\mu}\,,\qquad F^a_{\mu\nu}=\frac{\partial^2y^a}{\partial x^\mu \partial x^\nu}\,. \end{equation}
Multivector fields on ${\cal P}$ which have the local expression \eqref{semiholmvf}, but are not integrable, are usually called {\sl\textbf{semi-holonomic}}.

Bearing all this in mind, for a (pre)multicontact Lagrangian system $({\cal P},\Theta_\L,\omega)$
the Lagrangian field equations can be derived from a variational principle 
which is called the {\sl generalized Herglotz Principle}
and they can be stated alternatively as:
\ben[{\rm(1)}]
\item
The {\sl (pre)multicontact Lagrangian equations for holonomic sections},
$\bm{\psi}\colon M\to {\cal P}$, which are
\begin{equation}
\label{ELmcontact2}
\bm{\psi}^*\Theta_\Lag= 0 \,, \qquad
\bm{\psi}^*\innp{Y}\overline{\Omega}_\Lag= 0 \,, \quad \text{for every }Y\in\X({\cal P}) \,.
\end{equation}
\item
The {\sl (pre)multicontact Lagrangian equations for $\tau$-transverse and locally decomposable multivector fields}
${\bf X}_\L\in\vf^m({\cal P})$ which are
\begin{equation}
\label{fieldLcontact}
\innp{\bfX_\Lag}\Theta_\Lag=0 \,, \quad \innp{\bfX_\Lag}\overline{\Omega}_\Lag=0 \,,
\end{equation}
and the condition of $\tau$-transversality can be imposed by asking that $\innp{\bfX_\Lag}\omega=1$.
A $\tau$-transverse and locally decomposable $m$-multivector field solution to these equations is called a
{\sl\textbf{Lagrangian multivector field}}.
\een

In a natural chart of coordinates of ${\cal P}$, 
a $\tau$-transverse and locally decomposable $m$-multivector field ${\bf X}_\L\in\vf^m({\cal P})$ verifying the condition $\innp{\bfX}_\L\omega=1$ is given by
\begin{equation}
{\bf X}_\L= \bigwedge_{\mu=0}^{m-1}
\Big(\derpar{}{x^\mu}+(X_\L)_\mu^a\frac{\displaystyle\partial}{\displaystyle
\partial y^a}+(X_\L)_{\mu\nu}^a\frac{\displaystyle\partial}{\displaystyle\partial y^a_\nu}+(X_\L)_\mu^\nu\,\frac{\partial}{\partial s^\nu}\Big) \,.
\end{equation}
Taking into account \eqref{thetacoor1} and \eqref{Omegabar}, we have that equations \eqref{fieldLcontact} lead to
\begin{align}
0 &=
\displaystyle L + 
\frac{\partial L}{\partial y^a_\mu}\Big((X_\L)_\mu^a-y^a_\mu\Big)-(X_\L)_\mu^\mu\,,
\label{A-E-L-eqs4}
\\
0 &=
\displaystyle \Big((X_\L)_\mu^b-y^b_\mu\Big)
\frac{\partial^2L}{\partial y^b_\mu\partial s^\nu} \,,
\label{A-E-L-eqs2}
\\
\displaystyle 0&=\Big((X_\L)_\mu^b-y^b_\mu\Big)
\frac{\partial^2L}{\partial y^b_\mu\partial x^\nu} \,,
\label{A-E-L-eqs0}
\\
0 &=
\displaystyle \Big((X_\L)_\mu^b-y^b_\mu\Big)
\frac{\partial^2L}{\partial y^a_\nu\partial y^b_\mu}
\label{A-E-L-eqs1} \,,
\\
0 &=
\displaystyle
\Big((X_\L)_\mu^b-y^b_\mu\Big)
\frac{\partial^2 L}{\partial y^a\partial y^b_\mu}
+\frac{\partial L}{\partial y^a}- \parderr{L}{x^\mu}{y_\mu^a}
-\frac{\partial^2L}{\partial s^\nu\partial y^a_\mu}(X_\L)_\mu^\nu
\label{A-E-L-eqs3}
\\
\nonumber &\quad
-\frac{\partial^2L}{\partial y^b \partial y^a_\mu}(X_\L)_\mu^b
-\frac{\partial^2L}{\partial y^b_\nu\partial y^a_\mu}(X_\L)_{\mu\nu}^b
+\frac{\partial L}{\partial s^\mu}
\frac{\partial L}{\partial y^a_\mu}\,,
\end{align}
and a last group of equations which are identities when they are combined with the above ones.

If $X_\L$ is a semi-holonomic solution to \eqref{fieldLcontact}, then 
\beq
\label{semihol}
y^a_\mu=(X_\L)_\mu^a \,,
\eeq
and equations \eqref{A-E-L-eqs2}, \eqref{A-E-L-eqs0}, and \eqref{A-E-L-eqs1} hold identically, 
and \eqref{A-E-L-eqs4} and \eqref{A-E-L-eqs3} give 
\begin{align}
(X_\L)_\mu^\mu&= L \,,
\label{lagdef1}
\\
\frac{\partial L}{\partial y^a}- \parderr{L}{x^\mu}{y_\mu^a}
-\frac{\partial^2L}{\partial y^b \partial y^a_\mu}y_\mu^b
-\frac{\partial^2L}{\partial s^\nu\partial y^a_\mu}(X_\L)_\mu^\nu
-\frac{\partial^2L}{\partial y^b_\nu\partial y^a_\mu}(X_\L)_{\mu\nu}^b
&=-\frac{\partial L}{\partial s^\mu}
\frac{\partial L}{\partial y^a_\mu} \,.
\label{lagdef2}
\end{align}
Furthermore, if these semi-holonomic multivector fields ${\bf X}_\L$ are integrable
then, for their holonomic integral sections
\begin{equation} \bm{\psi}(x^\nu)=\Big(x^\mu,y^a(x^\nu),\derpar{y^a}{x^\mu}(x^\nu),s^\mu(x^\nu)\Big) \,, \end{equation}
equations \eqref{lagdef1} and \eqref{lagdef2}  transform into
\begin{align}
 \derpar{s^\mu}{x^\mu}&=L\circ{\bm{\psi}} \,,
 \label{ELeqs2}
 \\
\label{ELeqs1}
\frac{\partial}{\partial x^\mu}
\left(\frac{\displaystyle\partial L}{\partial
y^a_\mu}\circ{\bm{\psi}}\right)&=
\left(\frac{\partial L}{\partial y^a}+
\displaystyle\frac{\partial L}{\partial s^\mu}\displaystyle\frac{\partial L}{\partial y^a_\mu}\right)\circ{\bm{\psi}} \,,
\end{align}
which are precisely the coordinate expression of the Lagrangian equations \eqref{ELmcontact2} for holonomic sections.
Thus, equations \eqref{fieldLcontact} and \eqref{ELmcontact2} are equivalent in this way.
In particular, \eqref{ELeqs1}
are the so-called {\sl Herglotz--Euler--Lagrange field equations}.

In the case that $L$ is a regular Lagrangian, equations \eqref{A-E-L-eqs1} lead to the condition
of semi-holonomy \eqref{semihol}, since the Hessian matrix 
$\displaystyle\left(\frac{\partial^2L}{\partial y^b_\nu\partial y^a_\mu}\right)$ is regular.
Therefore, equations \eqref{lagdef2}
always have solution for the components $({\bf X}_\L)^b_{\mu\nu}$ of ${\bf X}_\L$
(although the system is undetermined, unless $m=1$).

Thus, we have proved that:

\begin{theorem}
Let $({\cal P},\Theta_\mathcal{L},\omega)$ be a multicontact Lagrangian system.
If ${\bf X}_\mathcal{L}$ is a semi-holonomic and integrable solution to the equations \eqref{fieldLcontact} (that is, a {\sc sopde}), its integral sections are the solutions to the multicontact Euler--Lagrange field equations for holonomic sections \eqref{ELmcontact2}.
These {\sc sopde}s are called the {\sl\textbf{Euler--Lagrange multivector fields}}
associated with $\mathcal{L}$.

In addition, if the Lagrangian system is regular then:
\begin{enumerate}[\rm (1)]
\item
The multicontact Lagrangian field equations
for multivector fields \eqref{vfH} 
have solutions on ${\cal P}$.
The solutions are not unique if $m>1$.
\item
The Lagrangian $m$-multivector fields ${\bf X}_\mathcal{L}$
solution to equations \eqref{vfH} are semi-holonomic.
\end{enumerate}
\end{theorem}

When $L$ is not regular and assuming that $({\cal P},\Theta_\L,\omega)$
is a premulticontact system, in general,
the equations \eqref{lagdef2}
have no solutions everywhere on ${\cal P}$,
since the compatibility of the system \eqref{lagdef2} depends on the rank of the Hessian matrix 
$\displaystyle\left(\frac{\partial^2L}{\partial y^b_\nu\partial y^a_\mu}\right)$. 
In the most favorable cases,
these field equations are compatible only on a submanifold of ${\cal P}$ which is obtained by applying a suitable constraint algorithm.
In any case, solutions to equations \eqref{fieldLcontact}
are not necessarily {\sc sopde}s and,
as a consequence, if they are integrable, 
their integral sections are not necessarily holonomic;
so this requirement must be imposed as an additional condition.
Hence, the final objective is to find the maximal submanifold 
${\cal S}_f$ of ${\cal P}$ where there are 
Euler--Lagrange multivector fields ${\bf X}_\L$ which are
solutions to the premulticontact Lagrangian field equations on ${\cal S}_f$ and are tangent to ${\cal S}_f$;
that is to say, where consistent solutions exist.
This procedure is explained in some of the applications developed
in Sections \ref{ex1} and \ref{ex2}.

\subsection{Multicontact Hamiltonian formalism: The (hyper)regular case}
\label{mhf}

For the Hamiltonian formulation of non-conservative (or action-dependent) first-order field theories,
first consider the so-called {\sl restricted multimomentum bundle} $J^{1*}\pi$,
whose natural coordinates are $(x^\mu,y^a,p_a^\mu)$ such that
$\omega=\d^mx$, and so 
$\dim\, J^{1*}\pi=n+m+nm$.
(This is the bundle where the Hamiltonian multisymplectic formalism of conservative field theories takes place
\cite{art:Roman09,CCI-91,LMM-96,EMR-99b,
EMR-00b,GMM-2022,HK-04,Kru2002,MS-98,Pau2002}).
Then, consider the manifold 
\begin{equation}
{\cal P}^* =J^{1*}\pi\times_M\Lambda^{m-1}(\Tan^*M) \,,
\end{equation}
which has natural coordinates $(x^\mu,y^a,p_a^\mu,s^\mu)$.
This bundle is identified with $J^{1*}\pi\times\Real^m$,
and so we shall do henceforth.
It is a bundle over $M$ with projection
$\widetilde{\tau}\colon{\cal P}^*\to M$.
\begin{equation}
\xymatrix{
&\ &  \  &{\cal P}^*=J^{1*}\pi\times_M\Lambda^{m-1}(\Tan^*M)  \ar[rrd]_{\widetilde\tau_1}\ar[lld]^{\varrho}\ar[ddd]_{\widetilde\tau}\ &  \   &
\\
&J^{1*}\pi \ar[ddrr]^{\bar{\kappa}^1}\ar[d]^{\kappa^1}\ & \ & \ & \ &\ar[ddll]_{\tau_o}\Lambda^{m-1}(\Tan^*M) 
\\
&E\ar[drr]^{\pi}\ & \ & \ & \ &  
 \\
&\ & \ &M \ & \ & 
}\nonumber
\end{equation}

Let $({\cal P},\Theta_\L,\omega)$ be a Lagrangian system,
with $\L=L\,\omega$.
The {\sl \textbf{Legendre map}} associated with the Lagrangian function $L\in\Cinfty({\cal P})$
is the map
${\cal FL}\colon {\cal P}\to {\cal P}^*$
locally given by
\beq
 {\cal FL}^*x^\nu = x^\nu \,, \qquad
 {\cal FL}^*y^a = y^a  \,, \qquad
 {\cal FL}^*p_a^\nu =\displaystyle\derpar{{L}}{y^a_\nu} \,, \qquad
 {\cal FL}^{\,*}s^\mu =s^\mu \,.
\label{FL1}
 \eeq
 The Lagrangian $L$ is regular if, and only if,
 ${\cal FL}$ is a local diffeomorphism,
and $L$ is said to be {\sl\textbf{hyperregular}} when ${\cal FL}$ is a global diffeomorphism.

Next, we consider the hyperregular case
(the regular case is the same but changing
${\cal P}$ and ${\cal P}^*$ by the corresponding open sets).
In this case ${\cal FL}({\cal P})={\cal P}^*$.
Then, the multicontact form $\Theta_\L\in\df^m({\cal P})$
can be projected by ${\cal FL}$ to an $m$-form in ${\cal P}^*$
and then there exists the {\sl \textbf{Hamiltonian $m$-form}} $\Theta_{\cal H}\in\df^m({\cal P}^*)$
such that $\Theta_\L={\cal FL}^*\Theta_{\cal H}$,
whose local expression is
\beq
\label{thetaHcoor}
\Theta_{\cal H}=
-p_a^\mu\d y^a\wedge\d^{m-1}x_\mu+H\,\d^m x+\d s^\mu\wedge \d^{m-1}x_\mu \,.
\eeq
Here $H\in\Cinfty({\cal P}^*)$ is the {\sl Hamiltonian function},
which is obtained as $H=({\cal FL}^{-1})^*E_\L$;
that is, its local expression is given by
\begin{equation} H=p^\mu_a({\cal FL}^{-1})^*y_\mu^a-({\cal FL}^{-1})^*L\,. \end{equation}
Then, $\Theta_{\cal H}$ is a variational multicontact form and hence $({\cal P}^*,\Theta_{\cal H},\omega)$
is called the {\sl\textbf{multicontact Hamiltonian system}} associated with the multicontact Lagrangian systems $({\cal P},\Theta_\L,\omega)$.
In this formalism, the {\sl dissipation form} is expressed as
\begin{equation}
\sigma_{\cal H}=\derpar{H}{s^\mu}\,\d x^\mu\,,
\end{equation}
and we can define the {\sl \textbf{Hamiltonian $(m+1)$-form}}
\begin{align}
\overline{\Omega}_{\cal H}&=\d\Theta_{\cal H}+\sigma_{{\cal H}}\wedge\Theta_{\cal H}=
\d\Theta_{\cal H}+\derpar{H}{s^\mu}\,\d x^\mu\wedge\Theta_{\cal H}\nonumber\\
&=
\d(-p_a^\mu\d y^a\wedge\d^{m-1}x_\mu+H\,\d^m x)+
\Big(\derpar{H}{s^\mu}\,p_a^\mu\,\d y^a-\derpar{H}{s^\mu}\,\d s^\mu\Big)\wedge\d^mx \,;
\label{OmegabarH}
\end{align}
which verifies that $\overline{\Omega}_\Lag={\cal FL}^*\overline{\Omega}_{\cal H}$.

For this multicontact Hamiltonian system, the field equations
can be stated alternatively as:
\begin{enumerate}[{\rm(1)}]
\item  
The \textsl{multicontact Hamilton--de Donder--Weyl equations for sections}
$\bm{\psi}\colon M\to{\cal P}^*$ are
\begin{equation}
\label{sect1H}
\bm{\psi}^*\Theta_{\cal H}= 0  \,, \qquad
\bm{\psi}^*\innp{Y}\overline{\Omega}_{\cal H}= 0 \,, \qquad \text{for every }\ Y\in\vf({\cal }P^*) \, .
\end{equation}
\item 
The \textsl{multicontact Hamilton--de Donder--Weyl equations for $\widetilde\tau$-transverse and locally decomposable multivector fields} ${\bf X}_{\cal H}\in\vf^m({\cal P}^*)$ are
\begin{equation}
\label{vfH}
\innp{\bfX_{\cal H}}\Theta_{\cal H}=0 \,, \qquad \innp{\bfX_{\cal H}}\overline{\Omega}_{\cal H}=0 \,,
\end{equation}
and the condition of $\widetilde\tau$-transversality can be imposed by asking that $\innp{\bfX_{\cal H}}\omega=1$.
\end{enumerate}

In natural coordinates, if
\beq
{\bf X}_{\cal H}=
\bigwedge_{\mu=0}^{m-1}\Big(\derpar{}{x^\mu}+ (X_{\cal H})^a_\mu\frac{\partial}{\partial y^a}+
(X_{\cal H})_{\mu a}^\nu\frac{\partial}{\partial p_a^\nu}+(X_{\cal H})_\mu^\nu\derpar{}{s^\nu}\Big)
\label{Hammv}
\eeq
is the expression of the $\widetilde\tau$-transverse and locally decomposable multivector field ${\bf X}_{\cal H}\in\vf^m({\cal P}^*)$, 
and they are solutions to equations \eqref{vfH},
taking into account the local expressions \eqref{thetaHcoor} and \eqref{OmegabarH},
these field equations lead to
\beq
(X_{\cal H})_\mu^\mu = 
p_a^\mu\,\frac{\partial H}{\partial p^\mu_a}-H \,,\qquad
(X_{\cal H})^a_\mu=\frac{\partial H}{\partial p^\mu_a} \,,\qquad
(X_{\cal H})_{\mu a}^\mu= 
-\left(\frac{\partial H}{\partial y^a}+ p_a^\mu\,\frac{\partial H}{\partial s^\mu}\right) \,,
\label{coor2}
\eeq
together with a last group of equations which are identities when the above ones are taken into account.
If ${\bf X}_{\cal H}$ are integrable, then their integral sections $\bm{\psi}(x^\nu)=(x^\mu,y^a(x^\nu),p^\mu_a(x^\nu),s^\mu(x^\nu))$
are the solutions to the equations \eqref{sect1H} which read as
\beq
\frac{\partial s^\mu}{\partial x^\mu} = \left(p_a^\mu\,\frac{\partial H}{\partial p^\mu_a}-H\right)\circ\bm{\psi}\,, \qquad 
\frac{\partial y^a}{\partial x^\mu}= \frac{\partial H}{\partial p^\mu_a}\circ\bm{\psi} \,,\qquad
\frac{\partial p^\mu_a}{\partial x^\mu} = 
-\left(\frac{\partial H}{\partial y^a}+ p_a^\mu\,\frac{\partial H}{\partial s^\mu}\right)\circ\bm{\psi} \,.
\label{coor1}
\eeq
These are the {\sl Herglotz--Hamilton--de Donder--Weyl equations}
for action-dependent field theories.
These equations are compatible in ${\cal P}^*$.

As ${\cal FL}$ is a diffeomorphism,
the solutions to the Lagrangian field equations for $({\cal P},\Theta_\L,\omega)$
are in one-to-one correspondence
to those of the Hamilton--de Donder--Weyl field equations for
$({\cal P}^*,\Theta_{\cal H},\omega)$.

\subsection{Multicontact Hamiltonian formalism: The singular case}
\label{Hamsing}

(See \cite{LGMRR-2022}).
For singular Lagrangians, the existence of
an associated Hamiltonian formalism is not assured
unless some minimal regularity conditions are assumed.
For instance, in the standard multisymplectic formulation of conservative field theories,
a sufficient condition is that the Lagrangian functions is {\sl almost-regular}
\cite{CCI-91,LMM-96,
EMR-00b}.
Similarly, in the multicontact formalism of action-dependent field theories, we say that
a Lagrangian $L\in\Cinfty({\cal P})$ is said to be {\sl\textbf{almost-regular}} if
(i) ${\cal P}_0^*= {\cal FL}({\cal P})$
is a submanifold of ${\cal P}^*$,
(ii) ${\cal FL}$ is a submersion onto its image,
and (iii) the fibers ${\cal FL}^{-1}({\rm p})$, for every ${\rm p}\in {\cal P}_0^*$,
are connected submanifolds of ${\cal P}$.
Nevertheless, even in these cases, the existence of a premulticontact structure
in ${\cal P}_0^*$ is not assured unless additional conditions are assumed,
as we will see in the examples.

Now, as a consequence of the definition of the
Legendre map ${\cal F}\L$ (see \eqref{FL1}),
we have that 
\begin{equation}
{\cal P}_0^*=P_0\times\Lambda^{m-1}(\Tan^*M)\simeq P_0\times\Real^m \,.
\end{equation}
It is a bundle over $M$ with projection
$\widetilde{\tau}_0\colon{\cal P}_0^*\to M$;
which is endowed with natural coordinates denoted
$(x^\mu,y^a,p^I,s^\mu)$, $1\leq I\leq \dim\,{\cal P}_0^*-n-2m$,
and are such that $\omega=\d^mx$.
\begin{equation}
\xymatrix{
&\ &  \  &{\cal P}_0^*=P_0\times_M\Lambda^{m-1}(\Tan^*M)  \ar[rrd]_{\upsilon}\ar[lld]^{\varrho_0}\ar[ddd]_{\widetilde\tau_0}\ &  \   &
\\
&P_0 \ar[ddrr]^{\bar{\kappa}_0^1}\ar[d]^{\kappa_0^1}\ & \ & \ & \ &\ar[ddll]_{\tau_o}\Lambda^{m-1}(\Tan^*M) 
\\
&E\ar[drr]^{\pi}\ & \ & \ & \ &  
 \\
&\ & \ &M \ & \ & 
}\nonumber
\end{equation}


We denote by ${\cal FL}_0\colon{\cal P}\to{\cal P}^*_0$ the restriction of ${\cal FL}$ to ${\cal P}_0^*$;
that is, ${\cal FL}=\jmath_0\circ{\cal FL}_0$,
where $\jmath_0\colon{\cal P}_0^*\hookrightarrow J^{1*}\pi$
denotes the canonical inclusion.
Then, for almost-regular Lagrangian systems, the form $\Theta_\Lag$
projects to ${\cal P}_0^*$ by ${\cal FL}_0$,
and then we have a Hamiltonian $m$-form $\Theta_{\cal H}^0\in\df^m({\cal P}_0^*)$
such that $\Theta_\Lag={\cal FL}_0^*\,\Theta_{\cal H}^0$
whose local expression is
\beq
\Theta_{\cal H}^0=
F_a^\mu\d y^a\wedge\d^{m-1}x_\mu+H_0\,\d^mx+\d s^\mu\wedge \d^{m-1}x_\mu \,,\label{thetaHcoor2}
\eeq
where $H_0\in\Cinfty({\cal P}_0^*)$ is now the Hamiltonian
function which is obtained as the projection of the Lagrangian energy $E_\Lag$ by ${\cal FL}_0$;
that is, $E_\mathcal{L}={\cal FL}_0^{\ *}\,H_0$. The functions $F_a^\mu\in\Cinfty({\cal P}_0^*)$ are determined by the condition $\Theta_\Lag={\cal FL}_0^*\,\Theta_{\cal H}^0$.
If $\Theta_{\cal H}^0$ is a premulticontact
form, then $({\cal P}_0^*,\Theta_{\cal H}^0,\omega)$ 
is said to be the {\sl\textbf{premulticontact Hamiltonian system}}
associated with the premulticontact Lagrangian system $({\cal P},\Theta_\L,\omega)$.

Let $(x^\mu,y^a,p^I,s^\mu)$ be adapted coordinates
for the premulticontact manifold
$({\cal P}_0^*,\Theta_{\cal H}^0,\omega)$
(see the Appendix \ref{append}). If $F_a^\mu$ are such that
\beq
\label{sufcon}
\displaystyle\derpar{F_a^\mu}{s^\nu}=0 \,,
\eeq
then
\begin{equation} \sigma_{\Theta_{\cal H}^0}=\derpar{H_0}{s^\mu}\,\d x^\mu\,,\end{equation}
and then the corresponding Hamiltonian $(m+1)$-form $\overline{\Omega}_{\cal H}^0\in\df^{m+1}({\cal P}_0^*)$ is
\begin{align}
    \overline{\Omega}_{\cal H}^0=&\ \d\Theta_{\cal H}^0+\sigma_{\Theta_{\cal H}^0}\wedge\Theta_{\cal H}^0=
    \ \d\Theta_{\cal H}^0+\derpar{H_0}{s^\mu}\,\d x^\mu\wedge\Theta_{\cal H}^0
    \nonumber \\
    =&\ \d(F_a^\mu\d y^a\wedge\d^{m-1}x_\mu+H_0\,\d^m x)-
    \Big(\derpar{H_0}{s^\mu}\,F_a^\mu\,\d y^a+\derpar{H_0}{s^\mu}\,\d s^\mu\Big)\wedge\d^mx \nonumber
    \\
    =&\ \derpar{F_a^\mu}{p^I}\,\d p^I\wedge\d y^a\wedge\d^{m-1}x_\mu+\derpar{F_a^\mu}{y^b}\,\d y^b\wedge\d y^a\wedge\d^{m-1}x_\mu+\derpar{H_0}{p^I}\,\d p^I\wedge\d^mx
    \nonumber\\
    & \ +\left(\derpar{H_0}{y^a}-\derpar{H_0}{s^\mu}\,F_a^\mu-\derpar{F_a^\mu}{x^\mu}\right)\d y^a\wedge\d^mx \,.
    \label{OmegabarH0}
\end{align}

For this system the field equations for sections $\bm\psi\colon M\to{\cal P}_0^*$ and for
$\widetilde\tau_0$-transverse, locally decomposable
multivector fields ${\bf X}_{{\cal H}_0}\in\vf^m({\cal P}_0^*)$
are, respectively,
\beq
\label{sect1Hb}
\bm{\psi}^*\Theta_{\cal H}^0= 0  \,, \qquad
\bm{\psi}^*\innp{Y}\overline{\Omega}_{\cal H}^0= 0 \,, \quad \text{for every }\ Y\in\vf({\cal }P^*) \, .
\eeq
and
\begin{equation}
\label{vfHb}
\innp{\bfX_{{\cal H}_0}}\Theta_{\cal H}^0=0 \,, \qquad \innp{\bfX_{{\cal H}_0}}\overline{\Omega}_{\cal H}^0=0 \,;
\end{equation}
where the condition of $\widetilde\tau_0$-transversality is imposed 
again by requiring that $\innp{\bfX_{{\cal H}_0}}\omega=1$.
In general, these equations are not compatible on ${\cal P}_0^*$ 
and a constraint algorithm must be implemented in order to find a final constraint submanifold ${\cal P}_f^*\hookrightarrow{\cal P}_0^*$
(if it exists) where there are integrable multivector fields solutions to the field equations on ${\cal P}_f^*$ and being tangent to ${\cal P}_f^*$.
This algorithm is explained in some of the applications developed
in Sections \ref{ex1} and \ref{ex2}.

If ${\bf X}_{{\cal H}_0}\in\vf^m({\cal P}^*_0)$ is a $\widetilde\tau_0$-transverse and locally decomposable multivector field  whose local expression is
\begin{equation}
{\bf X}_{{\cal H}_0}=
\bigwedge_{\mu=0}^{m-1}\Big(\derpar{}{x^\mu}+ (X_{{\cal H}_0})^a_\mu\frac{\partial}{\partial y^a}+
(X_{{\cal H}_0})^I_\mu\frac{\partial}{\partial p^I}+(X_{{\cal H}_0})_\mu^\nu\derpar{}{s^\nu}\Big) \,,
\end{equation}
and it is a solution to equations \eqref{vfHb};
taking into account the local expressions \eqref{thetaHcoor2}
and \eqref{OmegabarH0}, these field equations lead to
\begin{align}
0&=(X_{{\cal H}_0})^a_\mu\,F^\mu_a+H_0+(X_{{\cal H}_0})_\mu^\mu \,,
\label{Hameq10}
\\
0&=
-\derpar{F_a^\mu}{p^I}(X_{{\cal H}_0})^I_\mu+
\left(\derpar{F_b^\mu}{y^a}-\derpar{F_a^\mu}{y^b}\right)(X_{{\cal H}_0})^b_\mu +\derpar{H_0}{y^a}-\derpar{H_0}{s^\mu}\,F_a^\mu-\derpar{F_a^\mu}{x^\mu} \,,
\label{Hameq20}
  \\ 
0 &= \derpar{F_a^\mu}{p^I}(X_{{\cal H}_0})^a_\mu+\derpar{H_0}{p^I}\,,
\label{Hameq30}
\end{align}
together with a last group of equations which are identities when the above ones are taken into account.
If $\bm{\psi}(x^\nu)=(x^\mu,y^a(x^\nu),p^I(x^\nu),s^\mu(x^\nu))$
is an integral section of ${\bf X}_{{\cal H}_0}$, 
then it is a solution to the equations \eqref{sect1Hb} which read as
\begin{align}
0&=(F^\mu_a\circ\bm\psi)\derpar{y^a}{x^\mu}+(H_0\circ\bm\psi)+\derpar{s^\mu}{x^\mu}
\\
0&=
\Big(\derpar{F_a^\mu}{p^I}\circ\bm\psi\Big)\derpar{p^I}{x^\mu}+
\left(\Big(\derpar{F_b^\mu}{y^a}-\derpar{F_a^\mu}{y^b}\Big)\circ\bm\psi\right)\derpar{y^b}{x^\mu}+\Big(\derpar{H_0}{y^a}-\derpar{H_0}{s^\mu}\,F_a^\mu-\derpar{F_a^\mu}{x^\mu}\Big)\circ\bm\psi \,,
  \\ 
0 &= \Big(\derpar{F_a^\mu}{p^I}\circ\bm\psi\Big)\derpar{y^a}{x^\mu}+\Big(\derpar{H_0}{p^I}\circ\bm\psi\Big) \,.
\end{align}

In general, these equations
have no solutions everywhere on ${\cal P}_0^*$.
In the most favorable cases,
the Hamiltonian field equations are compatible only on a submanifold of ${\cal P}^*_f\subseteq{\cal P}^*_0$ which is obtained by applying a suitable constraint algorithm.
This algorithm is explained in some of the applications developed
in Sections \ref{ex1} and \ref{ex2}.

\section{Application to general Lagrangians}
\label{ex1}

As first applications of this (pre)multicontact formulation,
we consider, in general, the cases of field theories described by quadratic or affine Lagrangians. 

\subsection{Quadratic (regular and singular) Lagrangians}\label{sec:quadratic}

Many field theories in physics are described by quadratic-type Lagrangians \cite{GMS-97,Ka98},
\begin{equation} L(x^\mu,y^a,y^a_\mu) = 
\frac12 f^{\mu\nu}_{ab}(x^\alpha,y^c)\,y^a_\mu\, y^b_\nu-V(x^\alpha,y^c)\,. \end{equation}
They are usually called {\sl mechanical-type Lagrangians}, or {\sl modified kinetic energy Lagrangians} when the functions $f_{ab}^{\mu\nu}$ are not constant.
We will assume, without lost of generality, that $f^{\mu\nu}_{ab}=f^{\nu\mu}_{ba}$.
We study a modification of these Lagrangians
including a term that depends on ``action variables''.

\subsubsection{Lagrangian formalism}

As in Section \ref{mlf}, consider the bundle
${\cal P}\equiv J^1\pi\times\Real^m$
with coordinates $(x^\mu, y^a,y^a_\mu,s^\mu)$.
A {\sl quadratic Lagrangian} in ${\cal P}$ has the form,
\begin{equation}
\label{eq-quadlag}
    L(x^\mu,y^a,y^a_\mu,s^\mu) = 
\frac12 f^{\mu\nu}_{ab}(x^\alpha,y^c,s^\alpha)\,y^a_\mu\, y^b_\nu-V(x^\alpha,y^c,s^\alpha)\in\Cinfty({\cal P})  \,,
\end{equation}
where we have included a dependence on the additional variables $s^\mu$ to the functions $f^{\mu\nu}_{ab}$ and $V$. 
This Lagrangian is either regular or singular, depending on the regularity of the quadratic form whose matrix is
$\displaystyle\left(\frac{\partial^2L}{\partial y^a_\mu\partial y^b_\nu}\right)=f^{\mu\nu}_{ab}$.
The energy Lagrangian function is
\begin{equation}
E_\Lag=\frac{\partial L}{\partial y^a_\mu}\,y^a_\mu-L=\frac12 f^{\mu\nu}_{ab}y^a_\mu\, y^b_\nu+V\in\Cinfty({\cal P}) \,,
\end{equation}
the Lagrangian $m$-form \eqref{thetacoor1} is 
\begin{equation}
\Theta_{\mathcal{L}}=
-f_{ab}^{\mu\nu}y^b_\nu\,\d y^a\wedge\d^{m-1}x_\mu+E_\Lag\,\d^m x+\d s^\mu\wedge \d^{m-1}x_\mu\in\df^m({\cal P}) \,.
\end{equation}
and 
\begin{equation}
\sigma_{\Theta_{\mathcal{L}}}=-\left(\frac12\derpar{f^{\mu\nu}_{ab}}{s^\rho}\,y^a_\mu\,y^b_\nu-\derpar{V}{s^\rho}\right)\d x^\rho \,.
\end{equation}
Given a semi-holonomic $m$-multivector field,
\beq
{\bf X}_\L= \bigwedge_{\mu=0}^{m-1} X_\mu=
\bigwedge_{\mu=0}^{m-1}\Big(\derpar{}{x^\mu}+y_\mu^a\frac{\partial}{\partial y^a}+(X_\L)_{\mu\nu}^a\frac{\partial}{\partial y^a_\nu}+(X_\L)_\mu^\nu\,\frac{\partial}{\partial s^\nu}\Big)\in\vf^m({\cal P}) \,,
\label{mvflag-1}
\eeq
the Lagrangian equations \eqref{lagdef1} and \eqref{lagdef2}
take the form
\begin{align}
(X_\L)_\mu^\mu = & L\,,
\label{lagdef-0}
\\
\displaystyle
\frac12\derpar{f^{\mu\nu}_{bc}}{y^a}\,y^b_\mu\, y^c_\nu-\derpar{V}{y^a}
-\derpar{f^{\mu\nu}_{ab}}{x^\mu}y^b_\nu-\derpar{f^{\mu\nu}_{ac}}{y^b}\, y^c_\nu\, y^b_\mu-
\derpar{f^{\mu\alpha}_{ab}}{s^\nu}\,y_\alpha^b(X_\L)_\mu^\nu-
f_{ab}^{\mu\nu}(X_\L)_{\mu\nu}^b
= & \nonumber \\
\Big(\derpar{V}{s^\mu}-\frac12\derpar{f_{d c}^{\alpha\beta}}{s^\mu}\,y_\alpha^d\,y_\beta^c\Big)
f_{ab}^{\mu\nu}y_\nu^b\,. &
\label{lagdef-1}
\end{align}
For the integral (holonomic) sections
$\displaystyle \bm{\psi}(x^\nu)=\Big(x^\mu, y^a(x^\nu),\derpar{y^a}{x^\mu}(x^\nu),s^\mu(x^\nu)\Big)$ of ${\bf X}_\L$,
equations \eqref{lagdef-0} and \eqref{lagdef-1} read,
\begin{align}
\derpar{s^\mu}{x^\mu} = & L\,, \label{lagdef-00}\\
\displaystyle
\frac12\derpar{f^{\mu\nu}_{bc}}{y^a}\,\derpar{y^b}{x^\mu}\derpar{y^c}{x^\nu}-\derpar{V}{y^a}
-\derpar{f^{\mu\nu}_{ab}}{x^\mu}\derpar{y^b}{x^\nu}-\derpar{f^{\mu\nu}_{ac}}{y^b}\,\derpar{y^c}{x^\nu}\derpar{y^b}{x^\mu}-
\derpar{f^{\mu\alpha}_{ab}}{s^\nu}\,\derpar{y^b}{x^\alpha}\derpar{s^\nu}{x^\mu}-
f_{ab}^{\mu\nu}\frac{\partial^2y^b}{\partial x^\mu\partial x^\nu}
= & 
\nonumber\\
\Big(\derpar{V}{s^\mu}-\frac12\derpar{f_{d c}^{\alpha\beta}}{s^\mu}\,\derpar{y^d}{x^\alpha}\,\derpar{y^c}{x^\beta}\Big)
f_{ab}^{\mu\nu}\derpar{y^b}{x^\nu} \,, &
\label{lagdef-11}
\end{align}
which are the equations \eqref{ELeqs2} and \eqref{ELeqs1}
for this Lagrangian.

When the Lagrangian is regular, $\Theta_{\mathcal{L}}$ is a multicontact form and these equations are compatible.
In the singular case, if $\Theta_{\mathcal{L}}$ is a premulticontact form, equations \eqref{lagdef-1} could be incompatible; 
then compatibility constraints appear, and they define a submanifold of ${\cal P}$ where the equations have solutions.
Therefore, a constraint algorithm must be implemented as usual
in order to find the final constraint submanifold 
${\cal S}_f\subseteq{\cal P}$, if it exists, where there are consistent solutions.

As a particular situation, we can assume that
$\displaystyle\derpar{f^{\mu\nu}_{ab}}{s^\alpha}=0$ and that any coordinate has a velocity present, that is, for any $a$ there exist $\mu,\nu$ and $b$ such that $f^{\mu\nu}_{ab}\neq0$. Then, the equations \eqref{lagdef-1} reduce to
\begin{equation}
\frac12\derpar{f^{\mu\nu}_{bc}}{y^a}\,y^b_\mu y^c_\nu-\derpar{V}{y^a}
-\derpar{f^{\mu\nu}_{ab}}{x^\mu}y^b_\nu-\derpar{f^{\mu\nu}_{ac}}{y^b}\, y^c_\nu\, y^b_\mu-
f_{ab}^{\mu\nu}(X_\L)_{\mu\nu}^b
= \derpar{V}{s^\mu}\, f_{ab}^{\mu\nu}y_\nu^b \,.
\end{equation}
which, as mentioned above, can be compatible or not depending on the regularity of $L$; namely, on the quadratic form $f_{ab}^{\mu\nu}$.
For the integral (holonomic) sections of ${\bf X}_\L$,
these equations lead to
\begin{equation}\label{eq:quadLag}
\frac12\derpar{f^{\mu\nu}_{bc}}{y^a}\,\derpar{y^b}{x^\mu} \derpar{y^c}{x^\nu}-\derpar{V}{y^a}
-\derpar{f^{\mu\nu}_{ab}}{x^\mu}\derpar{y^b}{x^\nu}-\derpar{f^{\mu\nu}_{ac}}{y^b}\, \derpar{y^c}{x^\nu} \derpar{y^b}{x^\mu}-
f_{ab}^{\mu\nu}\frac{\partial^2y^b}{\partial x^\mu\partial x^\nu}
= \derpar{V}{s^\mu}\, f_{ab}^{\mu\nu}\derpar{y^b}{x^\nu} \,.
\end{equation}

\subsubsection{Hamiltonian formalism}

The Legendre map generated by the Lagrangian \eqref{eq-quadlag} gives the multimomenta
\beq
\label{quadLeg}
p_a^\mu=f_{ab}^{\mu\nu}\,y_\nu^b \,.
\eeq

In the hyperregular case this map is a diffeomorphism and these conditions give us the Hamiltonian function
\begin{equation}
H(x^\mu,y^a,p_a^\mu,s^\mu) = \frac12 f_{\mu\nu}^{ab}(x^\mu,y^a,s^\mu)\,p_a^\mu\,p_b^\nu+V(x^\mu,y^a,s^\mu)
\in\Cinfty({\cal P}^*)\,,
\end{equation}
where $\displaystyle (f_{\mu\nu}^{ab})=(f^{\mu\nu}_{ab})^{-1}$
in the sense $\displaystyle f_{\mu\nu}^{ab}\,f^{\nu\rho}_{bc}=\delta_c^a\,\delta_\mu^\rho$.
The Hamiltonian premulticontact $m$-form has the local expression \eqref{thetaHcoor}, and
\begin{equation}
\sigma_{\mathcal{H}}=\left(\frac12\derpar{f_{\mu\nu}^{ab}}{s^\rho}\,p_a^\mu\,p_b^\nu+\derpar{V}{s^\rho}\right)\d x^\rho \,.
\end{equation}
For a multivector field ${\bf X}_{\cal H}\in\vf^m({\cal P}^*)$
whose local expression is \eqref{Hammv},
the Hamilton--De Donder--Weyl equations \eqref{coor2} are
\begin{gather}
(X_{\cal H})_\mu^\mu = 
f_{\mu\nu}^{ab}\,p_a^\mu\,p_b^\nu-V \,,\qquad
(X_{\cal H})^a_\mu=f_{\mu\nu}^{ab}\,p_b^\nu\,,
\\
(X_{\cal H})_{\mu a}^\mu= 
-\left(\frac12
\derpar{f_{\nu\rho}^{bc}}{y^a}\,p_b^\nu\,p_c^\rho+
\derpar{V}{y^a}+\frac12\derpar{f_{\nu\rho}^{bc}}{s^\mu}\,p_a^\mu\,p_b^\nu\,p_c^\rho+\derpar{V}{s^\mu}\,p_a^\mu\right) \,.\end{gather}
The integral sections
$\bm{\psi}(x^\nu)=(x^\mu,y^a(x^\nu),p^\mu_a(x^\nu),s^\mu(x^\nu))$
of ${\bf X}_{\cal H}$
are the solutions to the equations \eqref{coor1} which in this case are
\begin{gather}
\frac{\partial s^\mu}{\partial x^\mu} = 
f_{\mu\nu}^{ab}\,p_a^\mu\,p_b^\nu-V \,,\qquad
\frac{\partial y^a}{\partial x^\mu}=f_{\mu\nu}^{ab}\,p_b^\nu\,,
\\
\frac{\partial p^\mu_a}{\partial x^\mu}= 
-\left(\frac12\derpar{f_{\nu\rho}^{bc}}{y^a}\,p_b^\nu\,p_c^\rho+
\derpar{V}{y^a}+\frac12\derpar{f_{\nu\rho}^{bc}}{s^\mu}\,p_a^\mu\,p_b^\nu\,p_c^\rho+\derpar{V}{s^\mu}\,p_a^\mu\right)\,.
\end{gather}
The singular case corresponds to the situation in which the matrix $(f_{ab}^{\mu\nu})$ is not regular.
If the rank of this matrix is constant everywhere on ${\cal P}$, then
the Lagrangian is almost-regular.
Therefore, the Legendre map is not a diffeomorphism
but a submersion onto its image,
and the equations \eqref{quadLeg} yield constraints defining
the submanifold ${\cal P}_0^*\subset{\cal P}^*$.
Then, we have the Hamiltonian function 
$H_0\in\Cinfty({\cal P}_0^*)$
and the Hamiltonian premulticontact $m$-form $\Theta_{\mathcal{H}}^0\in\df^m({\cal P}_0^*)$
given by \eqref{thetaHcoor},
and we are in a situation like the one described generically in Section \ref{Hamsing}.

\subsection{Affine Lagrangians}
\label{subsec:affine}

Some relevant classical field theories in physics such as
the \textsl{Einstein--Palatini} (or \textsl{metric-affine}) approach
to gravitation \cite{Capriotti,Einstein,Gaset:2018uwi}, or
Dirac fermion fields \cite{GMS-97} (among others),
are described by affine Lagrangians.
In natural coordinates, their general expression is
\begin{equation} L(x^\mu,y^b,y^b_\mu)=f^\alpha_a(x^\mu,y^b)\,y^a_\alpha-V(x^\mu,y^b)\,. \end{equation}
Next, we study a modification of these Lagrangians
which include a dependence on ``action variables''.

\subsubsection{Lagrangian formalism}

As above, consider the bundle
${\cal P}\equiv J^1\pi\times\R^m$
with coordinates $(x^\mu, y^a,y^a_\mu,s^\mu)$.
An {\sl affine Lagrangian} in ${\cal P}$ is a function of the form,
\begin{equation}
\label{eq-afflag}
L(x^\mu,y^b,y^b_\mu,s^\mu) = 
f^\alpha_a(x^\mu,y^b,s^\mu)\,y^a_\alpha-V(x^\mu,y^b,s^\mu)\in\Cinfty({\cal P})  \,,
\end{equation}
which is a singular Lagrangian since
$\displaystyle \frac{\partial^2L}{\partial y^b_\nu\partial y^a_\mu}=0$. 

Note that, as pointed out in Remark \ref{rem1}, not every Lagrangian function of the form \eqref{eq-afflag} yields a premulticontact structure. A particular case giving premulticontact structures are those affine Lagrangians such that $\displaystyle\derpar{f_a^\alpha}{s^\mu} = 0$, and there exist functions $J^b_\mu\in\mathcal{C}^\infty(\mathcal{P})$ such that
\begin{equation}\label{eq:cond2affine}
 \frac{\partial V}{\partial s^\mu}f^\mu_a+\frac{\partial V}{\partial y^a}+\frac{\partial f^\mu_a}{\partial x^\mu}=J^b_\mu\left(\frac{\partial f^\mu_a}{\partial y^b}-\frac{\partial f^\mu_b}{\partial y^a}\right).   
\end{equation}
A relevant example of an affine Lagrangian satisfying these conditions is the metric-affine Lagrangian for General Relativity (see Section \ref{ex:EP}). Hereafter, we assume that all the affine Lagrangians are of this kind.

The Lagrangian energy function is
\begin{equation}
E_\Lag=\frac{\partial L}{\partial y^a_\mu}y^a_\mu-L=V\in\Cinfty({\cal P}) \,,
\end{equation}
the Lagrangian $m$-form \eqref{thetacoor1} is
\begin{equation}
\Theta_{\mathcal{L}}=
-f_a^\mu\,\d y^a\wedge\d^{m-1}x_\mu+V\,\d^m x+\d s^\mu\wedge \d^{m-1}x_\mu\in\df^m({\cal P}) \,,
\end{equation}
and
\begin{equation}
\sigma_{\Theta_{\mathcal{L}}}=-\left(\derpar{f^\alpha_a}{s^\mu}\,y^a_\alpha-\derpar{V}{s^\mu}\right)\d x^\mu \,.
\end{equation}
For a semi-holonomic $m$-multivector field such as
\begin{equation}
{\bf X}_\L=\bigwedge_{\mu=0}^{m-1} X_\mu=
\bigwedge_{\mu=0}^{m-1}\left(\derpar{}{x^\mu}+y_\mu^a\frac{\partial}{\partial y^a}+(X_\L)_{\mu\nu}^a\frac{\partial}{\partial y^a_\nu}+(X_\L)_\mu^\nu\,\frac{\partial}{\partial s^\nu}\right)\in\vf^m({\cal P}) \,,
\label{mvflag}
\end{equation}
the Lagrangian equations \eqref{lagdef1} and \eqref{lagdef2} read
\begin{align}
(X_\L)_\mu^\mu &=L\,,
\label{lagdef0}
\\
\displaystyle
\derpar{f^\mu_b}{y^a}\,y^b_\mu-\derpar{V}{y^a}
-\derpar{f^\mu_a}{x^\mu}-\derpar{f^\mu_a}{y^b}\, y^b_\mu&=
\derpar{V}{s^\mu}
f_a^\mu\,.
\label{lagdef11}
\end{align}
For the integral (holonomic) sections
$\displaystyle \bm{\psi}(x^\nu)=\Big(x^\mu, y^a(x^\nu),\derpar{y^a}{x^\mu}(x^\nu),s^\mu(x^\nu)\Big)$ of ${\bf X}_\L$,
equations \eqref{lagdef0} and \eqref{lagdef11} lead to
\begin{align}
\label{lagdefsec0}
\derpar{s^\mu}{x^\mu} &=L\,,
\\
\label{lagdefsec1}
\displaystyle
\derpar{f^\mu_b}{y^a}\derpar{y^b}{x^\mu}-\derpar{V}{y^b}
-\derpar{f^\mu_a}{x^\mu}-\derpar{f^\mu_a}{y^b}\derpar{y^b}{x^\mu}&=
\derpar{V}{s^\mu}
f_a^\mu\,,
\end{align}
which are the equations \eqref{ELeqs2} and \eqref{ELeqs1}
for this system.

Due to the singularity of affine Lagrangians, equations \eqref{lagdef11} are constraints, which we denote by $\zeta_a$. They define the compatibility submanifold ${\cal S}_1\subset{\cal P}$.
Next, we impose the tangency of the multivector fields \eqref{mvflag}
which are solutions to the Lagrangian equations on ${\cal S}_1$.
These conditions are
\begin{equation}
X_\alpha(\zeta_a)\vert_{{\cal S}_1}=0 \,,
\end{equation}
and originate a system of equations giving relations among the component functions $(X_\L)_a^\nu$ and $(X_\L)_\mu^\nu$, and/or
new constraints. In the case where new constraints arise, the constraint algorithm continues as usual until obtaining the final constraint submanifold ${\cal S}_f\subseteq{\cal P}$.

\subsubsection{Hamiltonian formalism}
\label{subsubsection:affineHam}

The Legendre map generated by the Lagrangian \eqref{eq-afflag} gives the relations
\begin{equation}
p_a^\mu=f_a^\mu \,,
\end{equation}
which are {\sl primary constraints} defining the submanifold ${\cal P}_0^*\subset{\cal P}^*$;
thus ${\cal P}_0^*$ is diffeomorphic to $E$
and local coordinates in ${\cal P}_0^*$ are $(x^\mu,y^a,s^\mu)$.
The Lagrangian is almost-regular and, as stated above, we assume the conditions $\displaystyle\derpar{f^\mu_a}{s^\nu}=0$ and \eqref{eq:cond2affine}.

The Hamiltonian function is
\begin{equation}
H_0=V\in\Cinfty({\cal P}_0^*) \,.
\end{equation}
The Hamiltonian premulticontact $m$-form \eqref{thetaHcoor2} is
\begin{equation}
\Theta_{\mathcal{H}}^0=
-f_a^\mu\,\d y^a\wedge\d^{m-1}x_\mu+V\,\d^m x+\d s^\mu\wedge \d^{m-1}x_\mu\in\df^m({\cal P}_0^*) \,,
\end{equation}
and $(x^\mu,y^a,s^\mu)$ are adapted coordinates for
the premulticontact structure $(\Theta_{\mathcal{H}}^0,\omega)$.
Furthermore,
\begin{equation}\sigma_{{\mathcal{H}}^0}=\derpar{V}{s^\mu}\,\d x^\mu\,. \end{equation}
For a $\widetilde\tau_0$-transverse, locally decomposable
multivector field,
such as
\begin{equation}
{\bf X}_{\mathcal{H}_0}= \bigwedge_{\mu=0}^{m-1} X_\mu=
\bigwedge_{\mu=0}^{m-1}\Big(\derpar{}{x^\mu}+(X_{\mathcal{H}_0})_\mu^a\frac{\partial}{\partial y^a}+(X_{\mathcal{H}_0})_\mu^\nu\,\frac{\partial}{\partial s^\nu}\Big)\in\vf^m({\cal P}_0^*) \,,
\end{equation}
the Hamiltonian equations \eqref{Hameq10} and \eqref{Hameq20} lead to
\begin{align}
0&=-(X_{{\cal H}_0})^a_\mu\,f^\mu_a+V+(X_{{\cal H}_0})_\mu^\mu \,,
 \label{coor2b0}
\\
0&=
\left(\derpar{f_a^\mu}{y^b}-\derpar{f_b^\mu}{y^a}\right)(X_{{\cal H}_0})^b_\mu +\derpar{V}{y^a}+\derpar{V}{s^\mu}\,f_a^\mu+\derpar{f_a^\mu}{x^\mu} \,,
\label{coor2b}
\end{align}
and equations \eqref{Hameq30} vanish identically.
For the integral sections
$\displaystyle \bm{\psi}(x^\nu)=\big(x^\mu, y^a(x^\nu),s^\mu(x^\nu)\big)$ of ${\bf X}_{\mathcal{H}_0}$,
equations \eqref{coor2b0} and \eqref{coor2b} read,
\begin{align}
0&=-\derpar{y^a}{x^\mu}\,f_a^\mu+ V+\derpar{s^\mu}{x^\mu} \,, \\
0&=
\left(\derpar{f_a^\mu}{y^b}-\derpar{f_b^\mu}{y^a}\right)\derpar{y^b}{x^\mu} +\derpar{V}{y^a}+\derpar{V}{s^\mu}\,f_a^\mu+\derpar{f_a^\mu}{x^\mu} \,.
\end{align}

The compatibility of equations \eqref{coor2b} depends on the rank of the matrix $\displaystyle \left(\derpar{f_a^\mu}{y^b}-\derpar{f_b^\mu}{y^a} \right) $.
If this system is incompatible, then compatibility constraints could appear,
defining a submanifold of ${\cal P}^*_0$ where the equations have solutions,
and therefore the constraint algorithm must be implemented as usual
until finding the final constraint submanifold 
${\cal P}_f^*\subseteq{\cal P}^*_0$ where consistent solutions exist.

\section{Application to particular theories}
\label{ex2}

In this section, we study some specific classical field theories in physics described by quadratic and affine Lagrangians. We will keep a generic expression for the Lagrangian's coefficients. Due to this generality, not all possible values will lead to stable or physically sound systems. Some particular instances with an interesting interpretation will be considered.

\subsection{One-dimensional wave equation with first-order terms and external force}\label{ex:wave}

We study a one-dimensional wave equation, which can be considered an infinitely long vibrating string.
Consider the coordinates $(t,x)$ for the time and the space. Denote by $y$ the separation of a point in the string from its equilibrium point, and hence $y_t$ and $y_x$  denote the multivelocities associated with the two independent variables. The Lagrangian function for this system is
\begin{equation}\label{eq:Lagrangian-vibrating-string}
L(y,y_t,y_x)=\frac{1}{2}\rho y_t^2-\frac{1}{2}\tau y_x^2\,,
\end{equation}
where $\rho$ is the linear mass density of the string and $\tau$ is the tension of the string. We assume that these quantities are constant. The Euler--Lagrange equation for this Lagrangian density is
\begin{equation} 
\derpar{^2y}{t^2}=c^2\derpar{^2y}{x^2}\,, 
\end{equation}
where $c^2 = \dfrac{\tau}{\rho}$ is the square of the propagation speed of the wave. This last equation is the one-dimensional wave equation.

\subsubsection{Lagrangian formalism}

This is a particular case of a hyperregular quadratic Lagrangian.

To model a vibrating string with linear damping, we can modify the Lagrangian function \eqref{eq:Lagrangian-vibrating-string} so that it becomes a multicontact Lagrangian \cite{LGMRR-2022}. The fiber bundle $\tau\colon{\cal P}\to \R^2$ has adapted coordinates $(t,x,y,y_t,y_x,s^t,s^x)$, and the volume form of the base space $\R^2$ reads $\omega = \d t\wedge\d x$. The modified Lagrangian function $L$ reads
\begin{equation}\label{eq:Lagrangian-vibrating-string-dis}
L(t,x,y,y_t,y_x,s^t,s^x)=\frac{1}{2}\rho y_t^{\,2}-\frac{1}{2}\tau y_x^{\,2}+as^t-b\,\frac{\rho}{\tau}\,s^x+\rho yf(t,x)\,,
\end{equation}
where $a,b\in\mathbb{R}$ are constants, and $f(t,x)$ represents an external force acting on the string. 
It is a regular Lagrangian.
The Lagrangian energy associated with the Lagrangian \eqref{eq:Lagrangian-vibrating-string-dis} is
\begin{equation}
E_\L=\frac{1}{2}\rho y_t^{\,2}-\frac{1}{2}\tau y_x^{\,2}-as^t+b\,\frac{\rho}{\tau}\,s^x-\rho yf(t,x)\,.
\end{equation}
The Lagrangian multicontact two-form \eqref{thetacoor1} is now
\begin{equation}
\Theta_{\mathcal{L}}=
-\rho y_t\,\d y\wedge\d x-\tau y_x\,\d y\wedge\d t+E_\Lag\,\d t\wedge\d x+\d s^t\wedge \d x-\d s^x\wedge\d t \,.
\end{equation}
and
\begin{equation} \sigma_{\Theta_{\mathcal{L}}} = -\parder{L}{s^t}\d t - \parder{L}{s^x}\d x = -a\d t + b\frac{\rho}{\tau}\d x\,. \end{equation}
The Lagrangian equations  \eqref{lagdef-00} for a holonomic section
$$ \bm{\psi}(t,x)=\left(t,x,y(t,x),\derpar{y}{t}(t,x),\derpar{y}{x}(t,x),s^t(t,x),s^x(t,x)\right)
$$
read
\begin{align}
\derpar{s^t}{t}+\derpar{s^x}{x} &=L\,,
\\
\derpar{^2y}{t^2}-\frac{\tau}{\rho}\derpar{^2y}{x^2}&=a\derpar{y}{t}+b\derpar{y}{x}+f(t,x) \,.
\label{eq:vibrating-string-dampings}
\end{align}
which are the equations \eqref{ELeqs2} and \eqref{ELeqs1}
for this Lagrangian. 
In particular, equation \eqref{eq:vibrating-string-dampings} is the one-dimensional wave equation with an external force $f(t,x)$ 
and two first-order terms accounting for the non-conservativity.

\subsubsection{Hamiltonian formalism}

The fiber bundle $\widetilde{\tau}\colon{\P^\ast}\to \R^2$ has adapted coordinates $(t,x,y,p^t,p^x,s^t,s^x)$. 
The Legendre map $\F\L\colon\P\to\P^\ast$ associated with the Lagrangian \eqref{eq:Lagrangian-vibrating-string-dis} is given by
\begin{equation} 
\F\L(t,x,y,y_t,y_x,s^t,s^x) = (t,x,y, p^t, p^x, s^t, s^x)\,, 
\end{equation}
where $p^t = \rho y_t$ and $p^x = -\tau y_x$. 
The Legendre map is a global diffeomorphism since the Lagrangian function is hyperregular.
The Hamiltonian function is
\begin{equation} H = \frac{1}{2\rho}(p^t)^2 - \frac{1}{2\tau}(p^x)^2 - a s^t + b\,\frac{\rho}{\tau}\,s^x-\rho yf(t,x)\,. \end{equation}
The Hamiltonian $m$-form \eqref{thetaHcoor} is
\begin{align}
    \Theta_{\mathcal{H}}&=
-p^t\,\d y\wedge\d x+p^x\,\d y\wedge\d t+\d s^t\wedge \d x-\d s^x\wedge\d t
\\
&\left(\frac{1}{2\rho}(p^t)^2 - \frac{1}{2\tau}(p^x)^2 - a s^t + b\,\frac{\rho}{\tau}\,s^x-\rho yf(t,x)\right)\,\d t\wedge\d x \,,
\end{align}
and now
\begin{equation} \sigma_{\Theta_{\mathcal{H}}} = \parder{H}{s^t}\d t + \parder{H}{s^x}\d x = -a\d t + b\frac{\rho}{\tau}\d x\,. \end{equation}
For a section $\bm{\psi}(t,x) = (t,x,y(t,x),p^t(t,x),p^x(t,x),s^t(t,x),s^x(t,x))$, 
the Herglotz--Hamilton--De Donder--Weyl equations \eqref{coor1} read
\begin{gather}
    \parder{s^t}{t} + \parder{s^x}{x} = \frac{1}{2\rho}(p^t)^2 - \frac{1}{2\tau}(p^x)^2 + a s^t - b\frac{\rho}{\tau}s^x + \rho y f(t,x)\,, \\
    \parder{y}{t} = \frac{1}{\rho}p^t\,,\qquad 
    \parder{y}{x} = -\frac{1}{\tau}p^x\,,\\
    \parder{p^t}{t} + \parder{p^x}{x} = \rho f(t,x) + a p^t - b\frac{\rho}{\tau}p^x \,.
\end{gather}
Combining the last three equations above, we obtain equation \eqref{eq:vibrating-string-dampings}.
Thus, the Lagrangian and Hamiltonian formalisms are equivalent.

\subsection{Klein--Gordon equation. The telegrapher equation}
\label{ex:K-G}

The {\sl Klein--Gordon equation}, 
\begin{equation}\label{klein-gordon}
(\Box + m^2) y\equiv\partial_\mu\partial^\mu y + m^2 y\equiv \parderr{y}{x_\mu}{x^\mu} + m^2 y = 0
\,,
\end{equation}
is one of the most important equations in classical and quantum field theory \cite[p.\,108]{IZ,LPAF2018}.
In it, $y(x^\mu)$ is a scalar field in the Minkowski space-time $\R^4$,
we denote $\displaystyle\partial^\mu\equiv g^{\mu\nu}\partial_\nu$, $m\geq 0$ is a constant parameter, and $\Box$ denotes de D´Alembert operator in $\R^4$.

To develop the multisymplectic formulation of this theory
\cite{GR-2024,GGR-2023},
we use space-time coordinates $(x^\mu)$, $\mu=0,\ldots,3$,
we denote $y$ the field variable, and $y_\mu$ the corresponding multivelocities.
Then, the multivelocity phase space $J^1\pi$ has natural coordinates $(x^\mu,y,y_\mu)$,
and hence, the Lagrangian function for the Klein--Gordon equation is
\begin{equation}
\label{klein-gordon-lagrangian}
L_0(x^\mu,y,y_\mu)=
\frac{1}{2}\,y_\mu\, y^\mu-\frac{1}{2} m^2 y^2 \,,
\end{equation}
(where $y^\mu=g^{\mu\nu}y_\nu$),
which, evaluated on sections $\ds\phi(x)=\Big(x^\mu,y(x),\derpar{y}{x^\mu}(x)\Big)$, gives 
$\ds L_0(\phi)=\frac{1}{2}\,\partial_\mu y\,\partial^\mu y-\frac{1}{2}m^2y^2$.
This Lagrangian function can be slightly modified to include a more generic potential,
$\displaystyle L_0=\frac{1}{2}\,y_\mu\,y^\mu-V(y)$; however, we stick to the simplest case.
Note that, in any case, it is a quadratic hyperregular Lagrangian (see Section \ref{sec:quadratic}).

\subsubsection{Lagrangian formalism}

Consider now the bundle $\tau\colon{\cal P}\to \R^4$,
with adapted coordinates $(x^\mu,y,y_\mu,s^\mu)$, 
and the volume form $\omega=\d x^0\wedge\dotsb\wedge\d x^3\equiv\d^4x$ on $\R^4$.
The contactified Lagrangian $L\in\Cinfty({\cal P})$ we propose is given by
\begin{equation}
\label{klein-gordon-lagrangian-dis}
L(x^\mu,y,y_\mu,s^\mu)=L_0(x^\mu,y,y_\mu)+\gamma_\mu s^\mu = 
\frac{1}{2}\,y_\mu y^\mu-\frac{1}{2} m^2 y^2+\gamma_\mu s^\mu
\,,
\end{equation}
where $\gamma\equiv(\gamma_\mu)\in\R^4$ is a constant vector.
It is a regular Lagrangian.
Its associated Lagrangian energy is
\begin{equation}
E_\L=  \frac{1}{2}\, y_\mu\, y^\mu
+\frac{1}{2}\, m^2 y^2-\gamma_\mu s^\mu\,.
\end{equation}
The Lagrangian multicontact $4$-form \eqref{thetacoor1} is now
\begin{align}
\Theta_\L&= y^\mu\d y\wedge\d x_\mu + E_\L\d^4 x + \d s^\mu\wedge \d x_\mu \,.
\end{align}
 Now, $\sigma_{\Theta_\L} = -\gamma_\mu \d x^\mu$. 
The equations \eqref{lagdef-11} for holonomic sections
$\displaystyle \bm{\psi}(x^\nu)=\Big(x^\mu,y(x^\nu),\derpar{y}{x^\mu}(x^\nu),s^\mu(x^\nu)\Big)$ 
are
\begin{align}
\derpar{s^\mu}{x^\mu} &=L\,,
\\
m^2y + g^{\mu\nu}\frac{\partial^2y}{\partial x^\mu\partial x^\nu} &= g^{\mu\nu}\gamma_\nu\parder{y}{x_\mu}  \,.\label{eq:KG-dampings}
\end{align}
and they are the equations \eqref{ELeqs2} and \eqref{ELeqs1}
for this Lagrangian. Denoting $\gamma^\mu=g^{\mu\nu}\gamma_\nu$, we can write equation \eqref{eq:KG-dampings} as
\begin{equation}
\label{eq:KG-dampings-2}
\parderr{y}{x_\mu}{x^\mu} + m^2 y = \gamma^\mu \parder{y}{x^\mu}\,,
\end{equation}
which is the Klein--Gordon equation with additional first-order terms.

\begin{remark}
For simplicity, in this example, we have considered the Minkowski metric and $\gamma_\mu$ constants. However, a similar procedure can be performed for a generic metric $g_{\mu\nu}=g_{\mu\nu}(x^\nu)$ and functions $\gamma_\mu = \gamma_\mu(x^\nu)$, thus obtaining the equation
\begin{equation} \parderr{y}{x_\mu}{x^\mu} + m^2 y + \parder{g_{\mu\nu}}{x^\mu}\derpar{y}{x^\nu} = \gamma^\mu\parder{y}{x^\mu}\,. \end{equation}
\end{remark}

\paragraph{Telegrapher's equation}

An interesting application of this modified Klein--Gordon equation \eqref{eq:KG-dampings-2} is that we can derive from it the so-called
{\sl telegrapher's equation} (see \cite[p.\,306]{HaBu} and \cite[p.\,653]{Salsa}),
which describes the current and voltage on a uniform electrical transmission line:
\begin{equation}
    \begin{dcases}
        \parder{V}{x} = -L\parder{I}{t} - RI\,,\\
        \parder{I}{x} = -C\parder{V}{t} - GV\,,
    \end{dcases}
\end{equation}
where $V$ is the voltage, $I$ is the current, $R$ is the resistance, $L$ is the inductance, $C$ is the capacitance, and $G$ is the conductance.
This system can be decoupled, obtaining the system
\begin{equation}
    \begin{dcases}
        \parder{^2V}{x^2} = LC\parder{^2V}{t^2} + (LG + RC)\parder{V}{t} + RGV\,,\\
        \parder{^2I}{x^2} = LC\parder{^2I}{t^2} + (LG + RC)\parder{I}{t} + RGI\,.\\
    \end{dcases}
\end{equation}
Note that the two equations in the system above are identical, and also known as telegrapher's equations. Both of them can be written as
\begin{equation}\label{eq:telegraph-equation}
    \Box y + \gamma\parder{y}{t} + m^2 y = 0\,,
\end{equation}
where $\Box$ is the d'Alembert operator in 1+1 dimensions, and $\gamma$ and $m^2$ are adequate constants. Written this way, we can see the telegrapher equation as a modified Klein--Gordon equation. More precisely, taking $\gamma_\mu = (-\gamma, 0, 0, 0)$ in \eqref{eq:KG-dampings-2}, we obtain the telegrapher equation \eqref{eq:telegraph-equation}.

\subsubsection{Hamiltonian formalism}

The adapted coordinates of the fiber bundle $\widetilde{\tau}\colon{\P^\ast}\to \R^2$ are $(x^\mu,y,p^\mu,s^\mu)$. 
The Legendre map $\F\L\colon\P\to\P^\ast$ associated with the Lagrangian \eqref{klein-gordon-lagrangian-dis} is
\begin{equation} 
\F\L(x^\mu,y,y_\mu,s^\mu)=(x^\mu,y,p^\mu,s^\mu)\,, 
\end{equation}
with $p^\mu= y_\mu$. 
It is a diffeomorphism since the Lagrangian function is hyperregular.
The Hamiltonian function is
\begin{equation}
H=\frac{1}{2} p^\mu p_\mu+\frac{1}{2} m^2 y^2-\gamma_\mu s^\mu\,,
\end{equation}
the contact Hamiltonian $m$-form \eqref{thetaHcoor} is
\begin{align}
\Theta_\H&= p^\mu\d y\wedge\d x_\mu + H\,\d^4 x + \d s^\mu\wedge \d x_\mu \,,
\end{align}
and now $\sigma_{\Theta_\H}=-\gamma_\mu \d x^\mu$.
The Herglotz--Hamilton--De Donder--Weyl equations \eqref{coor1} for sections $\bm{\psi}(x^\nu)=(x^\mu,y(x^\nu),p^\mu(x^\nu),s^\mu(x^\nu))$ 
are
\begin{gather}
	\parder{s^\mu}{x^\mu} = \frac{1}{2}p^\mu p_\mu - \frac{1}{2}m^2y^2 + \gamma_\mu s^\mu\,, \\
    \parder{y}{x^\mu} = p_\mu \,,\qquad
    \parder{p^\mu}{x^\mu} = - m^2 y + \gamma_\mu p^\mu \,.
\end{gather}
Combining the last two equations above, we obtain equation \eqref{eq:KG-dampings-2}.
Thus, the Lagrangian and Hamiltonian formalisms are equivalent.

As pointed out at the end of the previous section, taking $\gamma_\mu = (-\gamma, 0, 0, 0)$, we recover the telegrapher's equation \eqref{eq:telegraph-equation} as a particular case of the Klein--Gordon equation with damping.

\subsection{Maxwell's electromagnetism}\label{ex:EM}

Action-dependent Maxwell's equations have been studied in the context of electromagnetism in matter. A variational derivation was first done in \cite{LPAF2018} and later formalized using $k$-contact geometry in \cite{GasMar21,GRR-2022}.  The damped/forced electromagnetic waves have been studied in \cite{LPAF2018} and some applications to materials are explored in \cite{GasMar21}. We present the multicontact description based on \cite{LGMRR-2022}.

\subsubsection{Lagrangian formalism}

We use the description of Electromagnetism in terms of principle bundles, that is, as a Yang-Mills theory. Consider the principle bundle $P \rightarrow M$ with structure group $U(1)$ over a four-dimensional space-time $M$. The associated bundle of connections is denoted as $\pi:C\rightarrow M$ (see \cite{CasMu} for more details). The multicontact Lagrangian formalism takes place in the space $\mathcal{P}=J^1\pi\times_M\Lambda^{3}(\Tan^*M) $, 
with local coordinates $(x^\mu, A_\mu,A_{\mu,\nu},s^\mu)$ such that $\omega=\d x^0\wedge \d x^1\wedge \d x^2\wedge\d x^3$. 
For this example, we consider the electromagnetic Lagrangian with a linear dissipation term:
\begin{equation}
L =  -\frac{1}{4\mu_0}g^{\alpha\mu}g^{\beta\nu}F_{\mu\nu}F_{\alpha\beta}-A_\alpha J^\alpha-\gamma_\alpha s^\alpha\,,
\end{equation}
where $F_{\mu\nu}=A_{\nu,\mu}-A_{\mu,\nu}$ is the electromagnetic tensor field, $g^{\mu\nu}$ is a constant metric on $M$ with signature $(-,+,+,+)$, $J^\alpha$ and $\gamma_\alpha\in \Cinfty(M)$ are smooth functions (for $0\leq\alpha\leq 3$) and $\mu_0$ is a constant \cite{GasMar21}. Notice that it is a singular Lagrangian.

The Lagrangian energy is
\begin{equation}
 E_\mathcal{L} = -\frac{1}{4\mu_0}g^{\mu\nu}g^{\alpha\beta}F_{\beta\nu}F_{\alpha\mu}+A_\alpha J^\alpha+\gamma_\alpha s^\alpha\,,
\end{equation}
and the Lagrangian $4$-form is
\begin{equation}
\Theta_\mathcal{L} =\frac{1}{\mu_0}g^{\alpha\beta}g^{\mu\nu}F_{\beta\nu}\d A_\alpha\wedge\d ^3x_\mu + E_{\mathcal{L}} \d ^4x+\d s^\mu\wedge\d ^3x_\mu\,.
\end{equation}

As the Lagrangian is singular, we should not expect the system to be multicontact, but it could be premulticontact if some extra compatibility conditions are met (see \ref{multicontactbundle}). To check it, we need to compute the characteristic and Reeb distributions:
\begin{equation}
\mathcal{C}=\left<\frac{\partial}{\partial A_{\mu,\nu}}+\frac{\partial}{\partial A_{\nu,\mu}}\right>_{\mu,\nu=0,1,2,3}
   \,,\qquad
\mathcal{D}^\mathfrak{R}=\left<\frac{\partial}{\partial A_{\mu,\nu}}+\frac{\partial}{\partial A_{\nu,\mu}},\frac{\partial}{\partial s^\mu}\right>_{\mu,\nu=0,1,2,3}\,.
\end{equation}
We see that rank$(\mathcal{C})=10$ and rank$(\mathcal{D}^\mathfrak{R})=14$, which are compatible with a space-time of dimension $m=4$ and $k=4$. The last condition holds because $\displaystyle \innp{\frac{\partial}{\partial s^\mu}}\Theta_{\mathcal{L}}=\d^3x_\mu$. Therefore, it is a premulticontact Lagrangian system. The corresponding dissipation form is  $\sigma_{\Theta_{\mathcal{L}}}=\gamma_\alpha\d x^\alpha$. 

For the integral holonomic sections $\displaystyle\bm{\psi}(x^\mu)=\Big(x^\mu,A_\alpha(x^\mu),A_{\alpha,\beta}(x^\mu)=\frac{\partial A_\alpha}{\partial x^\beta},s^\nu(x^\mu)\Big)$, equations \eqref{lagdef-11} read,
\begin{equation}
    \begin{dcases}
    \frac{\partial s^\mu}{\partial x^\mu} = L
    \,,
    \\
    \ \mu_0 J^\mu =g^{\nu\sigma}g^{\mu\alpha}\left(\frac{\partial F_{\sigma\alpha}}{\partial x^\nu}+\gamma_\nu F_{\sigma\alpha}\right) \,. 
    \end{dcases}
\label{ContactEOM}
\end{equation}

To recover an expression which we can interpret as an electromagnetic field in matter, we set $\displaystyle g^{\mu\nu} = \frac{1}{\sqrt{1+\chi_m}}\text{diag}\Big(-(1+\chi_e)(1+\chi_m),1,1,1 \Big)$, where $\chi_e$ and $\chi_m$ are the electric and magnetic susceptibilities respectively. Moreover, considering $x=0=ct$ and denoting $\displaystyle\gamma_\mu = \left(\frac{\gamma}{c},\pmb{\gamma}\right)$ and $\displaystyle J^\mu = \left(c\rho,\textbf{j}\right)$, the last equation of \eqref{ContactEOM} reads in vector notation as \cite{GasMar21}
\begin{align}
\rho&=(1+\chi_e)\epsilon_0\big(\nabla\cdot\textbf{E}+\pmb{\gamma}\cdot\textbf{E}\big)\,,
    \\
\textbf{j}&=-(1+\chi_e)\epsilon_0\left(\frac{\partial\textbf{E}}{\partial t}+\gamma\textbf{E}\right)+\frac{1}{(1+\chi_m)\mu_0}\left(\nabla\times\textbf{B}+\pmb{\gamma}\times\textbf{B}\right) \,.
\end{align}
When  $\gamma_\nu = 0$, we recover the {\sl Gauss Law (for electric fields)} and the {\sl Amp\`ere--Maxwell Law} for linear materials. The other Maxwell's equations, {\sl Gauss Law (for magnetic fields)} and {\sl Faraday--Henry--Lenz Law}, are the same as in the non-action-dependent case since they just state that the curvature of the connection is closed.

If $J^\mu=0$, the Lagrangian and the equations are invariant under the transformation $\ds A_\mu\rightarrow A_\mu+\frac{\partial f}{\partial x^\mu}$, for any smooth function $f$ in $M$. To compute  the dissipated quantities associated with this gauge group, we first need the infinitesimal symmetry:
\begin{equation}
    Y=\frac{\partial f}{\partial x^\mu}\frac{\partial}{\partial A_\mu}+\frac{\partial^2 f}{\partial x^\mu x^\nu}\left(\frac{\partial}{\partial A_{\mu\nu}}+\frac{\partial}{\partial A_{\nu\mu}}\right).
\end{equation}
Then, the dissipated quantity is
\begin{equation}
\innp{Y}\Theta_\mathcal{L}=\frac{1}{\mu_0}g^{\alpha\beta}g^{\mu\nu}F_{\beta\nu}\frac{\partial f}{\partial x^\alpha} \d ^3x_\mu=\xi^\mu \d ^3x_\mu\,. 
\end{equation}
These are the same functions $\xi^\mu$ described in \ref{eq:dissLaw1}. The conservation laws are replaced by the so-called dissipation laws \ref{eq:dissLaw1}. The geometric formulation using multicontact geometry is, for solutions $\psi$,
\begin{equation}
    \psi^*\d\left(\xi^\mu\wedge \d ^3x_\mu\right)=\psi^*\left(\sigma_{\Theta_\mathcal{L}}\wedge\left( \xi^\mu \d ^3x_\mu\right)\right),
\end{equation}
which are equivalent to the second set of field equations \ref{ContactEOM}.

\subsection{Metric-affine gravity}\label{ex:EP}

Gravity is one of the first field theories where action-dependent sources were considered. There are two promising applications. First, to describe dark energy as a non-conservative effect produced by an action-dependent term in the Lagrangian. Second, to describe open gravitational systems, like those emitting gravitational waves, as an ``effective'' field theory that is action-dependent. 

The earliest proposal was \cite{LPAF2017}, adding a linear action term to the Einstein--Hilbert action. Due to difficulties in Herglotz's variational principle for field theories, the model was later upgraded in \cite{Lazo2022}. This variational principle was clarified in \cite{GLMR-2022} and \cite{GasMas2022}, where the field equations are derived geometrically to ensure covariance. 
Another approach is to consider cosmology as a dynamical system with a scaling symmetry (or dynamical similarity)~\cite{Sloan:2020taf,Sloan:2021hwx,Sloan:2022exs}.  The presence of a symmetry of this kind provides the system with a natural contact structure. For a general approach, see \cite{Sloan:2024kzb}.

The metric-affine model (or Einstein--Palatini) is a first-order singular field theory for General Relativity. A multisymplectic formulation of the model has been developed in several works (see, for instance, \cite{Capriotti, Gaset:2018uwi, Capriotti2, Ibort:2016xoo}). The metric-affine action is equivalent to the Einstein--Hilbert action in the final constraint surface, after performing a gauge reduction (see \cite{Gaset:2018uwi} for a geometrical description). This property makes it an interesting non-trivial example for action-dependent field theories.

Given the relevance of action-dependent gravity, here we present the multicontact formalism for the metric-affine action with a linear term in the action.

\subsubsection{Lagrangian formalism}

The configuration bundle is $\pi\colon{\rm E}\rightarrow M$, 
where $M$ is a connected orientable 4-dimensional manifold representing space-time, with volume form denoted as $\omega\in\df^4(M)$, and
${\rm E}=\Sigma\times_MC(LM)$, where $\Sigma$ is the manifold of Lorentzian metrics on $M$, with signature $(-+++)$, 
and $C(LM)$ is the bundle of connections on $M$;
that is, linear connections in $\Tan M$.

Natural coordinates $(x^\mu)$ in $M$ are taken such that 
$\omega=\d x^0\wedge\dotsb\wedge\d x^3\equiv\d^4x$. 
The adapted fiber coordinates in $J^1\pi$
are $(x^\mu,g_{\alpha\beta},\Gamma^\nu_{\lambda\gamma},g_{\alpha\beta,\mu},\Gamma^\nu_{\lambda\gamma,\mu})$,
with $\mu,\nu,\gamma,\lambda=0,1,2,3$
and $0\leq\alpha\leq\beta\leq 3$. 
The functions $g_{\alpha\beta}$ are the components of the metric
and $\Gamma^\nu_{\lambda\gamma}$ are the Christoffel symbols of the connection. 
We do not assume torsionless connections and thus
$\Gamma^\nu_{\lambda\gamma}\neq\Gamma^\nu_{\gamma\lambda}$,
in general.
Let $\lambda=\lambda_\mu\d x^\mu$ be a $1$-form of $M$. The metric-affine Lagrangian is 
\begin{equation}
L_{\rm EP}={\tt g}R-\lambda_\mu s^\mu \,,
\end{equation}
where ${\tt g}=\sqrt{|\det(g_{\alpha\beta})|}$, 
$R=g^{\alpha\beta}R_{\alpha\beta}$ is the {\sl scalar curvature}, the components of the {\sl Ricci tensor} are
\begin{equation}
    R_{\alpha\beta}=
\Gamma^{\gamma}_{\beta\alpha,\gamma}-\Gamma^{\gamma}_{\gamma\alpha,\beta}+
\Gamma^{\gamma}_{\beta\alpha}\Gamma^{\sigma}_{\sigma\gamma}-
\Gamma^{\gamma}_{\beta\sigma}\Gamma^{\sigma}_{\gamma\alpha}\,,
\end{equation}
 which depend only on the connection, and
$g^{\alpha\beta}$ denotes the inverse matrix of $g$.

It is an affine Lagrangian, so we follow the general results presented in Section \ref{subsec:affine}. We have that:
\begin{align}  
f_\alpha^{\beta\gamma,\mu} &= \frac{\partial L}{\partial \Gamma^{\alpha}_{\beta\gamma,\mu}}={\tt g}\,(\delta_\alpha^\mu g^{\beta\gamma}-\delta_\alpha^\beta g^{\mu\gamma})\,,
\\
f^{\alpha\beta,\mu} &= \frac{\partial L}{\partial g_{\alpha\beta,\mu}}=0\,,
\\  
V &= {\tt g}\, g^{\alpha\beta}\left(\Gamma^{\gamma}_{\beta\sigma}\Gamma^{\sigma}_{\gamma\alpha}-\Gamma^{\gamma}_{\beta\alpha}\Gamma^{\sigma}_{\sigma\gamma}\right)+\lambda_\mu s^\mu\,.
\end{align}
This Lagrangian leads to a premulticontact Lagrangian because $\displaystyle\frac{\partial f^{\beta\gamma,\mu}_\alpha}{\partial s^\nu}=0$, and it fulfills condition \eqref{eq:cond2affine} as follows. We need two set of functions: $J_{\alpha\beta,\mu}$ with respect to the components of the metric $g_{\alpha\beta}$, and $J^\alpha_{\beta\gamma,\mu}$ with respect to the components of the connection $\Gamma^\alpha_{\beta\gamma}$. Their expressions can be computed explicitly (although they are not unique): 
\begin{align}
J_{\alpha\beta,\mu}&=\frac1{{\tt g}\,n(\rho\sigma)}\left(-\frac{1}{2}g_{\rho \sigma}g_{\alpha\beta}\delta_\gamma^\tau+\frac{1}{6}g_{\alpha\beta}g_{\gamma\sigma}\delta_\rho^\tau-\frac13g_{\alpha\gamma}g_{\beta\sigma}\delta_\rho^\tau+g_{\beta\sigma}g_{\alpha\rho}\delta_\gamma^\tau\right)\left(\lambda_\nu f^{\rho\sigma,\nu}_\tau+\frac{\partial V}{\partial 
 \Gamma^\tau_{\rho\sigma}}\right)\,,
 \\
 J^\alpha_{\beta\gamma,\mu}&=-\frac{4}{3{\tt g}\,n(\rho\sigma)}\delta^\alpha_\mu\left(g_{\beta\rho}g_{\gamma\sigma}\frac{\partial V}{\partial g_{\rho\sigma}}-\frac12g_{\beta\gamma} g_{\rho\sigma}\frac{\partial V}{\partial g_{\rho\sigma}}\right)\,.
\end{align}
The number $n(\rho\sigma)$ is $1$, if $\rho=\sigma$, and $2$, if $\rho\neq\sigma$. The Lagrangian premulticontact $4$-form \eqref{thetacoor1} is
\begin{equation}
\Theta_\Lag=  -f^{\beta\gamma,\mu}_{\alpha}\,\d \Gamma^{\alpha}_{\beta\gamma}\wedge \d^3x_{\mu}+V\,\d^4x
+\d s^\mu\wedge \d^3x_\mu\,,
\end{equation}
and $\sigma_{\Theta_{\mathcal{L}}} = \lambda_\mu \d x^\mu \,.$ For a holonomic section $\bm{\psi}:M\rightarrow J^1\pi\times_M\Lambda^{m-1}(\Tan^*M)$
\begin{equation}
\bm{\psi}(x^\nu)=\Big(x^\mu,g_{\alpha\beta}(x^\nu),\frac{\partial g_{\alpha\beta}}{\partial x^\mu}(x^\nu),\Gamma^\alpha_{\beta\gamma}(x^\nu),\frac{\partial \Gamma^\alpha_{\beta\gamma}}{\partial x^\mu}(x^\mu),s^\mu(x^\nu)\Big)
\end{equation}
the Lagrangian equations \eqref{ELeqs2} and \eqref{ELeqs1} 
are:
\begin{align}
\frac{\partial s^\mu}{\partial x^\mu}&=L\,, 
\\\label{eq:fun1}
\frac{\partial V}{\partial g_{\sigma\rho}}-
\frac{\partial \Gamma^\alpha_{\beta\gamma}}{\partial x^\mu} \frac{\partial f_{\alpha}^{\beta\gamma,\mu}}{\partial g_{\sigma\rho}}&=0 \,,
\\\label{eq:fun2}
\frac{\partial V}{\partial \Gamma^\alpha_{\beta\gamma} } +
\sum_{\rho\leq\sigma}\frac{\partial g_{\rho\sigma}}{\partial x^\mu} \frac{\partial f_{\alpha}^{\beta\gamma,\mu}}{\partial g_{\rho\sigma}} +\lambda_\mu f_{\alpha}^{\beta\gamma,\mu}&=0\,.
\end{align}
Equation \eqref{eq:fun1} has two distinct components. First, if we multiply by 
\begin{equation}
-\frac13g_{\epsilon\nu}g_{\zeta\gamma}\delta_\beta^\alpha+
g_{\zeta\gamma}g_{\epsilon\beta}\delta_\nu^\alpha +\frac13g_{\zeta\nu}g_{\epsilon\gamma}\delta_\beta^\alpha-
g_{\epsilon\gamma}g_{\zeta\beta}\delta_\nu^\alpha\,,
\end{equation}
and rearrange the terms, we obtain:
\begin{equation}
t^\alpha_{\beta\gamma}\equiv
T^\alpha_{\beta\gamma}-
\frac13\delta^\alpha_\beta T^\nu_{\nu\gamma}+\frac13\delta^\alpha_\gamma T^\nu_{\nu\beta}\,,
\end{equation}
 where $\displaystyle  T^\alpha_{\beta\gamma}=\frac12(\Gamma^\alpha_{\beta\gamma}-\Gamma^\alpha_{\gamma\beta})$. 
 These are the torsion constraints and have the same expression as the action-independent case \cite{Gaset:2018uwi}. Assuming these constraints, the equations \eqref{eq:fun2} can be rewritten as:
\begin{align}\label{eq:nonmetricity}
m_{\rho\sigma,\mu}\equiv g_{\rho\sigma,\mu}-
g_{\sigma\nu}\Gamma^\nu_{\mu\rho}-
g_{\rho\nu}\Gamma^\nu_{\mu\sigma}-
\frac{4}{3}g_{\rho\sigma}T^\nu_{\nu\mu}+ g_{\rho\sigma}\lambda_\mu\ \Rightarrow\ \left(\nabla^{\Gamma} g\right)_{\rho\sigma,\mu}=\frac{4}{3}g_{\rho\sigma}T^\nu_{\nu\mu}- g_{\rho\sigma}\lambda_\mu\,,\
\end{align}
which can be considered as a non-metricity condition. 

The covariant constraint algorithm requires us to compute the tangency conditions in all space-time directions. These are given by semi-holonomic locally decomposable multivector fields ${\bf X}_\Lag\in\vf^m(J^1\pi)$ whose integral sections are the solutions of the field equations. In this case,
\begin{align}
{\bf X}_\Lag &= \bigwedge_{\mu=0}^3 X_\mu\\
\nonumber&=\bigwedge_{\mu=0}^3\Big(\derpar{}{x^\mu}+g_{\alpha\beta,\mu}\frac{\partial}{\partial g_{\alpha\beta}}+X_{\alpha\beta\nu,\mu}\frac{\partial}{\partial g_{\alpha\beta,\nu}}+\Gamma^\alpha_{\beta\gamma,\mu}\frac{\partial}{\partial \Gamma ^\alpha_{\beta\gamma}}+X^\alpha_{\beta\gamma\nu,\mu}\frac{\partial}{\partial \Gamma ^\alpha_{\beta\gamma,\nu}}+X_\mu^\nu\,\frac{\partial}{\partial s^\nu}\Big) \,,
\end{align}
where $X_{\alpha\beta\nu,\mu},X^\alpha_{\beta\gamma\nu,\mu}$ and $X_\mu^\nu$ are smooth functions of $J^1\pi\times_M\Lambda^{m-1}(\Tan^*M)$. The tangency condition on the constraints leads to
\begin{align}
0=X_\mu(t^\alpha_{\beta\gamma})&=T^\alpha_{\beta\gamma,\mu}-
\frac13\delta^\alpha_\beta T^\nu_{\nu\gamma,\mu}+\frac13\delta^\alpha_\gamma T^\nu_{\nu\beta,\mu}\,,
\\
0=X_\tau(m_{\rho\sigma,\mu})&=X_{\rho\sigma\mu,\tau}-
g_{\sigma\nu,\tau}\Gamma^\nu_{\mu\rho}
-g_{\sigma\nu}\Gamma^\nu_{\mu\rho,\tau}
-g_{\rho\nu,\tau}\Gamma^\nu_{\mu\sigma}
-g_{\rho\nu}\Gamma^\nu_{\mu\sigma,\tau}\nonumber
\\
&\quad -\frac{4}{3}g_{\rho\sigma,\tau}T^\nu_{\nu\mu}
-\frac{4}{3}g_{\rho\sigma}T^\nu_{\nu\mu,\tau}
+g_{\rho\sigma,\tau}\lambda_\mu
+g_{\rho\sigma}\derpar{\lambda_\mu}{x^\tau}\,.
\end{align}
The first set of equations are new constraints. The subsequent tangency condition is:
\begin{align}
0 &= X_\tau(X_\mu(t^\alpha_{\beta\gamma}))
\nonumber\\
&=\frac12\left(
X^\alpha_{\beta\gamma\mu,\tau}-X^\alpha_{\gamma\beta\mu,\tau}\right)
+\frac16\left(
\delta^\alpha_\beta X^\nu_{\nu\gamma\mu,\tau}
-X^\nu_{\gamma\nu\mu,\tau}
+\delta^\alpha_\gamma X^\nu_{\nu\beta\mu,\tau}
-\delta^\alpha_\gamma X^\nu_{\beta\nu\mu,\tau}
\right)\,,
\end{align}
which are not constraints.

It is well-known that the trace of the torsion of the connection is a gauge freedom of the Metric-Affine gravity (see, for instance, \cite{Gaset:2018uwi,pons}). 
We can fix the gauge condition $\displaystyle T^\nu_{\nu\mu}=\frac34\lambda_\mu$, so that the metricity condition \eqref{eq:nonmetricity} holds, although the connection is not traceless. 
Alternatively, we can impose a torsionless connection, and the constraint \eqref{eq:nonmetricity} becomes:
\begin{equation}
\left(\nabla^{\Gamma} g\right)_{\rho\sigma,\mu}=-\lambda_\mu g_{\rho\sigma}\,,
\end{equation}
which can be thought of as a dissipation law for the components of the metric. Here $\nabla^{\Gamma}$ denotes the covariant derivative generated by the connection $\Gamma$. 
In particular, if $\lambda\neq0$, the connection $\Gamma$ is not the Levi-Civita connection of $g$. This fact signals that, in the action-dependent case, one cannot generically recover the Einstein--Hilbert Lagrangian from the metric-affine one, as it is the case when $\lambda$ vanishes \cite{Gaset:2018uwi}. 

To investigate this point, we compare the other equation \eqref{eq:fun1} and the equations derived from the Einstein--Hilbert Lagrangian. Denote by $R(\Gamma), R(g)$ the Ricci tensor of the connection $\Gamma$ and of the metric $g$ respectively. After minor algebraic effort, one can see that equation \eqref{eq:fun1} is equivalent to $R(\Gamma)_{\alpha\beta}=0$. Imposing, for instance, that $T^\nu_{\nu\mu}=0$, and considering the case where $\lambda$ is closed, it can also be written as:
\begin{align}
0=R(\Gamma)_{\alpha\beta}=R(g)_{\alpha\beta}-\frac12\left(\nabla^g_\beta\lambda_\alpha+\nabla^g_\alpha\lambda_\beta+g_{\alpha\beta}g^{\rho\sigma}\nabla^g_\sigma\lambda_\rho-\lambda_\alpha\lambda_\beta+g_{\alpha\beta}g^{\rho\sigma}\lambda_\rho\lambda_\sigma\right)\,,
\end{align}
where $\nabla^g$ is the covariant derivative generated by the Levi-Civita connection of $g$. This equation is different from the one derived from the Einstein--Hilbert Lagrangian with the same action-dependent term \cite{GasMas2022,Lazo2022}:
\begin{align}
    R(g)_{\alpha\beta}-\frac12\left(\nabla^g_\beta\lambda_\alpha+\nabla^g_\alpha\lambda_\beta+g_{\alpha\beta}g^{\rho\sigma}\nabla^g_\sigma \lambda_\rho+2\lambda_\alpha\lambda_\beta+g_{\alpha\beta}g^{\rho\sigma}\lambda_\rho\lambda_\sigma\right)=0\,,
\end{align}
at least in the points where $T^\nu_{\nu\mu}=0$. Thus, in general, the action-dependent metric--affine gravity is not equivalent to the action-dependent Einstein--Hilbert gravity, in the sense that the set of metrics that are solutions to the field equations is different.

An interpretation of this non-equivalence is the following. The metric-affine formulation has more solutions than the Einstein--Hilbert one, but it also contains a gauge symmetry. Each equivalent class of solutions contains exactly one solution of the Einstein--Hilbert field equations, and every solution to the latter is inside one equivalence class of the former. The gauge fixing that recovers this solution is the vanishing of the trace of the torsion, as already noticed by Einstein \cite{Einstein}. When we add an action-dependent term in the metric-affine Lagrangian that is no longer the case: in a class of solutions of the metric-affine Lagrangian, in general there is no solution to the Einstein--Hilbert Lagrangian with the same action-dependent term.

\subsubsection{Hamiltonian formalism}

As it is explained in Section \ref{subsubsection:affineHam}, the Legendre map induces the following relations on the multimomenta:
\begin{equation}
p_\alpha^{\beta\gamma,\mu}=\frac{\partial L}{\partial \Gamma^{\alpha}_{\beta\gamma,\mu}} = {\tt g}(\delta_\alpha^\mu g^{\beta\gamma}-\delta_\alpha^\beta g^{\mu\gamma})
\,, \qquad
p^{\alpha\beta,\mu}=\frac{\partial L}{\partial g_{\alpha\beta,\mu}}=0\,.
\end{equation}
These are restrictions that determine all the multimomenta as functions of the positions and define locally the submanifold ${\P^\ast}\subset\P$. Therefore, we can use  $(x^\mu,g_{\alpha\beta},\Gamma^\alpha_{\beta\gamma},s^\mu)$
as coordinates on ${\P^\ast}$. The Hamiltonian $4$-form $\Theta^0_{\mathcal{H}}$ has the same local expression as the Lagrangian one. The equations in coordinates are the same as in the Lagrangian case, but now the velocities are not coordinates of the system. The torsion constraint is also present in the Hamiltonian formalism, as it is projectable by the Legendre map. Furthermore, the non-metricity condition becomes an equation of motion.

\subsection{Burgers' equation from the heat equation}\label{ex:burguer}

Burgers' equation \cite{Ba-1915} is a remarkable nonlinear partial differential equation \cite{Salsa}. It appears in applied mathematics, physics, fluid dynamics, etc. It reads
\begin{equation}\label{eq:burgers}
   \derpar{u}{t}+ u\derpar{u}{x} = k\parder{^2u}{x^2}\,,
\end{equation}
where $t,x$ are the independent variables, $u = u(t,x)$ is the dependent variable, and $k\geq 0$ is a diffusion coefficient. Burgers' equation is closely related to the heat equation,
\begin{equation}\label{eq:heat}
  \derpar{u}{t} = k\parder{^2u}{x^2}\,,
\end{equation}
In this example we see that Burgers' equation can be obtained as a contactification of the heat equation (see also \cite{GGMRR-2019}).

\subsubsection{Lagrangian formalism}

Before contactifying the heat equation, we need a Lagrangian or Hamiltonian formulation of the heat equation. It is well known that the heat equation is not variational. However, it can be made variational by introducing an additional dependent variable $v$. 
The configuration bundle is $\pi:E\to \R^2$, where the coordinates in $\R^2$ and $E$ are $(t,x)$ and $(t,x;u,v)$, respectively. 
The phase bundle $J^1\pi$ has coordinates $(t,x; u, v, u_t, u_x, v_t, v_x)$. Consider the Lagrangian function $L_\circ\in\Cinfty(J^1\pi)$ given by
\begin{equation}\label{eq:lagrangian-heat}
    L_\circ = -ku_xv_x - \frac12(v u_t - uv_t)\,,
\end{equation}
whose Euler--Lagrange equations read
\begin{equation}\label{eq:EL-heat}
  \parder{u}{t} - k\parder{^2
  u}{x^2} = 0 \,, \qquad 
  \parder{v}{t} + k\parder{^2v}{x^2} = 0\,,.
\end{equation}
Since the second equation is linear homogeneous, it always has solutions, i.e. $v = 0$. Thus, there is a one-to-one correspondence between the solutions to the heat equation \eqref{eq:heat} and the solutions to the Euler--Lagrange equations \eqref{eq:EL-heat} with $v = 0$.

To perform the contactification, consider the fiber bundle $\tau:\mathcal{P}\to\R^2$, with adapted coordinates $(t,x; u,v, u_t,u_x,v_t,v_x,s^t,s^x)$, $\omega = \d t\wedge\d x$, and the singular Lagrangian function $L\in\Cinfty(\mathcal{P})$ given by
\begin{equation}\label{eq:lagrangian-burgers}
    L = L_\circ - \gamma us^x\,,
\end{equation}
where $L_\circ$ is the Lagrangian function given in \eqref{eq:lagrangian-heat} and $\gamma\in\R$ is a constant. Note that we have added a term that, unlike in previous examples, is not of the simple form $\gamma_\mu(x^\nu)s^\mu$. This additional term will give rise to the nonlinear term in the Burgers' equation. 
The Lagrangian energy is
\begin{equation}
\label{eq:BurgerLenergy}
E_\L=-ku_xv_x+\gamma u s^x \, ,
\end{equation}
and the Lagrangian $2$-form \eqref{thetacoor1} is
\begin{align}
\Theta_\L =& -kv_x\d u\wedge\d t-ku_x\d v\wedge\d t+\frac{1}{2}v\d u\wedge\d x-\frac12 u\d v\wedge\d x \nonumber \\
&+(-ku_xv_x+\gamma u s^x)\d t\wedge\d x+\d s^t\wedge\d x-\d s^x\wedge\d t   \, ,
\end{align}
and $\sigma_{\Theta_\L} = \gamma u\d x$. Note that $(\Theta_\L,\omega)$ define a 2-premulticontact structure on $\mathcal{P}$ with
$$ \mathcal{C}=\left< \parder{}{u_t},\parder{}{v_t}\right>
   \,,\qquad D^{\mathfrak{R}} = \left\langle \parder{}{u_t},\parder{}{v_t},\parder{}{s^t},\parder{}{s^x} \right\rangle\,. $$
For  holonomic sections
$\bm{\psi}(t,x)=\Big(t,x,u(t,x),v(t,x), \dparder{u}{t}(t,x),\dparder{u}{x}(t,x),\dparder{v}{t}(t,x),\dparder{v}{x}(t,x),s^t(t,x),s^x(t,x)\Big)$,
the Herglotz--Euler--Lagrange equations \eqref{ELeqs2} and \eqref{ELeqs1} are:
\begin{equation}\label{eq:EL-burgers}
    \parder{u}{t} -k\gamma u\parder{u}{x} = k\parder{^2u}{x^2}\,,\qquad \parder{v}{t} + k\gamma u\parder{v}{x} = -k\parder{^2v}{x^2} + \gamma s^x\,,\qquad \parder{s^t}{t} + \parder{s^x}{x} = L\,. 
\end{equation}
Note that there exists a correspondence between the solutions to Burgers' equation \eqref{eq:burgers} and the solutions $\Big(u,v, \dparder{u}{t},\dparder{u}{x},\dparder{v}{t},\dparder{v}{x},s^t,s^x\Big)$ to the Herglotz--Euler--Lagrange equations \eqref{eq:EL-burgers} for the Lagrangian \eqref{eq:lagrangian-burgers}, with $\gamma=-\frac1k$, $v = s^x = 0$ and $\dparder{s^t}{t} = L$.

It is important to highlight that, although the Lagrangian \eqref{eq:lagrangian-burgers} is singular, no compatibility constraints arise from the Herglotz--Euler--Lagrange equations.

\subsubsection{Hamiltonian formalism}

We have provided a Lagrangian formulation for both the heat equation and Burgers' equation.
Now, we develop the Hamilton--de Donder--Weyl formalism to the Lagrangian $L$ defined in \eqref{eq:lagrangian-burgers}. The Legendre map associated with the Lagrangian $L$ is the map
\begin{equation} \F L(t,x;u,v,u_t,u_x,v_t,v_x,s^t,s^x) = (t,x; u, v, p^t, p^x, q^t, q^x,s^t,s^x)\,, \end{equation}
where,
\begin{equation}
p^t = \parder{L}{u_t} = -\frac12v\,,\qquad p^x = \parder{L}{u_x} = -kv_x\,,\qquad q^t = \parder{L}{v_t} = \frac12u\,,\qquad q^x = \parder{L}{v_x} = -ku_x\,. 
\end{equation}
Hence, the image of the Legendre map $\mathcal{P}_0^\ast = \F L(\mathcal{P})$ is given by the two constraints
\begin{equation} p^t + \frac12v = 0\,,\qquad q^t - \frac12u = 0\,, \end{equation}
and the Lagrangian function $L$ is almost-regular.
We  use coordinates $(t,x;u,v,p^x,q^x,s^t,s^x)$ on $\mathcal{P}_0^\ast$. Hence, the Hamiltonian on $\mathcal{P}_0^\ast$ is
\begin{equation} 
H_0 = -\frac1k p^xq^x + \gamma us^x\,. 
\end{equation}
The Hamiltonian $2$-form \eqref{thetaHcoor2} is
\begin{align}
\Theta_\H^0 =& p^x\d u\wedge\d t+q^x\d v\wedge\d t+\frac{1}{2}v\d u\wedge\d x-\frac12 u\d v\wedge\d x \nonumber \\
&+\left(-\frac1k p^xq^x + \gamma us^x\right)\d t\wedge\d x+\d s^t\wedge\d x-\d s^x\wedge\d t   \, .
\end{align}
Note that $(\Theta_{\mathcal{H}}^0,\omega)$ define a 2-multicontact structure on $\mathcal{P}_0^*$. We have $\sigma_{\Theta_\H^0} = \gamma\,u\d x$.
For integral sections  
$\bm{\psi}(t,x)=(t,x,u(t,x),v(t,x),p^x(t,x),q^x(t,x),s^t(t,x),s^x(t,x))$,
the Herglotz--Hamilton--de Donder--Weyl equations \eqref{coor2} for this Hamiltonian:
\begin{gather}
    \parder{v}{t} = \parder{p^x}{x} + \gamma(s^x + up^x)\,,\qquad \parder{u}{t}  = - \parder{q^x}{x} -\gamma uq^x\,,\qquad \parder{u}{x} = -\frac1k q^x\,,\qquad \parder{v}{x} = -\frac1k p^x\,,
     \\
  \parder{s^t}{t} + \parder{s_x}{x} = p^x \parder{u}{x} + q^x\parder{v}{x} -\frac12 v\parder{u}{t} + \frac12 u\parder{v}{t} + \frac1k p^xq^x - \gamma u s^x  \,, \label{eq:burgers-ham-3}
\end{gather}
and there are no compatibility constraints.
Combining the first four equations, we obtain
\begin{align}
    \parder{u}{t} - \gamma ku\parder{u}{x} &= k\parder{^2u}{x^2}\,,\label{eq:burgers-ham-1}\\
    \parder{v}{t} + \gamma ku\parder{v}{x} &= -k\parder{^2v}{x^2} + \gamma s^x\label{eq:burgers-ham-2}\,.
\end{align}
If we set the value of the constant $\gamma$ to be $\gamma = -\dfrac{1}{k}$, equation \eqref{eq:burgers-ham-1} becomes Burgers' equation \eqref{eq:burgers}.

Note that Equation \eqref{eq:burgers-ham-3} admits solutions $(u,v,p^x,q^x,s^t,s^x)$ with $u$ being a solution to Burgers' equation \eqref{eq:burgers}, $v = p^x = s^t = s^x = 0$, and $q^x = -k\dparder{u}{x}$. As in the Lagrangian counterpart, no compatibility constraints arise.

\begin{remark}
    The heat equation can also be described via a Hamiltonian formalism by taking the Legendre map associated with the Lagrangian \eqref{eq:lagrangian-heat}.
\end{remark}

\section{Summary and outlook}

We have presented a multilayered introduction to action-dependent field theories. In a friendly presentation, we have discussed the interpretation of the ``action-variables'' and the field equations for regular Lagrangians \eqref{eq:HELActionDep} and Hamiltonians \eqref{eq:hamiltonintro}.

Necessitated by the intricacies of physically-motivated field theories, we have provided a quick presentation of the geometric framework underlying action-dependent field theory: multicontact geometry \cite{LGMRR-2022}. Geometric formulations are powerful techniques to study field theories, especially when they are not regular and/or symmetries are involved. Using the multicontact formalism, we have reviewed how to compute the dynamics of regular and singular Lagrangians. We have also shown how to construct the associated Hamiltonian formalism and how to apply the constraint algorithm.

We have developed a large collection of examples, illustrating how to derive the field equations and how to use the geometric structures. Both the Lagrangian and Hamiltonian approaches have been computed. The examples have been selected to showcase the influence of action-dependent terms on prototypical Lagrangians in physics. 

First, we have studied the generic cases of field theories described by quadratic (regular and singular) Lagrangians, 
and by affine Lagrangians. 
Second, we have analyzed some well-known theories for which the standard Lagrangians have been modified by adding terms depending on the ``action variables''. 
In particular, we have considered two regular systems: the one-dimensional wave equation 
and the Klein--Gordon equation (and the telegrapher equation as a particular case), 
and three singular ones:
Metric-affine gravity, 
Maxwell’s electromagnetism, 
and the Burgers equation obtained from the heat equation. 
(Two additional examples are discussed in the following appendix: the Bosonic string and $(2+1)$-dimensional gravity and Chern--Simons equation).

In the context of action-dependent field theories, the examples of metric-affine gravity, bosonic string, Chern--Simons, as well as the generic cases of quadratic and affine Lagrangians are new in the literature. The Klein--Gordon equation and Burger's equation were studied previously in the action-dependent context (see, for instance, \cite{GRR-2022} and \cite{GGMRR-2019} respectively), but here we have described them using the multicontact formalism for the first time. All in all, this work represents the largest increase of new instances of action-dependent field theories so far. 

The examples have been chosen to explore the behavior of action-dependent field theories. For instance, we observe that action-dependent terms have an intricate influence in General Relativity and produce new terms in the field equations that cannot be obtained by other kinds of particles. We have shown how equivalent Lagrangians lead to different field equations when an action-dependent term is added. Moreover, the non-conservative behavior of these theories makes them suitable for modeling open gravitational systems. These considerations make action-dependent gravity a promising field of research.

The study of the symmetries, with their associated dissipated quantities, remains a topic of great interest. The reduction of symmetries is currently being developed for dissipative mechanical systems~\cite{ACLSS}, but no similar procedure exists in the context of multicontact geometry yet. Another interesting object is the scaling symmetry or dynamical similarity. The recent work~\cite{Sloan:2024kzb} points out that usual systems with this special symmetry benefit from a description using contact geometry.

On a more topological note, the ``action variables'' provide a connection between the boundary and the bulk. One can foreshadow that the boundary terms play an important role in action-dependent field theories. Moreover, the dissipation form, 
that encodes the non-conservation, has a strong link to the cohomology of the manifold. How these elements relate to each other and their effects on the dynamics remain to be scrutinized.

Another open problem of the multicontact framework of action-dependent classical field theory is developing an space-time split formalism. This will provide a Hamiltonian formulation more akin to the commonly used one in theoretical physics. This approach will be presented in a future work.

Finally, multiple properties of multicontact geometry remain to be studied. To highlight two, Legendre submanifolds and the graded Jacobi algebra, which will pave the way to stronger geometric theorems to characterize action-dependent field theories.

\section*{Acknowledgments}

We acknowledge financial support of the 
{\sl Ministerio de Ciencia, Innovaci\'on y Universidades} (Spain), grants PID2021-125515NB-C21, PID2022-137909NB-C2, and RED2022-134301-T of AEI.
We also acknowledge of
the {\sl Secretary of University and Research of the Ministry of Business and Knowledge of the Catalan Government}, project 2021 SGR 00603,
and the financial support for research groups AGRUPS-2022 of the Universitat Polit\`ecnica de Catalunya (UPC).
M. de Le\'on also acknowledges financial support from the Severo Ochoa Programme for Centers of Excellence in R\&D, CEX-2023-001347-S.
X.~Rivas also acknowledges financial support of the Novee Idee 2B-POB II, project PSP: 501-D111-20-2004310 funded by the ``Inicjatywa Doskonałości-Uczelnia Badawcza'' (IDUB) program.

We truly appreciate the referee’s valuable feedback and observations, which have been a big help in improving our manuscript.


\appendix

\section{Other applications}
\label{apendA}

In this appendix, we analyze two other interesting physical theories modified with action-dependent terms, whose multicontact treatment is similar to some of the cases previously studied.
These are the Bosonic string, which is a regular theory, and the $(2 + 1)$-dimensional gravity and the Chern–Simons equation,
which is another affine (and then singular) case.

\subsection{Bosonic string theory}\label{ex:string}

(The multisymplectic formulation of the bosonic string has been recently developed in \cite{GR-2023,GR-2024,GGR-2023}.
See \cite{BBS} for more details on string theory).

In this theory, space-time is a $(d+1)$-dimensional manifold $M$
endowed with a space-time metric $G_{\mu\nu}$ with signature $(-+\dotsb +)$.
Local coordinates on $M$ are $x^\mu$ ($\mu=0,1,\ldots, d$)
The string worldsheet $\Sigma$ is a $2$-dimensional manifold, with local coordinates $\sigma^a$, with  $a=0,1$,
which is endowed with the volume form $\omega=\d^2\sigma$. 
The fields $x^\mu(\sigma)$ are scalar fields on $\Sigma$ given by the embedding maps $\Sigma \rightarrow M: \sigma^a\mapsto x^{\mu}(\sigma)$. 
Hence, the configuration bundle is $\pi:E=\Sigma\times M\rightarrow \Sigma$ and its sections are $\phi:\Sigma\rightarrow\Sigma\times M: \sigma^a \mapsto (\sigma^a,x^{\mu}(\sigma))$. 
The Lagrangian and Hamiltonian phase bundles $J^1\pi\rightarrow\Sigma$ and $J^{1*}\pi\rightarrow\Sigma$ have local coordinates $(\sigma^a,x^\mu,x_a^\mu)$ and $(\sigma^a,x^\mu,p^a_\mu)$, respectively.
On $J^1\pi$ we also have a $2$-form $\displaystyle g=\frac{1}{2}g_{ab} \d\sigma^a \wedge\d\sigma^b$,
whose pullback by jet prolongations of sections
$\displaystyle j^1\phi(\sigma^a)=\Big(\sigma^a,x^\mu(\sigma^a), \derpar{x^\mu}{\sigma^a}(\sigma^a)\Big)$
gives the induced metric on $\Sigma$,
\begin{equation}
\label{stringmetric}
(j^1\phi)^* g=h\equiv \frac{1}{2}h_{ab}\d\sigma^a\wedge\d\sigma^b\,, \quad \text{where} \quad h_{ab}=G_{\mu\nu}\frac{\partial x^\mu}{\partial \sigma^a}\frac{\partial x^\nu}{\partial\sigma^b}\,.
\end{equation}
The bosonic string theory is described by the {\sl Nambu--Goto Lagrangian},
\begin{equation}
L_0=-T \sqrt{-\det g}= -T\sqrt{-\det (G_{\mu\nu}x^{\mu}_a x^{\nu}_b)} \,,
\end{equation}
where $T$ is a constant called the {\sl string tension}.
This Lagrangian is regular since
\begin{equation}
\label{pbraneHessian}
\frac{\partial^2 L_0}{\partial x_a^\mu \partial x_b^\nu}=-T\sqrt{-\det g}\left[ G_{\mu\nu}g^{ba} -G_{\mu\alpha}G_{\rho\nu}x^\alpha_c x^\rho_d\left(g^{ba}g^{cd}+g^{bc}g^{ad}-g^{ca}g^{bd}\right)\right]\,,
\end{equation}
where $\displaystyle g^{ba}\equiv(g^{-1})^{ba}= \frac{1}{\det g}\epsilon^{bc}\epsilon^{ad}g_{d c}$, is a regular matrix everywhere.
Then, the Legendre map, which is a diffeomorphism,
allows us to translate the $2$-form $g$ to $J^{1*}\pi$ as
$\F\L_{\circ*}g=g\in\Omega^2(J^{1*}\pi)$.

\subsubsection{Lagrangian formalism}

For this theory, the bundle $\tau\colon{\cal P}\simeq J^1\pi\times\R^2\to\Sigma$
has adapted coordinates $(\sigma^a,x^\mu,x_a^\mu,s^a)$.
The contactified Lagrangian function $L\in\Cinfty({\cal P})$ we propose is
\begin{equation}
\label{bosonic_string}
L(\sigma^a,x^\mu,x_a^\mu,s^a)=L_0(\sigma^a,x^\mu,x_a^\mu)+\gamma_a s^a= 
-T\sqrt{-\det (G_{\mu\nu}x^{\mu}_a x^{\nu}_b)}+\gamma_a s^a=
-T \sqrt{-\det g}+\gamma_a s^a\,,
\end{equation}
where $\gamma\equiv(\gamma_a)\in\R^2$ is a constant vector,
and it is a regular Lagrangian function.
The Lagrangian energy is
\begin{equation}
E_\L \equiv \frac{\partial L}{\partial x_a^\mu}x_a^\mu - L = -T \sqrt{-\det g}\ (g^{ba}g_{ab}-1)-\gamma_a s^a =-T \sqrt{-\det g}- \gamma_a s^a\in\Cinfty({\cal P})\,,
\end{equation}
where the last equality follows because $g_{ab}$ is a $2\times2$ matrix in the case of the string.
The Lagrangian multicontact $2$-form \eqref{thetacoor1} is
\begin{align}
\Theta_\L &=
\frac{\partial L}{\partial x_a^\mu} \d x^\mu\wedge \d^1\sigma_a-E_\L \,\d^2\sigma + \d s^a\wedge \d^1\sigma_a \nonumber\\
&= -T\sqrt{-\det g} \,G_{\mu\nu}g^{ba}x_b^\nu\d x^\mu\wedge\d ^1\sigma_a-\left(T\sqrt{-\det g} + \gamma_a s^a\right)\d ^2\sigma + \d s^a\wedge \d^1\sigma_a \,, 
\end{align}
where $\displaystyle \d^1\sigma_a=\innp{\derpar{}{\sigma^a}}\d^2\sigma$.
In this case, we have $\sigma_{\Theta_\L} = -\gamma_a\d\sigma^a$.


For holonomic sections  
$\displaystyle\bm{\psi}(\sigma)=\left(\sigma^a,x^\mu(\sigma),\derpar{ x^\mu}{\sigma^a}(\sigma),s^a(\sigma)\right)$ 
the Lagrangian equations \eqref{lagdef1} and \eqref{lagdef2} lead to
\begin{align}
\derpar{s^a}{\sigma^a}& = L \,, 
\label{eq:stringfieldeqa0} \\
T \sqrt{-\det g}\,G_{\mu\nu}g^{ba}\gamma_a\derpar{x^\nu}{\sigma^b} &= \derpar{x^\rho}{\sigma^a}\left[\derpar{}{x^\rho}\left(\sqrt{-\det g}\, G_{\mu\nu}g^{ba}\derpar{x^\nu}{\sigma^b}\right)-\derpar{}{x^\mu}\left(\sqrt{-\det g}\, G_{\rho\nu}g^{ba}\derpar{x^\nu}{\sigma^b}\right)\right] \nonumber
\\
& \hspace{-4em} +\sqrt{-\det g}\left[ G_{\mu\nu}g^{ba} -G_{\mu\alpha}G_{\beta\nu}\left(g^{ba}g^{cd}+g^{cb}g^{ad}-g^{ca}g^{bd}\right)\derpar{x^\alpha}{\sigma^c} \derpar{x^\beta}{\sigma^d}\right] \frac{\partial^2x^\nu}{\partial\sigma^a\partial\sigma^b} \nonumber\\
& \hspace{-4em}
 + \left[\frac{1}{2}\sqrt{-\det g}\, g^{ba}\derpar{G_{\alpha\beta}}{x^\mu}\derpar{x^\alpha}{\sigma^a}\derpar{x^\beta}{\sigma^b}+\derpar{}{\sigma^a}\left(\sqrt{-\det g}\, G_{\mu\nu}g^{ba}\derpar{x^\nu}{\sigma^b}\right)\right] \,. \label{eq:stringfieldeqa1}
\end{align}
where \eqref{eq:stringfieldeqa1} are the Herglotz--Euler--Lagrange equations \eqref{ELeqs2} and \eqref{ELeqs1} which, using \eqref{stringmetric} and after some non-trivial calculations, can be written as
\begin{align}
& T \sqrt{-\det g}\,G_{\mu\nu}g^{ba}\gamma_a\derpar{x^\nu}{\sigma^b} = -\sqrt{-\det h}\, h^{ba}\derpar{G_{\alpha\beta}}{x^\mu}\derpar{x^\alpha}{\sigma^a}\derpar{x^\beta}{\sigma^b}+\derpar{}{\sigma^a}\left(\sqrt{-\det h}\, G_{\mu\nu}h^{ba}\derpar{x^\nu}{\sigma^b}\right) \nonumber\\
& \qquad\qquad\qquad+\derpar{}{x^\rho}\left(\sqrt{-\det h}\, G_{\mu\nu}h^{ba}\derpar{x^\nu}{\sigma^b}\right)\derpar{x^\rho}{\sigma^a}
\\
&\qquad\qquad\qquad +\sqrt{-\det h}\left[ G_{\mu\nu}h^{ba} -G_{\mu\alpha}G_{\beta\nu}\left(h^{ba}h^{cd}+h^{cb}h^{ad}-h^{ca}h^{bd}\right)\derpar{x^\alpha}{\sigma^c}\derpar{x^\beta}{\sigma^d}\right]\frac{\partial^2x^\nu}{\partial\sigma^a\partial\sigma^b} \,.\nonumber
\end{align}

\subsubsection{Hamiltonian formalism}

The Hamiltonian formalism is developed starting from the Legendre map, $\F\L:{\cal P}\rightarrow {\cal P}^*$, which is
\begin{equation}
\F\L^*\sigma^a=\sigma^a \,, \qquad
\F\L^*x^\mu=x^\mu \,, \qquad
\F\L^*p^a_{\mu} = -T \sqrt{-\det g}\ G_{\mu\nu}g^{ba}x_b^{\nu} \,.
\end{equation}
This Legendre map is invertible as expected from the regularity of the Hessian matrix \eqref{pbraneHessian}.
The Hamiltonian function is 
\begin{equation}
H(\sigma^a,x^{\mu},p^a_{\mu}, s^a)\equiv p^a_{\mu}(\F\L^{-1})^* x_a^{\mu} - (\F\L^{-1})^*L \in\Cinfty({\cal P}^*)\,.
\end{equation}
Introducing the matrix $\Pi$ whose components are $\displaystyle \Pi^{ab}\equiv G^{\mu\nu}p^a_\mu p^b_\nu$,
then 
\begin{equation}
\F\L^*\Pi^{ab}=-T^2 \det g\ g^{ba} \
\Longleftrightarrow \ \F\L^*\det\Pi=(-T^2\det g)^2\det(g^{-1})=T^4\det g \,,
\end{equation}
from which it follows that
$\displaystyle (\F\L^{-1})^*x_b^\nu=-\frac{1}{T}\sqrt{-\det \Pi}\ G^{\mu\nu}\Pi_{ab} p^a_\mu$, 
and the Hamiltonian function can be written as
\begin{equation}
H(\sigma^a, x^\mu, p^a_\mu, s^a)= -\frac{1}{T}\sqrt{-\det\Pi} - \gamma_a s^a\,.
\end{equation}
where $\displaystyle \Pi_{ab}=\frac{1}{\det\Pi}\epsilon_{cb}\epsilon_{d a}\Pi^{cd}$.

The multicontact Hamiltonian $2$--form is
\begin{align}
\Theta_{\H} &= p^a_{\mu} \d x^{\mu}\wedge \d^1\sigma_a - H\,\d^2\sigma=p^a_\mu\d x^\mu\wedge\d^1\sigma_a+\left(\frac{1}{T}\sqrt{-\det \Pi} + \gamma_a s^a\right)\d^2\sigma \label{eq:stringOmegHa}\,, 
\end{align}
and $\sigma_{\Theta_\H}=-\gamma_a\,\d\sigma^a$.
For sections 
$\bm{\psi}(\sigma^b)=(\sigma^a,x^\mu(\sigma^b),p^a_\mu(\sigma^b),s^a(\sigma^b))$ 
the field equations become
\begin{align}
& \parder{s^a}{\sigma^a} = \frac{\sqrt{-\det\Pi}}{T}\left( 1 - \Pi_{ba} G^{\mu\nu}p_\mu^ap_\nu^b \right) + \gamma_a s^a = \frac{\sqrt{-\det\Pi}}{T}\left( 1 - \Pi_{ba} \Pi^{ab} \right) + \gamma_a s^a\,,  \\
& \displaystyle \derpar{x^\mu}{\sigma^a} = -\frac{\sqrt{-\det \Pi}}{T}\Pi_{ba}G^{\mu\nu}p^b_\nu\,,\\ 
&\displaystyle \derpar{p^a_{\mu}}{\sigma^a} = \frac{\sqrt{-\det \Pi}}{2T}\Pi_{ba} \derpar{G^{\rho\alpha}}{x^\mu}p^a_\rho p^b_\alpha + \gamma_a p^a_\mu\,, 
\end{align}
which are the corresponding Herglotz--Hamilton--De Donder--Weyl equations for the bosonic string.
As usual,  the Legendre map allows us to transform these equations into \eqref{eq:stringfieldeqa0} and \eqref{eq:stringfieldeqa1} along with the holonomy condition and thus the Lagrangian and Hamiltonian formalisms are equivalent.

\subsection{\texorpdfstring{$(2+1)$}{}-dimensional gravity Chern--Simons equation}\label{ex:C-S}

(See \cite{GR-2023,GGR-2023} for the multisymplectic formulation of this theory).
Gravity with cosmological constant $\lambda$ in a $2+1$-dimensional space-time $M$ is developed in the tetrad formalism using the \textit{vierbein} $e^a_\mu$
given by $g_{\mu\nu}=e^a_\mu e^b_\nu \eta_{ab}$
($\mu=0,1,2$, $a=0,1,2$),
and the Hodge dual spin connection $\omega^a_\mu=\frac{1}{2}\epsilon^{abc}\omega_{\mu bc}$,
which are treated as the variational fields of the theory as in \cite{WittenGravity}.

\subsubsection{Lagrangian formalism}

The configuration bundle $\pi:E\rightarrow M$ has coordinates $(x^\mu, e^a_\mu, \omega_\rho^c)$. The multicontact Lagrangian formalism is established on the space $\mathcal{P}=J^1\pi\times_M\Lambda^{2}(\Tan^*M) $, 
with coordinates $(x^\mu,\, e^a_\mu,\, \omega_\rho^c,\, e^a_{\sigma,\mu},\, \omega^c_{\sigma,\rho},\,s^\alpha)$ such that the volum form is $\omega=\d x^0\wedge \d x^1\wedge \d x^2$. 
The Lagrangian of the Chern--Simons theory \cite{WittenGravity} with a linear dissipation term is given by
\begin{equation}
L(x^\mu, e^a_\mu, \omega_\rho^c, e^a_{\sigma,\mu}, \omega^c_{\sigma,\rho},s^\alpha)=
2\epsilon^{\mu\nu\rho}\eta_{ac}e^a_\mu\omega^c_{\nu,\rho}+\epsilon^{\mu\nu\rho}\epsilon_{abc}\Big(e^a_\mu\omega^b_\nu \omega^c_\rho+\frac{1}{3}\lambda e^a_\mu e^b_\nu e^c_\rho\Big)-\gamma_\alpha s^\alpha,
\end{equation}
which is an affine Lagrangian, where $\lambda$ is the cosmological constant and $\gamma_a\in\R$. 
The Lagrangian energy is
\begin{equation}
E_\L=-\epsilon^{\mu\nu\rho}\epsilon_{abc}\Big(e^a_\mu\omega^b_\nu \omega^c_\rho+\frac{1}{3}\lambda e^a_\mu e^b_\nu e^c_\rho\Big)+\gamma_\alpha s^\alpha \,,
\end{equation}
the Lagrangian $2$-form \eqref{thetacoor1} is
\begin{equation}
\Theta_\L=2\epsilon^{\mu\nu\rho}\eta_{ac}e^a_\mu\text{d}\omega^c_\rho\wedge\text{d}^2x_\nu+\left[-\epsilon^{\mu\nu\rho}\epsilon_{abc}\Big(e^a_\mu \omega^b_\nu \omega^c_\rho + \frac{1}{3}\lambda e^a_\mu e^b_\nu e^c_\rho\Big)+\gamma_\alpha s^\alpha\right]\text{d}^3x+\d s^\mu\wedge \d^{m-1}x_\mu\,, 
\end{equation}
which is a premulticontact form, and now $\sigma_{\Theta_\L}=\gamma_\mu\d x^\mu$.

We can derive directly the equations of motion for multivector fields using the derived equations for affine Lagrangians \eqref{lagdef0} and \eqref{lagdef11}. 
We have two sets of variables, the vierbein $e^a_\mu$ and the spin connection $\omega^a_\mu$. 
So, we have the semi-holonomic multivector fields on $\P$ expressed as
\begin{align} 
{\bf X}_\L &= \bigwedge_{\sigma=0}^2 X_\sigma\nonumber\\
&=
\bigwedge_{\sigma=0}^2\left(\derpar{}{x^\sigma}+e^a_{\mu,\sigma}\frac{\partial}{\partial e^a_\mu}+\omega^c_{\nu,\sigma}\frac{\partial}{\partial \omega^c_{\nu}}+F^a_{\mu\rho\sigma}\frac{\partial}{\partial e^a_{\mu,\rho}}+G^c_{\nu\rho\sigma}\frac{\partial}{\partial \omega^c_{\nu,\rho}}+X^\mu_\sigma\frac{\partial}{\partial s^\mu}\right) \,.
\end{align}
Then, the Lagrangian field equations \eqref{lagdef1} and \eqref{lagdef2} are
\begin{align}\label{eq:c-s}
X^\mu_\mu&=L\,,
\\
{\zeta_1}^\mu_{a}\equiv\epsilon^{\mu\nu\rho}\left[\eta_{ac}\omega^c_{\nu,\rho}+\frac12\epsilon_{abc}\left(\omega^b_\nu\omega^c_\rho+\lambda e^b_\nu e^c_\rho\right)\right]&=0\,,
\\
{\zeta_2}^\nu_{c}\equiv\epsilon^{\mu\nu\rho}\left[ \eta_{ac} e^a_{\mu,\rho}-\epsilon_{abc}e^a_\mu\omega^b_\rho-\gamma_\rho\eta_{ac}e^a_\mu\right]&=0\,.
\end{align}
The two last equations are constraints because they do not depend on the second derivatives of the fields, and they define the submanifold $\P_1\subset\P$. The tangency conditions are:
\begin{align}
X_\sigma({\zeta_1}^\mu_{a})=\epsilon^{\mu\nu\rho}\left[\eta_{ac}G^c_{\nu,\rho\sigma}+\epsilon_{abc}\left(\omega^b_\nu\omega^c_{\rho,\sigma}+\lambda e^b_\nu e^c_{\rho,\sigma}\right)\right]\Big\vert_{\P_1}&=0\,,
\\
X_\sigma({\zeta_2}^\nu_{c})=\epsilon^{\mu\nu\rho}\left[ \eta_{ac}F^a_{\mu,\rho\sigma}-\epsilon_{abc}e^a_{\mu,\sigma}\omega^b_\rho-\epsilon_{abc}e^a_\mu\omega^b_{\rho,\sigma}-\frac{\partial \gamma_{\rho}}{\partial x^\sigma}\eta_{ac}e^a_{\mu}-\gamma_\rho\eta_{ac}e^a_{\mu,\sigma}\right]\Big\vert_{\P_1}&=0\,.
\end{align}
These are compatible equations for ${F_e}^a_{\mu,\rho\sigma}$ and ${F_\omega}^c_{\nu,\rho\sigma}$, so the constraint algorithm finishes.

For the holonomic sections
\begin{equation} \bm{\psi}(x^\mu)=\Big(x^\mu, e^a_\mu(x^\mu), \omega_\rho^c(x^\mu), \derpar{e^a_{\sigma}}{x^\mu}(x^\mu), \derpar{\omega^c_{\sigma}}{x^\rho}(x^\mu),s^\alpha(x^\mu)\Big) \end{equation}
of ${\bf X}_\L$,
the equations \eqref{eq:c-s} become:
\begin{align}
\derpar{s^\mu}{x^\mu}=L\,,\qquad
{\zeta_1}^\mu_{a}(\psi)=0\,,
\qquad
{\zeta_2}^\nu_{c}(\psi)=0\,.
\end{align}
As expected, the last two equations are constraints. The tangency conditions lead to the following equations for sections:
\begin{align}
\epsilon^{\mu\nu\rho}\left[\eta_{ac}\frac{\partial^2{\omega}^c_{\nu}}{\partial x^\rho\partial x^\sigma}+\epsilon_{abc}\left(\omega^b_\nu\frac{\partial\omega^c_{\rho}}{\partial x^\sigma}+\lambda e^b_\nu \frac{\partial e^c_{\rho}}{\partial x^\sigma}\right)\right]\Big\vert_{\P_1}&=0\,,
\\
\epsilon^{\mu\nu\rho}\left[ \eta_{ac}\frac{\partial^2{e}^a_{\mu}}{\partial x^\rho\partial x^\sigma}-\epsilon_{abc}\frac{\partial e^a_{\mu}}{\partial x^\sigma}\omega^b_\rho-\epsilon_{abc}e^a_\mu\derpar{\omega^b_{\rho}}{x^\sigma}-\frac{\partial \gamma_{\rho}}{\partial x^\sigma}\eta_{ac}e^a_{\mu}-\gamma_\rho\eta_{ac}\derpar{e^a_{\mu}}{x^\sigma}\right]\Big\vert_{\P_1}&=0\,.
\end{align}

\subsubsection{Hamiltonian formalism}

For the multicontact Hamiltonian formalism we have $\mathcal{P}^*=J^{1*}\pi\times_M\Lambda^{2}(\Tan^*M) $, 
with coordinates $(x^\mu,\, e^a_\mu,\, \omega_\rho^c,\, p_a^{\mu\sigma},\, \pi^c_{\nu\rho},\,s^\alpha)$.
The image of the Legendre transformation leads to the following constraints:
\begin{equation}
    p_a^{\mu\sigma}=\frac{\partial L}{\partial e^a_{\mu,\sigma}}=0\,,\quad \pi_c^{\nu\rho}=\frac{\partial L}{\partial \omega^c_{\nu,\rho}}=2\epsilon^{\mu\nu\rho}\eta_{ac}e^a_\mu\,.
\end{equation}
These constraints define a submanifold ${\cal P}_0^*\subset{\cal P}^*$ and, as they are lineal relationships of the fields, the Lagrangian is almost-regular. Moreover, as they do not depend on $s^\mu$, we can use the coordinates $(x^\mu,e^a_\mu,\omega^c_\nu,s^\mu)$ on ${\cal P}_0^*$. 

The Hamiltonian function is 
\begin{equation}
H_0=-\epsilon^{\mu\nu\rho}\epsilon_{abc}\Big(e^a_\mu\omega^b_\nu \omega^c_\rho+\frac{1}{3}\lambda e^a_\mu e^b_\nu e^c_\rho\Big)+\gamma_\alpha\d s^\alpha \,,
\end{equation}
the Hamiltonian $3$-form is
\begin{align}
\Theta_{\mathcal{H}}^0=2\epsilon^{\mu\nu\rho}\eta_{ac}e^a_\mu\text{d}\omega^c_\rho\wedge\text{d}^2x_\nu-\left[\epsilon^{\mu\nu\rho}\epsilon_{abc}\Big(e^a_\mu\omega^b_\nu \omega^c_\rho+\frac{1}{3}\lambda e^a_\mu e^b_\nu e^c_\rho\Big)-\gamma_\alpha\d s^\alpha\right]\text{d}^3x+\d s^\mu\wedge \d^{m-1}x_\mu\,,
\end{align}
which is a premulticontact form and $\sigma_{\Theta_\H^0}=\gamma_\mu\,\d x^\mu$.

For an integral section
$\bm{\psi}(x^\mu)=\Big(x^\mu, e^a_\mu(x^\mu), \omega_\rho^c(x^\mu),s^\alpha(x^\mu)\Big)$,
the above equations become:
\begin{align}
\derpar{s^\mu}{x^\mu}&=L\,,
\\
\epsilon^{\mu\nu\rho}\left[\eta_{ac}\derpar{\omega^c_{\nu}}{x^\rho}+\frac12\epsilon_{abc}\left(\omega^b_\nu\omega^c_\rho+\lambda e^b_\nu e^c_\rho\right)\right]&=0\,,
\\
\epsilon^{\mu\nu\rho}\left[ \eta_{ac} \derpar{e^a_{\mu}}{x^\rho}-\epsilon_{abc}e^a_\mu\omega^b_\rho-\gamma_\rho\eta_{ac}e^a_\mu\right]&=0\,.
\end{align}

\section{Multicontact and premulticontact structures}
\label{append}

See \cite{LGMRR-2022} for an exhaustive exposition on multicontact and premulticontact structures and their application to describe
non-conservative (or action-dependent) first-order field theories.

Let $P$ be a manifold with $\dim{P}=m+N$ and $N\geq m\geq 1$, 
and two forms $\Theta,\omega\in\df^m(P)$ with constant rank.
If ${\cal D}\subset\Tan P$ is a regular distribution and $\Gamma({\cal D})$ denotes the set of sections of ${\cal D}$
(vector fields on ${\cal D})$); for every $k\in\Nat$, let
\begin{equation}
{\cal A}^k({\cal D})=\{ \alpha\in\df^k(P)\ \vert\ 
\innp{Z}\alpha=0\,;\, \forall Z\in\Gamma({\cal D})\}=
\{ \alpha\in\df^k(P)\ \vert\ 
\Gamma(\cal D)\subset\ker\,\alpha \} \,,
\end{equation}
where $\ker\alpha=\{ Z\in\vf(P)\, \vert\, \innp{Z}\alpha=0\}$.

The {\sl\textbf{Reeb distribution}} ${\cal D}^{\mathfrak{R}}\subset\Tan P$
associated with the couple $(\Theta,\omega)$ is defined,
defined, at every point ${\rm p}\in P$, as
\begin{equation}
{\cal D}_{\rm p}^{\mathfrak{R}}=\big\{ v\in(\ker\omega)|_{\rm p}\ \mid\ \innp{v}\dd\Theta_{\rm p}\in {\cal A}^{m}_{\rm p}(\ker\omega)\big\} \,.
\end{equation}
The set of sections of the Reeb distributions is denoted $\mathfrak{R}=\Gamma({\cal D}^{\mathfrak{R}})$
and its elements $R\in\mathfrak{R}$ are called
{\sl\textbf{Reeb vector fields}}:
\beq
\label{Reebdef}
\mathfrak{R}=\{ R\in\Gamma(\ker\omega)\,|\, \innp{R}\dd\Theta\in {\cal A}^{m}(\ker\omega)\} \,.
\eeq
Observe that  $\ker\omega\cap\ker\dd\Theta\subset{\cal D}^\mathfrak{R}$.
If $\omega$ is a closed form,
then $\mathfrak{R}$ is involutive.

\begin{definition}
\label{premulticontactdef}
The couple $(\Theta,\omega)$ is a {\sl \textbf{premulticontact structure}} if $\omega$ is a closed form and, for $0\leq k\leq N-m$,
we have that:
\begin{enumerate}[{\rm(1)}]
\item\label{prekeromega}
${\rm rank}\,\ker\omega=N$.
\item\label{prerankReeb}
${\rm rank}\,{\cal D}^{\mathfrak{R}}=m+k$.
\item\label{prerankcar}
${\rm rank}\left(\ker\omega\cap\ker\Theta\cap\ker\d\Theta\right)=k$.
\item \label{preReebComp}
${\cal A}^{m-1}(\ker\omega)=\{\innp{R}\Theta\ \vert \ R\in \mathfrak{R}\}$,
\end{enumerate}
The triple $(P,\Theta,\omega)$ is a {\sl \textbf{premulticontact manifold}}
and $\Theta$ is a {\sl \textbf{premulticontact form}} on $P$. 
The distribution
$\mathcal{C}\equiv\ker\omega\cap\ker\Theta\cap\ker\d\Theta$
is the {\sl \textbf{characteristic distribution}} of $(P,\Theta,\omega)$.

If $k=0$, the couple $(\Theta,\omega)$ is a {\sl \textbf{multicontact structure}},
$(P,\Theta,\omega)$ is a {\sl \textbf{multicontact manifold}}
and $\Theta$ is a {\sl \textbf{multicontact form}} on $P$.
\end{definition}

Given a (pre)multicontact manifold $(P,\Theta,\omega)$,
there exists a unique $1$-form
$\sigma_\Theta\in\df^1(P)$,
which is called the {\sl \textbf{dissipation form}}, verifying that
\begin{equation}
\sigma_\Theta\wedge\innp{R}\Theta=\innp{R}\dd\Theta \,,\quad \text{for every }R\in\mathfrak{R} \,.
\end{equation}
Then, we define the operator
\begin{align}
\bd:\df^k(P)&\longrightarrow \df^{k+1}(P)
\nonumber \\
\beta &\longmapsto\bd\beta=\dd \beta+\sigma_\Theta\wedge\beta\,.
\end{align}

\begin{theorem}
\label{cor:coordenadas}
Around every point ${\rm p}\in P$ of a premulticontact manifold $(P,\Theta,\omega)$, there exists a local chart of {\rm adapted coordinates} 
$(U;x^1,\dots,x^m,u^1\dots,u^{N-m-k},s^{1}\dots,s^{m},w^{1},\dots,w^{k})$ such that
\begin{align}
\ker\omega\vert_U&=\left<\frac{\partial}{\partial u^1},\dots, \frac{\partial}{\partial u^{N-m-k}},\frac{\partial}{\partial s^1},\dots, \frac{\partial}{\partial s^m},\frac{\partial}{\partial w^1},\dots, \frac{\partial}{\partial w^{k}}\right> \,, \\   D^\mathfrak{R}\vert_U&=\left<\frac{\partial}{\partial s^1},\dots, \frac{\partial}{\partial s^m},\frac{\partial}{\partial w^1},\dots, \frac{\partial}{\partial w^{k}}\right> \,, \\    \mathcal{C}\vert_U&=\left<\frac{\partial}{\partial w^1},\dotsc, \frac{\partial}{\partial w^{k}}\right>\,.
\end{align}
\end{theorem}

On these charts, the coordinates $(x^\mu)$
can be chosen in such a way that the form $\omega$ can be expressed as 
$\omega\vert_U=\dd x^1\wedge\dots\wedge\dd x^m\equiv\dd^m x$.

If $(P,\Theta,\omega)$ is a multicontact manifold, in the above chart of coordinates,
there exists a unique local basis
$\{ R_\mu\}$ of $\mathfrak{R}$ such that 
\begin{equation}
\innp{R_\mu}\Theta=\dd^{m-1}x_\mu\,.
\end{equation}
Moreover, $[R_\mu,R_\nu]=0$.
These vector fields $R_\mu\in\mathfrak{R}$ are the {\sl\textbf{local Reeb vector fields}} of the multicontact manifold  $(P,\Theta,\omega)$
in the chart $U\subset P$.
Furthermore, from \eqref{Reebdef}, we have that there are local functions $\Gamma_\mu\in \Cinfty(U)$ associated with the basis $\{ R_\mu\}$ which are given by
\begin{equation}
\innp{R_\mu}\d\Theta = \Gamma_\mu\,\omega\,, \quad \forall \mu\,,
\end{equation}
because $\innp{R_\mu}\d\Theta\in{\cal A}_{\rm p}^m(\ker\omega)=\langle\omega\rangle$.
As a consequence, the dissipation form can be locally expressed on these charts as 
\begin{equation}
\sigma_\Theta=\Gamma_\mu\dd x^\mu \,, 
\end{equation}
because $\sigma_\Theta\wedge\dd^{m-1}x_\mu=\Gamma_\mu\,\omega=\Gamma_\mu\, \dd^{m}x$, for every $\mu$.

If $(P,\Theta,\omega)$ is a premulticontact manifold, then
there exist local vector fields
$\{ R_\mu\}$ of $\mathfrak{R}$ such that $\mathfrak{R}=\left<R_\mu\right>+ \mathcal{C}$ and $\innp{R_\mu}\Theta=\dd^{m-1}x_\mu$. They are unique up to a term in the characteristic distribution. Moreover $[R_\mu,R_\nu]\in\Gamma(\mathcal{C})$.

Observe that, using the adapted coordinates,
the local Reeb vector fields are $\displaystyle R_\mu=\derpar{}{s^\mu}$.

The local structure of (pre)multicontact manifolds is stated in the following:

\begin{proposition}
Every (pre)multicontact manifold $(P,\Theta,\omega)$ is locally diffeomorphic 
to a fiber bundle $\tau\colon P\to M$,
where $M$ is an orientable manifold with volume form $\omega_{_M}$, and $\omega=\tau^*\omega_{_M}$.
\end{proposition}

This is the canonical model for (pre)multicontact manifolds 
and it is the situation which is interesting in field theories. 

\begin{definition}
\label{multicontactbundle}
\ben
\item
The couple $(\Theta,\omega)$ is a {\sl \textbf{multicontact bundle structure}} 
and $(P,\Theta,\omega)$ is said to be
a {\sl \textbf{multicontact bundle}} if: 
\begin{enumerate}[{\rm(1)}]
\item
${\rm rank}\,{\cal D}^{\mathfrak{R}}=m$.
\item
$\ker\omega\cap\ker\Theta\cap\ker\d\Theta=\{0\}$.
\item 
${\cal A}^{m-1}(\ker\omega)=\{\innp{R}\Theta\,,\ R\in \mathfrak{R}\}$.
\end{enumerate}
\item
The couple $(\Theta,\omega)$ is a {\sl \textbf{premulticontact bundle structure}}
and $(P,\Theta,\omega)$ is said to be
a {\sl \textbf{premulticontact bundle}} if, for $0< k\leq N-m$,
we have that:
\begin{enumerate}[{\rm(1)}]
\item\label{prerankReebbundle}
${\rm rank}\,{\cal D}^{\mathfrak{R}}=m+k$.
\item\label{prerankcarbundle}
${\rm rank}\left(\ker\omega\cap\ker\Theta\cap\ker\d\Theta\right)=k$.
\item \label{preReebCompbundle}
${\cal A}^{m-1}(\ker\omega)=\{\innp{R}\Theta\,,\ R\in \mathfrak{R}\}$,
\end{enumerate}
\een
\end{definition}

Furthermore, for field theories, the (pre)multisymplectic structures  satisfy an additional requirement:

\begin{definition}
If $(P,\Theta,\omega)$ is a (pre)multicontact manifold such that
\begin{equation}
\innp{X}\innp{Y}\Theta = 0 \,, \quad
\text{for every } X,Y\in\Gamma(\ker\omega) \,,
\end{equation}
then $(\Theta,\omega)$ is said to be a 
{\sl \textbf{variational (pre)multicontact structure}} and
$(P,\Theta,\omega)$ is a {\sl \textbf{variational (pre)multicontact manifold}}.
\end{definition}

The terminology comes from the fact that this last condition  is what is imposed to the multicontact form to ensure that the theory is variational and hence the field equations derive from a Lagrangian (see \cite{GLMR-2022}).

\bibliographystyle{abbrv}


\begin{thebibliography}{99}

{\small

\bibitem{Atiyah}
M. F. Atiyah,
``Convexity and commuting Hamiltonians'',
{\sl Bull. London Math.
Soc.} \textbf{14} (1), (1982).
(\url{ https://doi.org/10.1112/blms/14.1.1})

\bibitem{MW}
J. Marsden, A. Weinstein,
``Reduction of symplectic manifolds with symmetry'', {\sl Rep. Math. Phys.}, {\bf 5}(1) (1974) 121--130. 
(\url{https://doi.org/10.1016/0034-4877(74)90021-4}).

\bibitem{AM-78}
R.~Abraham and J.E.~Marsden,
{\it Foundations of Mechanics} ($2nd$ ed.),
Benjamin--Cummings, Redwood City CA, 1987.
(ISBN-10: 080530102X).

\bibitem{Ar}
{\rm V.I. Arnold},
{\it Mathematical Methods of Classical Mechanics},
Graduate Texts in Mathematics {\bf 60}. Springer-Verlag, New York, 1989.
(\url{https://doi.org/10.1007/978-1-4757-2063-1}).

\bibitem{N2020}
N. Román-Roy
``A summary on symmetries and conserved quantities of autonomous Hamiltonian systems'', {\sl J. Geom. Mech.}, {\bf 12}(3) (2020) 541-551. 
(\url{https://doi.org/10.3934/jgm.2020009}).

\bibitem{EMRV}
A. Echeverría-Enriquez; M. C. Muñoz-Lecanda; N. Román-Roy; C. Victoria-Monge,
``Mathematical foundations of geometric quantization'', {\sl Extracta Mathematicae}, {\bf 13} (2), (1998). 

\bibitem{LD-1996}
M. de León, D.M. de Diego
``On the geometry of non‐holonomic Lagrangian systems'', {\sl J. Math. Phys.}, {\bf 37}, (1996) 3389–3414. 
(\url{https://doi.org/10.1063/1.531571}).

\bibitem{Go-69}
C. Godbillon,
{\it G\'eom\'etrie Diff\'erentielle et M\'ecanique Analytique},
Hermann, Paris, 1969.
(ISBN 10: 2705656588).

\bibitem{LR-85}
M. de Le\'on and P.R. Rodrigues,
{\it Generalized Classical Mechanics and Field Theory. A Geometrical Approach of Lagrangian and Hamiltonian Formalisms Involving Higher Order Derivatives},
{\sl North-Holland Math. Studies} {\bf 112} 
{\sl Notes Pure Math.} {\bf 102}. 
North-Holland Publishing Co., Amsterdam, 1985.

\bibitem{LM-87}
P.~Libermann and C.~Marle.
{\it Symplectic Geometry and Analytical Mechanics}.
D. Reidel, Dordrecht, 1987.
(\url{https://doi.org/10.1007/978-94-009-3807-6}).

\bibitem{LR-89}
M. de Le\'on and P.R. Rodrigues,
{\it Methods of Differential Geometry in Analytical Mechanics},
{\sl North-Holland Math. Studies} {\bf 158} . 
North-Holland Publishing Co., Amsterdam, 1989.

\bibitem{holm1}
D.D. Holm, 
{\em Geometric Mechanics. Part I}.
Imperial College Press, London, 2011. 
(ISBN: 978-1848161955).

\bibitem{MR-2024}
M.C. Mu\~noz-Lecanda and N. Rom\'an-Roy,
{\it Geometry of Mechanics}, Advanced Textbooks in Mathematics, World Scientific (in press), 2024. 
(\url{https://doi.org/10.1142/q0490}).

\bibitem{Bravetti2017}
A.~Bravetti,
\newblock {``Contact Hamiltonian dynamics: the concept and its use''},
\newblock {\sl Entropy} {\bf 19}(10) (2017) 535.
(\url{https://doi.org/10.3390/e19100535}).

\bibitem{CG-2019}
J.~Cari{\~{n}}ena and P.~Guha,
\newblock ``Nonstandard Hamiltonian structures of the Li\'enard equation
and contact geometry'',
\newblock {\sl Int. J. Geom. Meth. Mod. Phys.} {\bf 16}(supp 01) (2019) 1940001.
(\url{https://doi.org/10.1142/S0219887819400012}).

\bibitem{BCT-2017}
A.~Bravetti, H.~Cruz, and D.~Tapias,
\newblock {``Contact Hamiltonian mechanics''},
\newblock {\sl Ann. Phys. (N.Y.)} {\bf 376} (2017) 17--39.
(\url{https://doi.org/10.1016/j.aop.2016.11.003}).

\bibitem{Lainz2018}
M.~de~Le{\'{o}}n and M.~La\'inz-Valc{\'{a}}zar ,
\newblock {``Contact Hamiltonian systems''},
\newblock {\sl J. Math. Phys.} {\bf 60}(10) (2019) 102902.
(\url{https://doi.org/10.1063/1.5096475}).

\bibitem{BLMP-2020}
A. Bravetti, M. de Le\'on, J.C. Marrero, and E. Padr\'on,
``Invariant measures for contact Hamiltonian systems: symplectic sandwiches with contact bread'',
{\sl J. Phys. A: Math. Gen.} {\bf 53}(45) (2020) 455205.
(\url{https://doi.org/10.1088/1751-8121/abbaaa}).

\bibitem{GG-2022}
K. Grabowska and J. Grabowski,
``A geometric approach to contact Hamiltonians and contact Hamilton--Jacobi theory'',
{\sl J. Phys. A: Math. Theor.} {\bf 55}(43) (2022) 435204.
(\url{https://doi.org/10.1088/1751-8121/ac9adb}).

\bibitem{GGMRR-2019b}
J.~{Gaset}, X.~{Gr\`acia}, M.C.~{Mu\~noz-Lecanda}, X.~{Rivas}, and N.~{Rom\'an-Roy},
``New contributions to the Hamiltonian and Lagrangian contact formalisms for dissipative mechanical systems and their symmetries'',
{\sl Int. J. Geom. Meth. Mod. Phys.} {\bf 17}(6) (2020) 2050090.
(\url{https://doi.org/10.1142/S0219887820500905}).

\bibitem{LGGMR-2022}
M.~de~Le{\'{o}}n, J. Gaset, X. Gr\`acia, M.C. Mu\~noz-Lecanda, X. Rivas,
``Time-dependent contact mechanics'',
{\sl Monatsh. Math.} {\bf 201}:1149--1183 (2023).
(\url{https://doi.org/10.1007/s00605-022-01767-1}).

\bibitem{RT-2023}
X. Rivas and D. Torres,
``Lagrangian--Hamiltonian formalism for cocontact systems'',
{\sl J. Geom. Mech.} {\bf 15}(1) (2023) 1--26.
(\url{https://doi.org/10.3934/jgm.2023001}).

\bibitem{LLLR-2022}
M.~de León, M.~Laínz, A.~López-Gordón, X.~Rivas,
``Hamilton--Jacobi theory and integrability for autonomous and non-autonomous contact systems'', {\sl J. Geom. Phys.} {\bf 187} (2023) 104787.
(\url{https://doi.org/10.1016/j.geomphys.2023.104787}).

\bibitem{GRL-2023}
J. Gaset, X. Rivas, and A. López-Gordón,
``Symmetries, conservation and dissipation in time-dependent contact systems'', 
{\sl Fortsch. Phys.} {\bf 71}(8-9) (2023) 2300048.
(\url{https://doi.org/10.1002/prop.202300048}).

\bibitem{CCM-2018}
F.~M. Ciaglia, H.~Cruz, and G.~Marmo,
\newblock {``Contact manifolds and dissipation, classical and quantum''},
\newblock {\sl Ann. Phys. (N.Y.)} {\bf 398} (2018) 159--179.
(\url{https://doi.org/10.1016/j.aop.2018.09.012}).

\bibitem{He-1930}
G. Herglotz, ``Ber\"uhrungstransformationen'', Lectures at the University of Gottingen, 1930.

\bibitem{Her-1985}
G. Herglotz, 
{\em Vorlesungen \"uber die Mechanik der Kontinua}. 
Teubner-Archiv zur Mathematik~{\bf 3};
Teubner, Leipzig, 1985.

\bibitem{Geiges2008}
H.~Geiges,
\newblock {\em {An Introduction to Contact Topology}},
\newblock Cambridge University Press, Cambridge, 2008.

\bibitem{Kholodenko2013}
A.L. Kholodenko,
\newblock {\em {Applications of Contact Geometry and Topology in Physics}},
\newblock World Scientific, Singapore, 2013.

\bibitem{Banyaga2016}
A.~Banyaga and D.F. Houenou,
\newblock {\em {A Brief Introduction to Symplectic and Contact Manifolds}},
\newblock World Scientific, Singapore, 2016.

\bibitem{Gc-73}
P.L. Garc\'ia,
``The Poincar\'e--Cartan invariant in the calculus of variations'',
{\sl Symp. Math.} {\bf 14} (1973) 219--246.

\bibitem{GS-73}
H. Goldschmidt and S. Sternberg,
``The Hamilton--Cartan formalism in the calculus of variations'',
{\sl Ann. Inst. Fourier Grenoble} {\bf 23}(1) (1973) 203--267.

\bibitem{Kij1979}
J. Kijowski and W.M. Tulczyjew, 
``Symplectic Framework for Field Theories'',
{\sl Lect. Notes Phys.} {\bf 170}, Springer-Verlag, Berlin-New York, 1979.
(\url{https://doi.org/10.1007/3-540-09538-1}).

\bibitem{Ald1980}
V. Aldaya and J.A. de Azc\'arraga,
``Geometric formulation of classical mechanics and field theory'',
{\sl Riv. Nuovo Cim.} {\bf 3}(1)
(1980) 1--66. (\url{https://doi.org/10.1007/BF02906204}).

\bibitem{book:Saunders89}
D.J. {Saunders}, 
{\it The Geometry of Jet Bundles}, London Math.
  Soc., {\sl Lect. Notes Ser.} {\bf 142}, Cambridge Univ. Press,
  Cambridge, New York, 1989.
 (\url{https://doi.org/10.1017/CBO9780511526411}).

\bibitem{EMR-96}
A. Echeverr\'\i a-Enr\'\i quez, M.C. Mu\~noz-Lecanda, and N. Rom\'an-Roy,
``Geometry of Lagrangian first-order classical field theories'',
{\sl Fortsch. Phys.} {\bf 44} (1996) 235--280.
(\url{https://doi.org/10.1002/prop.2190440304}).

 \bibitem{art:Echeverria_Munoz_Roman98}
A. Echeverr\'\i a-Enr\'\i quez, M.C. Mu\~noz-Lecanda, and N. Rom\'an-Roy,
``Multivector fields and connections: Setting Lagrangian equations in field theories'',
{\sl J. Math. Phys.} \textbf{39}(9) (1998) 4578--4603. 
(\url{https://doi.org/10.1063/1.532525}).

\bibitem{LMS-2004}
M. de Le\'on, D. Mart\'in de Diego, A. Santamar\'ia-Merino, 
``Symmetries in classical field theory'',
{\sl Int. J. Geom. Meth. Mod. Phys.} {\bf 1}(5) (2004) 651--710.
(\url{https://doi.org/10.1142/S0219887804000290}).

\bibitem{art:Roman09}
N. Rom\'{a}n-Roy,
``Multisymplectic Lagrangian and Hamiltonian formalisms of classical field theories'',
\textsl{Symm. Integ. Geom. Meth. Appl.} 
\textbf{5} (2009) 100. 
(\url{https://doi.org/10.3842/SIGMA.2009.100}).
 
\bibitem{LeSaVi2016}
M.~de~Le\'on, M.~Salgado, S.~Vilari\~no,
\emph{Methods of Differential Geometry in Classical Field Theories: $k$-Symplectic and $k$-Cosymplectic Approaches},
World Scientific, Hackensack, 2016.
(\url{http://doi.org/10.1142/9693}).

\bibitem{CCI-91}
J.F. Cari\~nena, M. Crampin, and L.A. Ibort,
``On the multisymplectic formalism for first order field theories'',
{\sl Diff. Geom. Appl.} {\bf 1}(4) (1991) 345--374.
(\url{https://doi.org/10.1016/0926-2245(91)90013-Y}).

\bibitem{LMM-96}
M. de Le\'on, J. Mar\'\i n-Solano, and J.C. Marrero,
``A Geometrical approach to classical field theories:  A constraint algorithm for singular theories'',
{\sl Proc.  New Develops. Diff. Geom.},
L. Tamassi, J. Szenthe eds., Kluwer Acad. Press, (1996) 291--312.
(\url{https://doi.org/10.1007/978-94-009-0149-0\_22}).

\bibitem{MS-98}
J.E. Marsden and S. Shkoller,
``Multisymplectic geometry, covariant Hamiltonians and nonlinear {\sc pde}s'',
{\sl Comm. Math. Phys.} {\bf 199}(2) (1998) 351--395.
(\url{https://doi.org/10.1007/s00220005050}).

\bibitem{EMR-99b}
A. Echeverr\'\i a-Enr\'\i quez, M.C. Mu\~noz-Lecanda, and N.
 Rom\'an-Roy, 
``Multivector field formulation of Hamiltonian
 field theories: Equations and symmetries'',
 {\sl J. Phys. A: Math. Gen.} {\bf 32}(48) (1999) 8461--8484.
 (\url{https://doi.org/10.1088/0305-4470/32/48/309}).

\bibitem{EMR-00b}
A. Echeverr\'ia-Enr\'iquez, M.C. Mu\~noz-Lecanda, and N. Rom\'an-Roy,
``Geometry of multisymplectic Hamiltonian first-order field theories'',
{\sl J. Math. Phys.} {\bf 41}(11)  (2000) 7402--7444.
 (\url{https://doi.org/10.1063/1.1308075}).

 \bibitem{Kru2002}
O. Krupkova,
``Hamiltonian field theory'',
{\sl J. Geom. Phys.} {\bf 43}(2--3) (2002) 93--132
(\url{https://doi.org/10.1016/S0393-0440(01)00087-0}).

\bibitem{Pau2002}
C. Paufler and H. R\"omer,
``Geometry of Hamiltonian $n$-vector fields in multisymplectic field theory'',
{\sl J. Geom. Phys} {\bf 44}(1) (2002)
52--69.
(\url{https://doi.org/10.1016/S0393-0440(02)00031-1}).

\bibitem{HK-04}
F. H\'elein and J. Kouneiher,
``Covariant Hamiltonian formalism for the calculus of variations with several variables: Lepage--Dedecker versus De Donder--Weyl'',
{\sl Adv. Theor. Math. Phys.} {\bf 8} (2004) 565--601.

\bibitem{GIMMSY}
M.J. Gotay, J. Isenberg, J.E. Marsden, and R. Montgomery, 
``Momentum maps and classical relativistic fields I: Covariant theory'', 
arXiv:physics/9801019 [math-ph] (2004).

\bibitem{GMM-2022}
F. Gay-Balmaz, J.C. Marrero, and N. Mart\'inez,
``A new canonical affine bracket formulation of Hamiltonian classical field theories of first-order'',
{\sl Rev. Real Acad. Cienc. Exact. Fis. Nat. Ser. A-Mat.} {\bf 118} 103 (2024). (\url{https://doi.org/10.1007/s13398-024-01603-1}).

\bibitem{book:binz}
E. {Binz}, J. {{{\' S}}niatycki}, and H. {Fischer}, 
\emph{Geometry of Classical Fields} ($3rd$ ed.), Elsevier, North Holland, 1988.

\bibitem{LGMRR-2022}
M.~de~Le{\'{o}}n, J. Gaset, M.C. Mu\~noz-Lecanda, X. Rivas, and N. Rom\'an-Roy,
``Multicontact formulation for non-conservative field theories'',
\newblock {\sl J. Phys. A: Math. Theor.} {\bf 56}(2) (2023) 025201.
(\url{https://doi.org/10.1088/1751-8121/acb575}).

\bibitem{Vi-2015}
L. Vitagliano,
``$L_\infty$-algebras from multicontact geometry'',
{\sl Diff. Geom. Appl.} {\bf 59} (2015) 147--165.
 (\url{https://doi.org/10.1016/j.difgeo.2015.01.006}).

\bibitem{Geor-2003}
B. Georgieva, R. Gunther, and T. Bodurov,
``Generalized variational principle of Herglotz for several independent variables. First Noether-type theorem'',
{\sl J. Math. Phys.} {\bf 44} (2003) 3911.
 (\url{https://doi: 10.1063/1.1597419}).

\bibitem{GasMas2022}
J. Gaset and A. Mas, 
``A variational derivation of the field equations of an action-dependent Einstein--Hilbert Lagrangian'', 
{\sl J. Geom. Mech.} {\bf 15}(1) (2023) 357--374. (\url{https://doi.org/10.3934/jgm.2023014}).
 
\bibitem{GLMR-2022}
J. Gaset, M. La\'inz, A. Mas, and X. Rivas,
``The Herglotz variational principle for dissipative field theories'', 
{\sl Geom. Mech.} {\bf 1}(2) (2024) 153--178.
(\url{https://doi.org/10.1142/S2972458924500060}).

 \bibitem{LPAF2018}
M.J. Lazo, J. Paiva, J.T.S. Amaral, and G.S.F. Frederico,
``An action principle for action-dependent Lagrangians: Toward an action principle to non-conservative systems'',
{\sl J. Math. Phys.} {\bf 59}(3) (2018) 032902.
(\url{https://doi.org/10.1063/1.5019936}).


\bibitem{GGMRR-2019}
J.~{Gaset}, X.~{Gr\`acia}, M.C.~{Mu\~noz-Lecanda}, X.~{Rivas}, and
  N.~{Rom\'an-Roy},
``A contact geometry framework for field theories with dissipation'',
\newblock {\sl Ann. Phys.} {\bf 414} (2020) 168092.
(\url{https://doi.org/10.1016/j.aop.2020.168092}).

\bibitem{GGMRR-2020}
J.~{Gaset}, X.~{Gr\`acia}, M.C.~{Mu\~noz-Lecanda}, X.~{Rivas}, and N.~{Rom\'an-Roy},
``A $k$-contact Lagrangian formulation for nonconservative field theories'',
{\sl Rep. Math. Phys.} {\bf 87}(3) (2021) 347--368.
(\url{https://doi.org/10.1016/S0034-4877(21)00041-0}).

\bibitem{Ri-2022}
X. Rivas, 
``Nonautonomous $k$-contact field theories'', 
{\sl J. Math. Phys.} {\bf 64}(3) (2023) 033507.
(\url{https://doi.org/10.1063/5.0131110}).

\bibitem{LRS-2024}
J. de Lucas, X. Rivas, and T. Sobczak,
``Foundations on $k$-contact geometry''.
 arXiv:2409.11001v1 [math.DG] (2024).

\bibitem{Gotay}
M.~{Gotay}.
``A multisymplectic framework for classical field theory and the calculus of variations II: space+time decomposition'',
{\sl Diff. Geom. Appl.} {\bf 4} (1991) 375--390.
(\url{https://doi.org/10.1016/0926-2245(91)90014-Z}).

\bibitem{dD-1930}
T. de Donder, 
\emph{Th\'eorie Invariantive du Calcul des Variations}.
Gauthier-Villars, Paris, 1930.

\bibitem{Weyl-1935}
H. Weyl,
``Geodesic fields in the calculus of variation for multiple integrals'',
{\sl Ann. Math.} {\bf 36} (1935) 607--629.
 (\url{https://doi.org/10.2307/1968645}).

\bibitem {Gimmsy2}
M.J. Gotay, J. Isenberg, J.E. Marsden, 
``Momentum maps and classical relativistic fields II: canonical analysis of field theories''. 
arXiv:0411032v1 [math-ph] (2004).

\bibitem{AB-1992}
A.M. Abrahams, D. Bernstein, D. Hobill, E. Seibel, and L. Smarr,
``Numerically generated black-hole spacetimes: Interaction with gravitational waves'', {\sl Phys. Rev. D} {\bf 45}(10)
(1992) 3544--3558. (\url{https://doi.org/10.1007/978-3-642-95732-1_2}).

\bibitem{cook-2000}
G.B. Cook,
\newblock {``Initial Data for Numerical Relativity''},
\newblock {\sl Living Rev. Relativ.} {\bf 3}(1): 5 (2000) .
(\url{https://doi.org/10.12942/lrr-2000-5. PMC 5660886. PMID 29142501}).

\bibitem{ErrastiDiez:2023gme}
V.~Errasti D\'\i{}ez, M.~Maier and J.~A.~M\'endez-Zavaleta,
``Constraint characterization and degree of freedom counting in Lagrangian field theory'',
{\sl Phys. Rev. D} \textbf{109}(2) (2024).
(\url{https://doi.org/10.1103/PhysRevD.109.025010}).

\bibitem{AA-78}
V. Aldaya and J.A. de Azc\'arraga,
``Vector bundles, $rth$-order Noether invariants and canonical symmetries in Lagrangian field theory'',
{\sl J. Math. Phys.} {\bf 19}(9) (1978) 1876--1880.
(\url{https://doi.org/10.1063/1.523905}).

 \bibitem{FS-2012}
M. Forger and B.L. Soares,
``Local symmetries in gauge theories in a finite-dimensional setting'',
{\sl J. Geom. Phys.} {\bf 62}(9) (2012) 1925--1938. 
(\url{https://doi.org/10.1016/j.geomphys.2012.05.003}).

\bibitem{GPR-2016}
J. Gaset, P.D. Prieto-Mart\'inez, and N. Rom\'an-Roy,
``Variational principles and symmetries on fibered multisymplectic manifolds'',
{\sl Comm. Math.} {\bf 24}(2) (2016) 137-152.
(\url{https://doi.org/10.1515/cm-2016-0010}).

 \bibitem{RWZ-2016}
L. Ryvkin, T. Wurzbacher, and M. Zambon,
``Conserved quantities on multisymplectic manifolds'',
{\sl J. Austral Math. Soc.} {\bf 108}(1) (2020) 120--144.
(\url{https://doi.org/10.1017/S1446788718000381}).

\bibitem{GR-2023}
A. Guerra IV and N. Rom\'an-Roy,
``More insights into symmetries in multisymplectic field theories'',
{\sl Symmetry} {\bf 2023}, 15, 390 (2023).
(\url{https://doi.org/10.3390/sym15020390}).

\bibitem{GR-2024}
A. Guerra IV and N. Rom\'an-Roy,
``Canonical lifts in multisymplectic De Donder-Weyl Hamiltonian field theories'',
{\sl J. Phys. A: Math. Theor.} {\bf 57}(33) (2024) 335203.
(\url{https://doi.org/10.1088/1751-8121/ad6654}).

\bibitem{GasMar21}
J. Gaset and A. Mar\'in-Salvador, 
``Application of Herglotz's variational principle to electromagnetic systems with dissipation'', 
{\sl Int. J. Geom. Meth. Mod. Phys.} {\bf 19}(supp 10) (2022) 2250156. (\url{https://doi.org/10.1142/S0219887822501560}).

\bibitem{Lazo2022}
J. Paiva, M.J. Lazo, and V.T. Zanchin,
``Generalized nonconservative gravitational field equations from Herglotz action principle'',
{\sl Phys. Rev. D} {\bf 105}(12) (2022)
124023.
(\url{https://doi.org/10.1103/PhysRevD.105.124023}).

\bibitem{LPF2019}
M.J. Lazo, J. Paiva, and  G.S.F. Frederico,
``Noether theorem for action-dependent Lagrangian functions: conservation laws for non-conservative systems'',
{\sl Nonlinear Dyn.} {\bf 97} (2019) 1125--1136.
(\url{https://doi.org/10.1007/s11071-019-05036-z}).

\bibitem{C1981} M. Crampin, “On the differential geometry of the Euler--Lagrange equations, and
the inverse problem of Lagrangian dynamics”, 
{\sl J. Phys. A: Math. Gen. } {\bf 14}(1981)  2567–
2575.
(\url{https://doi.org/10.1088/0305-4470/14/10/012}).

\bibitem{LGL-2021}
M. de Le\'on, J. Gaset, and M La\'inz-Valc\'azar, 
``Inverse problem and equivalent contact systems'', 
{\sl J. Geom. Phys.} {\bf  176} (2022) 104500. 
(\url{https://doi.org/10.1016/j.geomphys.2022.104500}).

\bibitem{GGKM-2024}
K. Grabowska, J. Grabowski, M. Kuś and G. Marmo, 
``Contactifications: a Lagrangian description of compact Hamiltonian systems'', 
{\sl J. Phys. A: Math.Theor.} {\bf  57} (2024) 395204.
(\url{https://doi.org/10.1088/1751-8121/ad75d8}).

\bibitem{Sloan:2024kzb}
D.~Sloan,
``Dynamical similarity in field theories'',
{\sl Class. Quant. Grav.}, 
 \textbf{42}, (2025) 045001.

\bibitem{Ca96a}
F. Cantrijn, L.A. Ibort, and M. de Le\'on,
``Hamiltonian structures on multisymplectic manifolds'',
{\sl Rend. Sem. Mat. Univ. Pol. Torino} \textbf{54}(3) (1996) 225--236.
(\url{https://www.researchgate.net/publication/233917958}).

\bibitem{IEMR-2012}
L.A. Ibort, A. Echeverr\'ia-Enr\'iquez, M.C Mu\~noz-Lecanda, and N. Rom\'an-Roy,
``Invariant forms and automorphisms of locally homogeneous multisymplectic manifolds'',
{\sl J. Geom. Mech.} {\bf 4}(4) (2012) 397--419.
(\url{https://doi.org/10.3934/jgm.2012.4.397}).

\bibitem{Aldaya_Azcarraga78_2}
V. Aldaya and J.A. de Azcarraga,
``Variational principles on $rth$ order jets of fiber bundles in field theory'', 
\textsl{J. Math. Phys.} {\bf 19}(9) (1978) 1869--1875.
(\url{https://doi.org/10.1063/1.523904}).

\bibitem{GMS-97}
G. Giachetta, L. Mangiarotti, and G. Sardanashvily,
{\it New Lagrangian and Hamiltonian Methods in Field Theory},
 World Scientif\/ic Publishing Co., Inc., River Edge, NJ, 1997.
 (\url{https://doi.org/https://doi.org/10.1142/2199}).

 \bibitem{Ka98}
I.V. Kanatchikov, 
``Canonical structure of classical field theory in the polymomentum phase space'',
{\sl Rep. Math. Phys.} {\bf 4}(1) (1998) 49--90.
(\url{https://doi.org/10.1016/S0034-4877(98)80182-1})

\bibitem{Capriotti}
S. Capriotti,
``Differential geometry, Palatini gravity and reduction'',
{\sl J. Math. Phys.} {\bf 55}(1) (2014) 012902.
(\url{https://doi.org/10.1063/1.4862855}).

\bibitem{Einstein}
A.~Einstein, 
``Einheitliche Fieldtheorie von Gravitation und Elektrizität'', 
{\sl Pruess. Akad. Wiss.} {\bf 414} (1925); A. Unzicker and T. Case, ``Translation of Einstein’s attempt of a unified field theory with teleparallelism'',
arXiv:physics/0503046 [physics.hist-ph] (2005).

\bibitem{Gaset:2018uwi}
J. Gaset and N. Rom\'an-Roy,
``New multisymplectic approach to the Metric-Affine (Einstein-Palatini) action for gravity'',
{\sl J. Geom. Mech.} {\bf 11}(3) (2019) 361--396. 
(\url{https://doi.org/10.3934/jgm.2019019}).

\bibitem{IZ}
C. Itzykson and J.B. Zuber,
{\it Quantum Field Theory},
McGraw-Hill, New York, 1980.
(ISBN: 978-0486445687).

\bibitem{GGR-2023}
J. Gomis, A. Guerra IV, and N. Rom\'an-Roy,
``Multisymplectic constraint analysis of scalar field theories, Chern-Simons gravity, and bosonic string theory'',
{\sl Nucl. Phys. B} {\bf 987} (2023) 116069.
(\url{https://doi.org/10.1016/j.nuclphysb.2022.116069}).

\bibitem{HaBu}
W.H. Hayt Jr. and J.A. Buck, 
{\it Engineering Electromagnetics} (6th ed.),
McGraw-Hill, New York, 2018. 

\bibitem{Salsa} 
S. Salsa, 
{\it Partial Differential Equations in Action} 
(3rd ed.), Springer, Switzerland, 2016.
(\url{https://doi.org/10.1007/978-3-319-15093-2}).

\bibitem{GRR-2022}
X.~{Gr\`acia}, X.~{Rivas}, and N.~{Rom\'an-Roy}.
``Skinner--Rusk formalism for $k$-contact systems'',
{\sl J. Geom. Phys.} {\bf 172} (2022) 104429.
(\url{https://doi.org/10.1016/j.geomphys.2021.104429}).

\bibitem{CasMu} 
M. Castrill\'on-L\'opez, J. Mu\~noz-Masqu\'e, ``The geometry of the bundle of connections'', {\sl Math. Z.} {\bf 236}(4) (2001) 797--811. 
(\url{https://doi.org/10.1007/PL00004852}).

\bibitem{LPAF2017}
M.J. Lazo, J. Paiva, J.T.S. Amaral, and G.S.F. Frederico,
``Action principle for action-dependent Lagrangians toward nonconservative gravity: Accelerating universe without dark energy'',
{\sl Phys. Rev. D} {\bf 95}(101501) (2017) (\url{https://doi.org/10.1103/PhysRevD.95.101501})

\bibitem{Sloan:2020taf}
D.~Sloan,
``New action for cosmology'',
{\sl Phys. Rev. D} \textbf{103}(4) (2021) 043524.
(\url{https://doi.org/10.1103/PhysRevD.103.043524})

\bibitem{Sloan:2021hwx}
D.~Sloan,
``Scale symmetry and friction'',
{\sl Symmetry} \textbf{13}(9) (2021) 1639.
 (\url{https://doi.org/10.3390/sym13091639})

\bibitem{Sloan:2022exs}
D.~Sloan,
``Herglotz action for homogeneous cosmologies'',
{\sl Class. Quant. Grav.} \textbf{40}(11) (2023) 115008.
 (\url{https://doi.org/10.1088/1361-6382/accef6})

\bibitem{Ibort:2016xoo}
A.~Ibort and A.~Spivak,
``On a covariant Hamiltonian description of Palatini's gravity on manifolds with boundary'',
arXiv:1605.03492 [math-ph] (2016).

\bibitem{Capriotti2}
S. Capriotti,
``Unified formalism for Palatini gravity'',
{\sl Int. J. Geom. Meth. Mod. Phys.} {\bf 15}(3) (2018) 1850044.
(\url{https://doi.org/10.1142/S0219887818500445}).

\bibitem{pons}
N. Dadhich and J. M. Pons,
``On the equivalence of the Einstein--Hilbert and the Einstein--Palatini formulations of general relativity for an arbitrary connection'', 
{\sl Gen. Rel. Grav.}, {\bf 44} (2012), 2337--2352.(\url{https://doi.org/10.1007/s10714-012-1393-9}).

\bibitem{Ba-1915}
H. Bateman,
``Some recent researches on the motion of fluids'', {\sl Monthly Weather Rev.} {\bf 43}(4) (1915) 163--170.
(\url{http://dx.doi.org/10.1175/1520-0493(1915)43<163:SRROTM>2.0.CO;2}).

\bibitem{ACLSS}
{\rm A. Anahory Simoes, L. Colombo, M. de León, M. Salgado, and S. Souto},
``Symmetry reduction and reconstruction in contact geometry and Lagrange--Poincaré--Herglotz equations'',
arXiv:2408.06892 [math-ph].

\bibitem{BBS}
K. Becker, M. Becker, and J.H. Schwarz,
{\it String Theory and M-Theory. A Modern Introduction},
Cambridge Univ. Press, Cambridge, 2006.
(\url{https://doi.org/10.1017/CBO9780511816086}).

\bibitem{WittenGravity}
E. Witten
``$(2+1)$-dimensional gravity as an exactly soluble system'',
{\sl Nucl. Phys. B} {\bf 311}(1) (1988) 46--78.
(\url{https://doi.org/10.1016/0550-3213(88)90143-5}).



}
\end{thebibliography}

\end{document}